\documentclass[12pt]{article}

\newenvironment{sciabstract}{%
\begin{quote} \bf}
{\end{quote}}

\usepackage{color, colortbl}
\usepackage{times}
\usepackage{graphicx}
 \usepackage{epstopdf}
\usepackage{subfig}
\usepackage{amsmath,stmaryrd,xparse,empheq}
\usepackage{lineno}
\usepackage{cite}
\newcommand{\SM}{Supplementary Materials}
\newcommand{\SN}{Supplemental Notes}
\newcommand{\MM}{Methods}
\newcommand{\bS}{{\bf S}}
\newcommand{\bF}{{\bf F}}
\newcommand{\bU}{{\bf U}}
\newcommand{\by}{{\bf y}}
\newcommand{\bx}{{\bf x}}
\newcommand{\bz}{{\bf z}}
\newcommand{\la}{{\lambda}}
\newcommand{\beq}{\begin{equation}}
\newcommand{\eeq}{\end{equation}}
\newcommand{\eps}{\varepsilon}
\newcommand{\scd}{Colorbar: ratio of deformed to undeformed density.}
\newcommand{\tred}{\textcolor{black}} 
\newcommand{\tblue}{\textcolor{black}} 
\definecolor{Gray}{gray}{0.9}
\DeclarePairedDelimiter{\jumpop}{\llbracket}{\rrbracket}

\NewDocumentCommand{\dgal}{sO{}m}{
    \IfBooleanTF{#1}
    {\dgalext{#3}}
    {\dgalx[#2]{#3}}
}

\NewDocumentCommand{\dgalext}{m}{
    \sbox0{
        \mathsurround=0pt 
        $\left\{\vphantom{#1}\right.\kern-\nulldelimiterspace$
    }
    \sbox2{\{}
    \ifdim\ht0=\ht2
    \{\kern-.625\wd2 \{#1\}\kern-.625\wd2 \}
    \else
    \left\{\kern-.7\wd0\left\{#1\right\}\kern-.7\wd0\right\}
    \fi
}

\NewDocumentCommand{\dgalx}{om}{
    \sbox0{\mathsurround=0pt$#1\{$}
    \sbox2{\{}
    \ifdim\ht0=\ht2
    \{\kern-.625\wd2 \{#2\}\kern-.625\wd2 \}
    \else
    \mathopen{#1\{\kern-.7\wd0 #1\{}
    #2
    \mathclose{#1\}\kern-.7\wd0 #1\}}
    \fi
}

\newcommand{\beginsupplement}{
         \setcounter{page}{1}
        \renewcommand{\thepage}{S\arabic{page}}
        \setcounter{table}{0}
        \renewcommand{\thetable}{S\arabic{table}}
        \setcounter{figure}{0}
        \renewcommand{\thefigure}{S\arabic{figure}}
        \renewcommand{\thesection}{S\arabic{section}} 
        \setcounter{equation}{0}
        \renewcommand{\theequation}{S\arabic{equation}}
     }

\makeatletter
\newcommand*\titleheader[1]{\gdef\@titleheader{#1}}
\AtBeginDocument{%
  \let\st@red@title\@title
  \def\@title{%
    \bgroup\normalfont\large\centering\@titleheader\par\egroup
    \vskip1.0em\st@red@title}
}
\makeatother

\title{\bf Cells exploit a phase transition  to \textcolor{black}{mechanically remodel the fibrous extracellular matrix}}
\titleheader{\textit{J. R. Soc. Interface} \textbf{18}: 20200823  \; https://doi.org/10.1098/rsif.2020.0823}
\author{Georgios Grekas,$^{1,*}$ Maria Proestaki,$^2$ Phoebus Rosakis,$^{3,4,*}$ Jacob Notbohm,$^2$\\ Charalambos Makridakis$^{3,4,5}$ \& Guruswami Ravichandran$^{6}$}

\begin{document}

\baselineskip24pt

\maketitle

\normalsize{\noindent$^{1}$Aerospace Engineering and Mechanics, University of Minnesota, Minneapolis, USA}\\
\normalsize{$^{2}$Department of Engineering Physics, University of Wisconsin-Madison, USA}\\
\normalsize{$^{3}$Department of Mathematics and Applied Mathematics, University of Crete, Greece}\\
\normalsize{$^{4}$Institute of Applied \& Computational Mathematics, Foundation for Research \& Technology-Hellas, Heraklion, Greece}\\
\normalsize{$^{5}$ Department of Mathematics, MPS, University of Sussex, Brighton, United Kingdom}\\
\normalsize{$^{6}$Division of Engineering and Applied Science, California Institute of Technology}

\begin{sciabstract}
\noindent Through mechanical forces, biological cells \textcolor{black}{remodel the surrounding collagen network, generating striking  deformation patterns}. Tethers---tracts of high densification and fiber alignment---form between cells, thinner bands emanate from cell clusters. While tethers facilitate cell migration and communication, how they form is unclear. Combining modelling, simulation and experiment, we show that tether formation is a densification phase transition of the extracellular matrix, caused by buckling instability of network fibers under cell-induced compression, featuring unexpected similarities with martensitic microstructures. Multiscale averaging yields a two-phase, bistable continuum energy landscape for fibrous collagen, with a densified/aligned second phase. Simulations predict strain discontinuities between the undensified and densified phase, which localises within tethers as experimentally observed. In our experiments, active particles induce similar localised patterns as cells. This shows how cells exploit an instability to mechanically remodel the extracellular matrix simply by contracting, thereby facilitating mechanosensing, invasion and metastasis. 
\end{sciabstract}

\numberwithin{equation}{section}  
\clearpage
\section{Introduction}{Biological cells remodel the extracellular matrix (ECM) either biochemically, producing and degrading collagen fibers, or mechanically, by exerting mostly contractile forces \cite{remodegrad}. Mechanical remodelling  of fibrous collagen ECM by cellular forces results in distinctive deformation patterns extending over long distances.} If the ECM were a typical elastic material, deformations due to cells contracting would decay with distance within a few cellular diameters \cite{Rudnicki2013,Notbohm2015}. Instead, the  observations of Weiss \cite{weiss} and Harris \& Stopak \cite{harris1981fibroblast,stopak1982connective} 
led the way \cite{korff1999,vader2009strain,shi2014rapid} in  showing dramatic spatial patterns of densification and fiber alignment, localised within tether-like bands joining distant cell clusters 
(Fig.~\ref{f1a}). Tethers were recently observed between individual cells \cite{vader2009strain,Notbohm2015} as well (Fig.~\ref{f1b}).  In addition,  radial hair-like emanations from each cluster of cells (Figs~\ref{f1a}, \ref{f1c}) were observed.These patterns are a central feature of \textcolor{black}{mechanical ECM remodelling} and its role in cell motility, invasion and intercellular communication: individual cells leave their cluster and move along a tether to neighbouring clusters \cite{harris1981fibroblast,stopak1982connective,vader2009strain,shi2014rapid}; single fibroblasts joined by a tether grow appendages along it toward each other \cite{Notbohm2015}. Cancer cells preferentially invade along densified regions of the ECM with high fiber alignment \cite{yu2011forcing,early,aligned,stroma,compressive,prov1,prov2,provdens}. \textcolor{black}{Specific  patterns of  ECM density and alignment  have been identified as biomarkers for breast cancer  \cite{stroma,prov1,prov2,provdens,aligned}.}

Along the axis of each tether, the ECM is stretched by as much as 40\%, but also compressed in the transverse direction to half or even a quarter of its original thickness \cite{shi2014rapid,ban2018}. Density can be 3 to 5  times higher within tethers than without. These deformations are not only severe, but also spatially localised within tethers \cite{harris1981fibroblast}, and along radial hairlike bands that issue from individual cell clusters \cite{stopak1982connective} (Fig.~\ref{f1c}). What is the mechanism underlying the formation of these patterns? Many studies focus on the behaviour of collagen in tension, and argue that alignment and densification are induced by tensile strain \cite{vader2009strain}, or are due to a stiffening nonlinearity of the stress-strain behaviour of the ECM in tension \cite{winer2009non,wang2014long}. At the same time, density within tethers exceeds the undeformed value by almost an order of magnitude \cite{shi2014rapid,ban2018}, which implies compressive strains at least  twice as large as tensile ones in magnitude (\SM: Compressive Stretch Estimate). While cells, by contracting, apply radial tensile forces to the ECM they adhere to, they also decrease their perimeter, thereby inducing compression in the circumferential direction \cite{Notbohm2015,Rosakis2015}. The ECM ligament between two cells, where a tether forms, is thus under axial tension and transverse compression, but the observed compressive strains are unexpectedly large \cite{shi2014rapid}. The degree of localisation of densification is also unexpected. Centimeter-scale tethers induced by multi-cell tissue explants have well defined boundaries, across which  density is virtually discontinuous \cite{harris1981fibroblast,stopak1982connective} (Fig.~\ref{f1a}). Numerous thinner radial bands issue from each explant. At scales comparable to fiber length, density localisation is gradual but still  pronounced \cite{Notbohm2015}, whereas  fewer radial bands emerge from each cell (Fig.~\ref{f1b}). Solitary cell clusters also produce radial bands (Fig.~\ref{f1c}) \cite{harris1981fibroblast,stopak1982connective}. Deformations are  strongly inhomogeneous in the angular coordinates and radial symmetry is broken. 
What is surprising in these observations is the persistent appearance of localised, inhomogeneous deformations, not only at the scale of fiber network inhomogeneity \cite{Notbohm2015}, but at macroscopic scales as well \cite{stopak1982connective}.  Current nonlinear continuum models do not seem to explain these patterns of localisation \cite{ban2018,wang2014long,Rosakis2015,malandrino2019}.

\textcolor{black}{Our model and simulations demonstrate that these phenomena are possible because of a material instability that stems from a special nonlinearity in mechanical behaviour of individual network fibers. Using  active particles, our experiments establish that these severe,  localised deformations can be brought about simply by mechanical forces, such as the ones due to cell contraction, without  involvement of biochemical factors, and they remain after forces are removed. This indicates that they are in fact a form of mechanical ECM remodelling \cite{compressive,early,remodelvisc}.}

The connection with instability was first established  through experiments and modelling  in fiber networks and open-cell foams \cite{Lakes1993} and later for fibrin \cite{Kim2015}. Rather like rubber bands, individual fibers support tensile forces and may stiffen with increasing tension \cite{van2006,hudson2010}, but buckle under compression, losing stiffness and eventually collapsing. Larger polyhedral groups of fibers buckle and collapse under compression \cite{Lakes1993}. These \tred{microscopic} buckling instabilities cause the appearance of \tred{macroscopic} bands of intense compressive deformation and high density, within which fibers are mostly buckled and compacted, alternating with regions of normal density and low compressive strain, where fibers are largely unbent and loosely arranged. These two region types (high- and low-density) are separated by sharp interfaces, across which strain and density jump discontinuously at the macroscopic scale.  This behaviour is suggestive of coexistence of a densified phase and an undensified phase. \tred{To understand this, we start from the buckling  behaviour of a single fiber, and use multiscale averaging  to obtain a higher dimensional energy landscape for general deformations. Surprisingly, this energy exhibits bistability in strain. The theory of mechanical, diffusionless, isothermal phase transitions predicts} the spontaneous  appearance of the distinctive patterns of localised densification and alignment that feature in ECM remodelling and serve as biomarkers for tumour invasion in fibrous ECM.

\section{Modelling the Energy Landscape of the Fibrous ECM} We model the nonlinear elastic energy landscape of  fibrous collagen ECM, starting from the relation  $S=S(\la)$ between force $S$ and effective stretch $\la=d/l_0$ for a single flexible elastic fiber; here $d$ is the distance between endpoints, $l_0$ is the relaxed (reference) length. See \MM: Model and Supplemental Fig.~\ref{sf1a} for details. We assume $S(\la)$ stiffens in tension (its slope increases with $\la>1$) as is common for biopolymers \cite{van2006,piechocka2010,hudson2010}.
  Aside from direct observation of buckling \cite{munster2013,kim2014,burkel2017}, very little is known about the post-buckling behaviour of individual collagen or fibrin fibers in compression \cite{simhadri2019}, as $\la$ decreases from $1$ toward $0$ (total collapse).
This depends on their  bending behaviour, which is difficult  to characterise,  because of  their inhomogeneous, hierarchical structure \cite{piechocka2010} as bundles of loosely connected fibrils. 
We choose $S(\la)$  as in  Fig.~\ref{f2a} so that there is stiffening in tension ($\la>1$) but loss of stiffness in compression ($\la<1$) because of buckling. An example  is $S(\la)=\mu(\la^3-1)$  (Fig.~\ref{f2a}).  The corresponding elastic energy of the fiber is then $w(\la)=\int_1^\la S(\la')d\la'$.
Suppose the undeformed 2D network has a uniform distribution of fiber orientations. When subjected to a 2D deformation, the  stretch $\la=\la(\theta)$  of a fiber making an angle $\theta$ with the principal axes of stretch in the reference configuration is $\la(\theta)=\sqrt{(\la_1 \cos\theta)^2
 + (\la_2 \sin\theta)^2}$, where $\la_1$, $\la_2$ are the principal stretches of the deformation (\SM: Principal Stretches). Summing over all orientation angles, following Treloar's approach \cite{treloar1975} and the virtual internal bond model \cite{vainchtein2002strain,gloria2014}, we find a continuum elastic 2D energy density of the network:
\begin{equation}
 \hat{W}(\la_1, \la_2) = \frac{1}{2\pi}\int_0^{2\pi} w \left(\sqrt{(\la_1 \cos\theta)^2
 + (\la_2 \sin\theta)^2} \right) d \theta,
 \label{int}
\end{equation}
see \MM: Model for more details and an explicit form Eq.~(\ref{W}) of the 2D energy. 

\paragraph{\tred{Instability in Uniaxial Compression.}}\tred{Surprisingly, the 2D energy density $\hat{W}$ has an instability, evident in Fig.~\ref{f2b}, the stress-strain relation in uniaxial compression $S_1(\la_1)=\partial {\bar W}(\la_1,1)/\partial\la_1$.} This relation  is not monotonic, but its slope becomes negative for $\la_1< 0.45$  (Fig.~\ref{f2b}). This is unexpected, since the original single fiber stress $S(\la)$ is monotone increasing. Nonetheless, with increasing compression as $\la_1$ decreases toward $0$,  the network loses stiffness because of buckling of more  fibers. Also, fibers reorient due to compression,  increasing the angle they make with the compression axis, so they support less load along the compression axis.  As a result, the stiffness eventually becomes negative (strain softening), triggering a compression instability. For a detailed explanation, see \SM: Stiffness Loss.

Physically, we expect that  increasing compression ($\la_1\to 0$) leads to densification, and fibers getting squeezed together, thereby resisting further compression and eventually restoring stability.  To account for \tred{resistance to crushing (due to  fiber volume), we  add a volume penalty term  to the energy \eqref{int} that  resists collapse to zero volume in the crushing limit as the volume ratio $J=\la_1\la_2\to 0$:
\begin{equation}W={\bar W}(\la_1, \la_2) =\hat{W}(\la_1, \la_2) +\Phi(\la_1\la_2),
 \label{pe}
\end{equation}
where the penalty function $\Phi(J)$  increases rapidly as the volume ratio $J=\la_1\la_2\to 0$ as shown in Supplemental 
 Fig.~\ref{sf1b}. We use the specific form given by Eq.~(\ref{Phi}). This restores positive stiffness at extreme compression as $\la$ approaches zero (Fig.~\ref{f2c}). Thus  the decreasing branch with negative slope in the $S_1$-$\la$ curve in  Fig.~\ref{f2c},  which is  unstable, is now sandwiched between two increasing branches. These correspond to two  stable phases: a low compression phase and a high compression, densified phase. At a suitable compressible stress $S_0<0$ (red line in Fig.~\ref{f2c} and Supplemental Fig.~\ref{sf3b}), there are three corresponding stretch states (red and blue dots in Figs~\ref{f2c}, \ref{sf3b}). The middle one is unstable. The  states $\la_d$ and $\la_u$ (red dots) are stable. These two states of compressive stretch $\la_d$ (densified) and $\la_u$ (undensified)  can coexist side by side in neighbouring parallel bands in the ECM, under the same compressive stress. Such banding deformations with alternating zones of high and low densification, separated by sharp interfaces normal to the compression axis, are observed in open-cell foams and fibrin \cite{Lakes1993,Kim2015}.} This bistable behaviour has much in common with that of shape-memory materials \cite{delaey,otsuka}.  

\paragraph{\tred{A Bistable Energy Density.}} In the model just described,  the densified phase is only stable under compressive stress \tred{and disappears upon removal of the latter}; this is observed in fully crosslinked collagen \cite{vader2009strain}. In uncrosslinked collagen, after compression is removed, part of the densified phase remains in the ECM \cite{vader2009strain,munster2013,vos2017,burkel2018,ban2018}. Various factors  contribute to this, e.g., adhesion and new crosslinking of fibers coming into contact due to densification,  subject to van der Waals, or other noncovalent types of attraction \cite{crosslinkreview,vos2017}. To  model this, we add a short range attractive adhesion potential $w_a(\la)$  (blue curve in Fig.~\ref{f2d}) to the single fiber buckling potential $w(\la)$  (green curve in Fig.~\ref{f2d}). \tred{This represents minimal modelling of new crosslinks developing during fiber proximity \cite{crosslinkreview,benetatoscrosslink,vavyloniscrosslink} that occurs during  collapse caused by buckling.  Crosslinks can be broken by sufficient stretching, but can be reformed by subsequent compression-induced fiber proximity \cite{reverscrossl}.
This is modelled here by the short-range adhesion potential $w_a(\la)$.} It renders fiber collapse energetically favourable at short distances of the endpoints ($\la$ near zero). The resulting single fiber potential $w_*(\la)=w(\la)+w_a(\la)$ is now a two-well potential  (red curve in Fig.~\ref{f2d}) with a new minimum corresponding to a collapsed fiber state ($\la=0$) that is stable under zero load. Replacing $w(\la)$ by $w_*(\la)$ in Eq.~(\ref{int}), we obtain a new bistable energy $W^*$ that has multiple  global minima
(Supplemental Figs~\ref{sf3a}, \ref{sf2a}) compared to the single minimum of the monostable energy $W$ (Supplemental  Figs~\ref{sf3b}, \ref{sf2b}). 
\tred{The new minimum, $\la_d$ in Fig.~\ref{sf3a}, corresponds to a crushed (densified) state that is now stable in the absence of compressive stress, modelling the situation where densely packed fibers (collapsed due to buckling) are held together by new crosslinks that develop between them, without the need of external compressive stress.}

\paragraph{\tred{Nonconvexity in 2D}}\tred{Next we consider the energy landscape in 2D. What sets both energy densities $W$ and $W^*$ apart from previous continuum models of fibrous ECM \cite{ban2018,Rosakis2015,wang2014long} is nonconvexity (Supplemental Figs~\ref{sf3}, \ref{sf2}) and the instability associated with it. This they share with non-convex nonlinear elastic models developed for  austenite-martensite phase transformations, twinning,  and the shape memory effect \cite{Knowles1978,Ball1987,Ball1999,Abeyaratne2006}. Physically, instability occurs because of stiffness loss and collapse in compression caused by fiber buckling at the microscopic level. Mathematically, this instability  is reflected in the fact that} both $W$ and $W^*$  suffer a loss of a property known as strong ellipticity \cite{ball1976,Knowles1976,Knowles1978} at some strains, whereby some higher dimensional measure of stiffness becomes negative \cite{rosakis1990}. See \SM: Ellipticity Loss, also Supplemental Figs~\ref{sf2d}, \ref{sf2e}.  

The bistable energy density  $W^*$  is a multi-well potential (Supplemental Figs~\ref{sf3a}, \ref{sf2a}).  \tred{This allows  the densified phase to be stable at zero stress, as has been observed in uncrosslinked collagen \cite{vader2009strain,munster2013,vos2017,burkel2018,ban2018}.}  Although the monostable energy density $W$ has a single minimum (Supplemental  Figs~\ref{sf3b}, \ref{sf2b}), the associated  Gibbs free energy $W(\bF)-\bS_0\cdot\bF$ is a double well potential for a special value of the  stress $\bS_0$, with minima similar to those of $W^*$ (Supplemental Fig.~\ref{sf2c}).  Thus under suitable compressive stress, the monostable energy exhibits bistable behaviour \tred{and coexistence of phases, but predicts the disappearance of the densified phase under zero stress, consistent with crosslinked collagen behaviour \cite{vader2009strain}}. 

\tred{For standard values of model parameters  used in our simulations (Supplemental Table S1), the minima (energy wells) of the 2D energy density $W^*(\la_1,\la_2)$ occur at the undeformed state $(\la_1,\la_2)=(1,1)$,  and at  the densified state $(\la^*_1,\la^*_2)=(0.2,1.06)$.} The latter  is a severe  compression with strain $\eps_1=\la^*_1-1=-0.8$, combined with a  small extension  $\eps_2=\la^*_2-1=0.06$ in the perpendicular direction. 
These two energy wells  correspond to the undensified phase  and the densified phase of the material, respectively. \tred{The densified phase involves a five-fold  increase in density with ratio $1/(\la^*_1\la^*_2)\approx4.7$}. 

\tred{The  energy minima  just described satisfy geometric compatibility conditions \cite{Ball1987} allowing the two corresponding strain states  to coexist in the material, separated by a coherent phase boundary of strain discontinuity \cite{Ball1999}. These conditions are described in \SM: Compatibility.  This means that bands of the densified state can coexist in equilibrium, side by side within the undensified state.}  An additional minimum at  $(\la_1,\la_2)=(0.45,0.45)$ is not compatible with the undeformed state $(1,1)$; see \SM: Compatibility. This explains why it is never encountered in our simulations.

\paragraph{\tred{Additional Energy Contributions.}} The elastic energy of a deformation $\by(\bx)$ is
\beq E\{\by\}=\int_\Omega W(\nabla\by(\bx))d\bx\label{w}\eeq
where $\Omega$ is the undeformed region occupied by the ECM.
Going from the discrete energy of a random fiber network of characteristic fiber size $\eps$ to the continuum energy in an asymptotic expansion \cite{bardenhagen1994,vainchtein2002strain} as $\eps\to 0$, one obtains Eq.~(\ref{w}) as the first term, followed by a higher gradient term quadratic in $\eps\nabla\nabla\by(\bx)$. We choose a simple form of isotropic higher gradient energy \cite{triantafyllidis1986,ball2011local}
\beq G_\eps\{\by\}= {\eps^2\over 2}\int_\Omega|\nabla\nabla\by(\bx)|^2 d\bx.\label{grad}\eeq

To model contractile  multi-cell clusters (explants \cite{harris1981fibroblast,stopak1982connective}, acini \cite{shi2014rapid,ban2018}), we let $\Omega$ contain initially circular cavities. At the boundary of each cavity, forces can be exerted onto the ECM.  \tred{In contrast to previous studies \cite{Notbohm2015,wang2014long,ban2018,ronceray2016fiber,Lesman}, cellular clusters are not assumed in our model to remain circular after deformation, and their centres are allowed to  move, since they are deformable,  and attached  to a deforming ECM.}   Accordingly, 
 each cellular cluster is represented by a distribution of linear springs, connecting each cavity boundary point to a central point which is free to move (Supplemental Fig.~\ref{sf1c}). This contributes to the energy a term 
 \beq C\{\by;\bz_1,\ldots,\bz_n\}=\sum_{i=1}^n\int_{\Gamma_i} {k\over 2}\Bigl[ |\by(\bx)-\bz_i|-(R-u_0)\Bigr]^2 ds_\bx,\label{spr}\eeq
 where $\Gamma_i$ is the boundary  of the $i^{\rm th}$ cavity ($i=1,\ldots,n$), $k$ is the spring stiffness,
 $\bz_i$ the variable centre position of the $i^{\rm th}$ cluster, $R$ the undeformed radius, and  $u_0$ the cellular spring contraction. \tred{Our soft model for clusters allows them to contract and exert radial forces to the ECM but change shape as well. The model mimics the radial contracting actomyosin network of  contractile cells, e.g., \cite{rape}. Simulations yield deformed cluster shapes consistent with observations: egg-shaped and pointed towards the tether (observation Fig.~\ref{f1a} reproduced from \cite{stopak1982connective}; simulations Fig.~\ref{f1e}). Our active particles are an order of magnitude stiffer than typical cellular clusters \cite{matzelle2003,acinimodulus}. They  are modelled as stiff spheres, so that they do not deviate much from sphericity; see \MM: Model for Active Particles, Eq. \eqref{spr2},  Supplemental Fig.~\ref{sf1d}, and \SM: Simulation Parameters.}

 The total energy  to be minimised is the sum of Eqs~(\ref{w}), (\ref{grad}) and (\ref{spr}):
 \beq\Phi\{\by;\bz_1,\ldots,\bz_n\}=E\{\by\}+G_\eps\{\by\}+C\{\by;\bz_1,\ldots,\bz_n\}.\label{ener}\eeq

 \section{Experiments and Simulations  of ECM Deformations Induced by Active Particles}
\tblue{How can we ensure that the densification observed in experiments  \cite{vader2009strain,shi2014rapid,ban2018} is not due to biochemical factors (e.g., cells depositing more fibers) but entirely mechanical? In these experiments,  cells are concentrated within well defined clusters while the densification takes place, and they only migrate away from these clusters after densification has occurred. So there are no cells (that could deposit new fibers) in the densified zones until the latter have fully developed.  That the densification is associated with compressive strain can be directly deduced from [\!\cite{shi2014rapid} supplemental video sm14], where the  evolving compression is recorded and the compressive strain can be estimated  from the deformation of gridlines deposited on the ECM.}
  
 \tblue{To further ascertain that tether formation and concomitant densification and strain localisation  can occur solely due to mechanical forces, not to other biochemical factors,  we embedded  active hydrogel microspheres \cite{burkel2017} into  Collagen I  instead of cells.} These PNIPAAm particles undergo a temperature-induced phase transition, causing their radius to contract by as much as 60\%  when heated above $32^\circ$C (\MM: Experiments with Active Particles).  Advantages of this are that we can control the amount of contraction via temperature, and even cause reverse re-expansion by cooling, without recourse to chemical means. See \MM: PNIPAAm Particle Generation  for more details.  \tblue{As we report below, the densification patterns caused by active particle compression are consistent with those caused by cell clusters  \cite{vader2009strain,shi2014rapid,ban2018}, providing strong evidence for their mechanical origin}.

After developing a  finite element scheme that can handle deformation gradient discontinuities (\MM: Numerical Method) we minimise the total energy,  Eq.~(\ref{ener}), with respect to the deformation field $\by(\bx)$ and the particle/cluster center positions $\bz_i$. The simulated ECM domain contains one or more initially circular cavities of radius $R$, representing  cell clusters,  or the active contracting particles in our experiments. Choosing $u_0>0$ in Eq.~(\ref{spr}) contracts the natural length of the springs comprising the cell cluster model from $R$ to $R-u_0$,  thus exerting contractile centripetal forces onto the cavity boundaries. For simulations of active particles we use Eq.~(\ref{spr2}) in the energy \eqref{ener}.

  \paragraph{\tred{Tethers.}}The most striking feature of our numerical solutions  involving two contracting cavities is the spontaneous appearance of a tether, a zone of high density joining the two clusters \cite{harris1981fibroblast,stopak1982connective,shi2014rapid,ban2018}, and thinner hairlike bands emerging from each cluster in the radial direction \cite{harris1981fibroblast,stopak1982connective}, tapering off and terminating  within the domain. \tred{This is the ubiquitous morphology reported previously (Figs~\ref{f1}a-\ref{f1c}) and encountered in our simulations for a wide range of model parameters (Figs~\ref{f1}d-\ref{f1f}).   In these simulations,}  within each tether and radial band, stretches are in the densified phase; outside they take values in the undensified phase. Density is discontinuous across the boundary of tethers and radial bands; the ratio of densities outside and inside the tether is in the range 3-5. Within each tether, there is tension along the tether axis and compression in the transverse direction. The compressive stretch is discontinuous across the tether boundary and as low as $20\%$ (compressive strain is as high as $80\%$); the tensile stretch is much smoother although it is higher within tethers, with tensile strains as high as $30\%$. This is consistent with stretch values observed in experiments \cite{shi2014rapid,ban2018}. \tred{The large discontinuous jump in compressive stretch} is related to the fact that energy bistability occurs in compression. \tred{Compatibility of discontinuous strains  plays an important role. Essentially, compatibility restricts possible discontinuities of strains (or stretches) across a phase boundary  to ensure that the displacement remains continuous across it. It allows the compressive  stretch normal to the tether boundary to be discontinuous across it, but the tangential tensile stretch is restricted  to be continuous, in order to ensure displacement continuity across a phase boundary (see \SM: Compatibility).} This is consistent with experiments [\cite{ban2018} Figs S2, S5] and is observed  in our simulations  (Figs~\ref{ff3a}, \ref{ff3b}).  

How close do two active particles have to be (Fig.~\ref{f3}), and how much must they contract in order to form a tether joining them?  \tred{We performed multiple simulations of a pair of particles. We independently varied particle contraction and  distance between particles. The simulations provided a separatrix curve of average particle radial strain versus distance between particles (blue curve in Fig.~\ref{f1h}). Above this curve, our model predicts that a tether forms joining the two particles;  below the curve no tether will form.  Data from our experiments  agreed with this prediction: blue points in Fig.~\ref{f1h} are data points from particle pairs with a  tether observed joining them, red points correspond to pairs without a  tether between  them. }

 \paragraph{\tred{Shape Change  and Particle Motion.}}
Departing from common practice \cite{Notbohm2015,wang2014long,ban2018,ronceray2016fiber,Lesman}, we allow  clusters to change shape and move during ECM deformation in our model. \tred{See Eq.~\eqref{spr},  also \MM: Model for Active Particles and Supplemental Fig.~\ref{sf1d}.}
 In experiments  \cite{harris1981fibroblast,stopak1982connective}, isolated explants remain roughly circular (Fig.~\ref{f1c}) after contraction. In contrast, neighbouring clusters connected by a tether lose circularity \cite{harris1981fibroblast,stopak1982connective} and become egg-shaped with pointed ends toward each other (Fig.~\ref{f1a}) due to tether tension. The parameter controlling shape change for clusters and particles is the stiffness $k$ in  \eqref{spr}, \eqref{spr2}. Active particles are an order of magnitude stiffer than cell clusters  \cite{matzelle2003,acinimodulus}. For low values of cellular spring stiffness, $k=1$ in  Eq.~\eqref{spr}, our model predicts similar shape changes with pointed end where the tether makes contact with the cluster
 (compare  2nd and 3d panels in Fig.~\ref{f1e} to shape of explants in Fig.~\ref{f1a}).  In order for simulated explants to  remain circular after contraction as typically assumed,  we must choose an artificially high stiffness  ($k=100$ in  Eq.~\eqref{spr}, 1st panel in Fig.~\ref{f1e}). See \SM: Simulation Parameters for stiffness values.
 
Contracting active particles in our experiments \tred{(Figs~\ref{f3a}, \ref{f3c})} and simulations \tred{(Figs~\ref{f3b}, \ref{f3d})} move towards each other by as much as 10\% of their original distance  (Fig.~\ref{f1g}), while remaining nearly spherical. These particles are roughly an order of magnitude stiffer than cell clusters. \tred{To simulate them, we use a high spring stiffness value $k=10$ (compared to $k=1$ for cell clusters) in \eqref{spr2}. }

 \paragraph{\tred{Tether Splitting.}}
 
\tred{ In our experiments, we observe differences in morphology of tethers between active particles. A tether is sometimes separated into thinner parallel bands (Fig.~\ref{f3a}). Other tethers are in the form of  uniform bands making full contact with particles (Fig.~\ref{f3c}). Mammary acini 
  \cite{vader2009strain,shi2014rapid,ban2018} are typically joined by uniform tethers. The tether continues into a layer of the densified phase enveloping each acinus. }
  
 Why this difference in morphologies? Unexpectedly the answer comes from our understanding of austenite-martensite transformations, where the Bain strain at the martensitic energy minimum \cite{delaey} is incompatible with zero-strain austenite \cite{Ball1987,Ball1999} (see \SM: Compatibility). Namely, these two minimal-energy strains cannot occur on either side of a phase boundary (strain discontinuity) without causing a mismatch in displacements. This forces splitting and tapering of twin bands in a crystal near an incompatible boundary \cite{james1995tipsplit}. An example of such split martensitic twins  is shown  in Fig.~\ref{f3e}.  Here also, energy minimisation compels  strains not to stray far from energy-density minima. \tred{The active particles in Fig.~\ref{f3a} contracted by $u_0/R=38$\%. The azimuthal stretch $\la_\theta=1-u_0/R=0.62$ imposed at the  particle boundary by contraction is incompatible with the stretch $\la^*_1=0.2$ corresponding to the densified-phase energy well. To avoid this mismatch while maintaining displacement continuity, the tethers splits into narrow bands to minimise contact with the particle boundary (\tred{Experiment}:Fig.~\ref{f3a}, simulation  \ref{f3b}). The particles of Fig.~\ref{f3c} contracted more ($52\%$). The resulting tether does not split, but is in full contact with the left particle (torn fibers near the right particle keep the tether from extending to the left particle boundary). A corresponding simulation (Fig.~\ref{f3d}) confirms the absence of tether splitting near the particles  for this level of contraction. 
 For even higher  contraction levels of $80\%$, possible for acini  \cite{vader2009strain,shi2014rapid,ban2018}  but not for active particles, the azimuthal stretch $\la_\theta\approx \la^*_1$ (is compatible with the energy minimal stretch $ \la^*_1$). In hypothetical simulations of this contraction level, the densified phase is in full contact and envelopes the particle because of enhanced compatibility (Fig.~\ref{f3g}). These observations support the hypothesis that tether morphology is decided by strain compatibility as in other types of coherent phase transitions (martensitic).}

 \paragraph{\tred{Contracting vs Expanding Particles.}} What causes the densified phase to split into thin radial bands around particles, \tblue{such as ``hairs'' emanating from clusters in Fig.~\ref{f1c} (reproduced from \cite{harris1981fibroblast,stopak1982connective}) and  in Fig.~\ref{f4b}?}  A contracting inclusion induces radial tension and circumferential compression. Since the energy is bistable in compression,  phase change tends to occur along the direction of compression, with phase boundaries normal to it, roughly along radial lines (Fig.~\ref{f4b}). Instead, an expanding particle would create radial compression, hence a circumferential phase boundary and densified ring enveloping the particle (Fig.~\ref{f4e}). The stark  contrast between the ECM's response to contracting and expanding particles is seen in our experiments  (contracting particle Fig.~\ref{f4b} vs expanding  Fig.~\ref{f4e}) and for the first time captured by a continuum model simulation (contracting Fig.~\ref{f4c} vs expanding particle  Fig.~\ref{f4f}). See also the densified ring surrounding a re-expanded particle in \tred{experiment Fig.~\ref{f5c}, simulation \ref{f5f})}. \textcolor{black}{Remarkably, this explains the sharply defined circumferential layer of densification typically observed surrounding expanding tumour spheroids; see, e.g.,  [\! \cite{compressive} Figs~1e,f].}

 \paragraph{\tred{Splitting Hairs. Numerical Mesh Dependence.}} \tred{As noted above, for contracting particles, the radial orientation of interfaces forces the layer of densification around each particle to split into roughly radial bands. Another factor that promotes even finer splitting is compatibility of strains, in order to lower the cost of matching displacements at the particle boundary. This is the same phenomenon as splitting of a tether into finer bands (see Tether Splitting above). These effects are responsible for the radial hair morphology around each particle (Figs~\ref{f1}, \ref{f3}). As a result, each contracting particle is surrounded by a mixture of the undensified and densified phases, the latter  occurring within thin, roughly radial, hairlike bands. These spatially fine phase mixtures establish a further connection with the nonlinear elastic theory of phase transitions  in crystals \cite{Ball1987,Ball1999}. In  multi-well energy  minimisation problems for martensitic transitions, it is known that  the energy cannot reach a global minimum because of strain incompatibility  \cite{Ball1987,dacorogna2007direct}.  Decreasing the energy is facilitated by  refinement: increasing the number of bands in alternating phases, while decreasing their thickness, in effect decreases the mismatch in displacements that costs energy to overcome.  In theory, the energy approaches an infimum only in the limit of infinite refinement. This explains observed finely twinned martensitic microstructures \cite{Ball1987,bhattacharya2003microstructure}. In computations \cite{luskin1996,healey2007} this phenomenon manifests itself as an increase in the number and spatial frequency of  twin boundaries with increasing computational finite element mesh resolution. Similarly, when we decrease the computational mesh size in our simulations,  the number of radial bands issuing from each cluster increases and their thickness decreases (Fig.~\ref{f1f}).}
 
To show that this \tred{finite element mesh} dependence is not an artifact of the numerical method, we add the higher-gradient term Eq.~(\ref{grad}) to the energy. This introduces a length scale proportional to $\eps$, an additional material parameter related to characteristic fiber length, bending stiffness and other fiber network parameters \cite{vainchtein2002strain}. For larger  $\eps$, there are fewer, thicker radial bands  (Fig.~\ref{f1d}). Above a critical value of $\eps$, they disappear, but the tether persists.  The presence of this term has a smoothening effect as it penalises high strain gradients; strain discontinuities are replaced by transition layers with thickness of order $\eps$ \cite{triantafyllidis1986,ball2011local}. \tred{This eliminates mesh dependence at numerical mesh sizes below  $\eps$, which controls the scale of the phase mixture. In practice we expect $\eps$ to be of the order of the free fiber length in the network.} Accordingly, many hairlike bands are experimentally observed (Fig.~\ref{f1a}, \ref{f1c}) issuing from millimetre-size explants  \cite{stopak1982connective},  (where $\eps$ is very small compared to explant size), while only a few  (Fig.~\ref{f1b}) from single micron-scale fibroblasts \cite{Notbohm2015}.

 \paragraph{\tred{Residual Densified Zones and Inelasticity.}}Do the densification patterns in the ECM  persist after cell-exerted contractile forces cease \cite{vader2009strain,ban2018}? \tred{In our experiments,} active particles (Fig.~\ref{f5a}) contracted upon heating to $39^\circ$C, causing densified tethers and radial hairs to appear (Fig.~\ref{f5b}), then  expanded to their original radius upon cooling back to $26^\circ$C (Fig.~\ref{f5c}) thus providing a controllable method of performing one---or several---loading/unloading cycles.  
 When  contraction was reversed \tred{in our experiments,} some residual tethers remained  (Fig.~\ref{f5c}),  but they were thinner and less prominent than ones appearing during particle contraction (Fig.~\ref{f5b}). Many radial bands issuing from particles largely disappeared. 

We performed \tred{contraction-expansion cycle simulations (Figs~\ref{f5d}-\ref{f5f})} of an active particle pair undergoing quasistatic, gradual contraction, followed by  gradual expansion to original size, using the bistable energy density (we matched initial  diameters, distance and contractile strains of both particles; see Table S1). Experimental observations  \tred{(Figs~\ref{f5a}-\ref{f5c})} confirmed by our model simulations \tred{(Figs~\ref{f5d}-\ref{f5f})} included: (i) a  residual   tether \tred{remaining upon re-expansion, which  is weaker, thinner and disjointed compared to the tether appearing during contraction} (ii) most radial bands disappearing, (iii) a new circumferential layer of densification appearing upon re-expansion on each particle boundary  \tred{(compare simulation Fig.~\ref{f5f} to experimental Fig.~\ref{f5c})}. 

Is plasticity of collagen necessary for tether/densification pattern formation \cite{vader2009strain,ban2018,malandrino2019}? To answer this, we also performed cycle simulations using the monostable energy density  $W$. Typical  tethers and radial hairs  appeared during particle contraction, but  disappeared upon re-expansion to original size. This shows that the appearance of  localised densification bands is due to the elastic microbuckling  instability, which is accounted for in the monostable energy density \tred{and does not require inelasticity}.  \tred{In our model,} whether these patterns persist once contractile forces are removed, depends on whether the densified phase is stable under zero stress. This is the case for the bistable energy density, with its equally stable minima, but not for the monostable energy density. 

This difference is consistent with \tred{experiments} in fully crosslinked collagen, where the densified phase is only observed during application of compressive stress \cite{vader2009strain}, versus uncrosslinked collagen, where part of the densified phase remains after compression is removed \cite{vader2009strain,munster2013,vos2017,burkel2018,ban2018}.  
    Analogously, some austenite-martensite transitions occur under stress only, with martensite disappearing upon its removal; this behaviour is called pseudoelasticity \cite{delaey,otsuka}, described by monostable-type  nonconvex elastic energy models. Under different circumstances, such as temperature, martensite persists after load removal, as captured by a bistable energy \cite{delaey,bhattacharya2003microstructure,abchujam}. \tred{The description of both the pseudoelastic and residual  transitions, and the present densification transition, does not require plasticity type models, as hysteresis and residual deformations are accounted for by phase transition models \cite{delaey,bhattacharya2003microstructure,abchujam}. This is demonstrated by our simulations Figs~\ref{f5d}-\ref{f5f}. }

 \paragraph{\tred{Tether-Crack Interaction and Tether Networks.}} Our model captures complex experiments of Shi et al \cite{shi2014rapid},  where a cut (crack) is made between two acini in order to interfere with tether formation \tred{(\tred{experiment:} dotted line in Fig.~\ref{f5g},  reproduced from \cite{shi2014rapid})}. The original tether disappears;  instead tethers form that bypass the crack by going around its corners \tred{ (arrows in Fig.~\ref{f5g}). Our simulation (Fig.~\ref{f5h}) captures this phenomenon, with predicted tethers bypassing the opened crack.}

Multiple contractile cell clusters or acini generate a network of tethers, which brings about extensive remodelling of the ECM. See., e.g., [\!\! \cite{shi2014rapid}, Fig.~1D].  \tred{A  simulation of our model with a hypothetical distribution of contractile acini  is shown in Fig.~\ref{f7}. In Fig.~\ref{f7b} the deformed positions of acini are superposed on the reference positions to demonstrate considerable motion of particles due to matrix deformation. This simulation  suggests that multiple acini contracting undergo more dramatic motion and distortion than isolated pairs of acini.}

\section{\tred{Discussion}} 

\tred{The  continuum model we have proposed differs from previous ones, because it is based on a nonlinear elastic, nonconvex strain energy density, that  features a regime of  instability in compression. This unstable strain regime separates two distinct stable strain regimes, the undensified and densified phases of the material. The macroscopic  instability stems  from the microbuckling  behaviour of individual fibers. It is the central player in the prediction of the densification patterns due to contractile cells and clusters. While various previous models (including our own)  have incorporated fiber buckling in different ways \cite{Notbohm2015,Rosakis2015,Lesman}, they have restricted attention to small deformations, and have not taken full account of  kinematic nonlinearity.  Other nonlinear elastic models allow large deformations but assume a phenomenological  behaviour that is stable to begin with \cite{wang2014long}.  Some  models have focused on tensile stiffening behaviour of network fibers \cite{winer2009non}. In most cases, the unstable strain regime involving  compression/densification is not probed by continuum or discrete models so far, e.g., by  discrete models  studying stress amplification and force chains \cite{ronceray2016fiber,Lesman}.  Other models investigate nonaffinity in network displacements due to factors such as network randomness, usually restricted to small deformations \cite{MacK}. The present model averages over this type of nonaffinity by coarse-graining from a discrete network to a continuum description, but identifies a more severe type of  nonaffine deformations: strain localisation due to a mechanical instability. These deformations are unusually large but observed in tethers.  This requires us to retain full nonlinear kinematics and to consider arbitrarily severe strains and large rotations.}

\tred{Plasticity has been proposed as a mechanism for tether formation, e.g., \cite{ban2018,malandrino2019}. On the other hand, based on experiments, Vader et al. \cite{vader2009strain} conclude that ``the reversibility of fiber alignment and gel densification, seen in crosslinked collagen samples, show that these effects are primarily elastic''. Our simulations with the monostable energy model agree with this conclusion for crosslinked collagen. It seems that this behaviour is in fact pseudoelastic \cite{delaey,otsuka}, namely due not to plasticity but to an elastic instability.  The macroscopic instability is caused by microbuckling and reflected by nonconvexity of the elastic energy landscape.}  

\tred{For uncrosslinked  networks, residual tether-like deformations are predicted by a plasticity model incorporating crosslinking  \cite{ban2018}, but also by the bistable version of our model. Nonetheless,  there are various characteristics observed in our experiments and in those of \cite{ban2018}, that are not captured by plasticity models. These include virtually discontinuous  compressive stretch (but smoothly varying tensile stretch) in the densified zones, and the split tether and split radial hairlike band microstructures seen in our experiments.  Our phase transition model does predict these numerically and explains them theoretically as arising from geometric incompatibility of energy minimising strains with particle  imposed deformations.  Our bistable energy model does account for new crosslink formation and captures the residual densification after force removal observed in our experiments. This is due to stability of the densified phase under zero stress, analogous to residual martensitic transformations associated with the shape memory effect that are not due to plasticity \cite{bhattacharya2003microstructure}.}

\tblue{Our model simulations tend to overestimate the length of radial hairlike bands (Figs~\ref{f3}-\ref{f5})} \tred{and the change in distance between particles,  Fig.~\ref{f1g}, caused by deformation from our experiments. On the other hand, the model only involves five constitutive parameters (Table S1), less than half the number employed in plasticity based models.}
 
 How is tether formation involved in aiding and abetting cell migration, invasion and intercellular communication?  After acini contract causing a tether to form,  individual cells from each acinus start migrating along the tether, towards the acinus at the other end \cite{harris1981fibroblast,stopak1982connective,vader2009strain,shi2014rapid}.  Isolated fibroblasts grow appendages toward each other along the tether that forms (Fig.~\ref{f1b}) as a result of their contraction \cite{Notbohm2015}.  Given this observed tendency of cells to approach each other, the advantage offered by the biphasic behaviour of the fibrous ECM becomes clear: to detect its neighbours, all a cell has to do is contract, uniformly in all directions,  without prior cues as to the direction of neighbours. The  automatic response of the ECM is to form sharply defined, densified  paths (tethers) leading directly to  nearby cells or clusters. Our model identifies this as a special feature of the instability-driven multiphase behaviour of the ECM, not possible in stable, single-phase elastic materials.

Two ECM-related factors identified as breast cancer risk biomarkers are  density \cite{breastdens,provdens} and fiber alignment \cite{aligned,prov1,prov2,stroma}. Both arise in our model and experiments, as a result of phase change brought about by buckling fiber collapse. The densified phase  in  our model and experiments, also in \cite{shi2014rapid,ban2018},  is characterised by a 3- to 5-fold increase in density, but also strong fiber alignment toward the direction of maximum stretch, as compression squeezes an originally uniform angular distribution of fibers away from the compressive direction  and toward the orthogonal tensile direction (see \SM: Fiber Alignment in the Densified Phase). Hence, within a  tether formed between contracting particles (Fig.~\ref{f5e}) the fiber angle distribution we predict, Eq.~(\ref{s2}), is as in  Fig.~\ref{sf4a}, with a peak along the tether axis, normal to the contracting spheroid boundary.  \tred{This compares well with} Fig.~\ref{sf4b} (reproduced from  \cite{prov1}) which shows a typical fiber distribution associated with TACS-3 \cite{prov1,prov2,stroma,aligned} or Tumour-Associated Collagen Signature 3 (aligned fibers normal to contractile tumour spheroid boundary) a biomarker for invasion of tumour cells along tracts of aligned fibers. Tether formation exhibits the defining features of TAC3; accordingly, when contractility is inhibited, tethers typical of TACS-3 are suppressed  \cite{prov2}.  At earlier stages when a tumour is growing and pressing against the surrounding ECM,  our simulation Figs~\ref{f4f},\ref{ff3c},\ref{ff3d} of expanding particles (Fig.~\ref{f4c}) predicts a fiber distribution   Fig.~\ref{sf4c} peaked in the direction tangent to the (expanding) boundary. This agrees qualitatively with collagen signature TACS-2 \cite{prov1,prov2,stroma,aligned} (fibers compacted and aligned parallel to a growing tumour boundary) with measured distribution Fig.~\ref{sf4d} (reproduced from  \cite{prov1}).

\textcolor{black}{Our work provides strong evidence that two different TACS  in ECM associated with tumour growth and  invasion are distinct manifestations of a phase change in the ECM, caused by tumour growth (TACS-2 \cite{prov1,stroma}) or tumour cell contraction (TACS-3) \cite{prov2}.}

\tblue{Other than serving as a track for cell migration, the aligned and densified collagen fibers within tethers can enhance  anisotropic transport of macromolecules and extracellular vesicles. This can enhance cancer-stroma communication and facilitate tumour progression \cite{gomez,jung}}.

Taken together, our experiments, model and simulations show that the densification patterns caused by contractile cells and cellular clusters remodelling fibrous  ECM exhibit many salient features of stress-induced phase transformations in solids, captured by nonlinear models featuring a nonconvex strain 
energy \cite{Knowles1978,Ball1987,Ball1999,Abeyaratne2006,ball1976,ball2011local,dacorogna2007direct,bhattacharya2003microstructure,luskin1996,healey2007,abchujam,james1995tipsplit}. Remarkably, we find that such nonconvex energies arise naturally in modelling the mechanics of fibrous biomaterials, once the microbuckling instability mechanism  is accounted for.  Our model is minimal, with a handful of  parameters,  and demonstrates the necessity of a paradigm shift: material instability has to be taken seriously if we are to understand the behaviour of  fibrous networks, such as collagen ECM. This is what enables the model  to capture the distinctive morphology \textcolor{black}{characteristic of mechanical remodelling  of the fibrous ECM for the first time,} in the form of cell-induced, two-phase, pattern-forming deformations. We expect it will provide new insights  into the role played by these singular deformations in cell migration, communication, \textcolor{black}{tumour cell invasion, and  alteration of mechanical ECM properties caused by cell-induced remodelling.}

 \section{Methods}

\subsection*{Experiments with Active Particles} 

\paragraph*{PNIPAAm Particle Generation.}
PNIPAAm particles were generated using two different recipes, one giving larger particles (diameter 150 $\mu$m) and the other giving smaller particles (diameter 75 $\mu$m). \tred{Although these sizes are greater than the size of a cell, they could match the size of a cell cluster or mammary acinus.  More importantly, the contraction of the particles create controllable localised forces that bring about the densified zones of interest to this study.}
To generate the larger particles, kerosene with 3.5\% Span 80 (Tokyo Chemical Industries) was degassed under vacuum for 1 hour. It was then maintained under nitrogen for 10 minutes before stirring at 350 rpm at 22$^\circ$C. A solution containing 1 g N-isopolyacrylamide (Sigma), 7.5 mL of 2\% bis-acrylamide solution (Bio-Rad), 0.05 g Ammonium Persulfate (Bio-Rad) and 1.5 mL of 1 $\times$ tris-buffered saline was prepared. 10 $\mu$l TEMED (Bio-Rad) was added, and the solution was immediately added to the kerosene. To generate the smaller particles, cyclohexane, rather than kerosene, with 3.5\% Span 80 was used as the solvent, and stirring occurred at 450 rpm instead of 350 rpm. A solution of 1 g N-isopolyacrylamide (Sigma), 3.75 mL of 2\% bis-acrylamide solution (Bio-Rad), 0.05 g Ammonium Persulfate (Bio-Rad), 1.5 mL of 1 $\times$ tris-buffered saline and 3.75 mL deionised water was prepared. Again, 10 $\mu$l TEMED (Bio-Rad) was added, and the solution was immediately combined with the cyclohexane. For both large and small particles, the solutions were stirred for 30 min, and then the PNIPAAm particles were allowed to settle overnight. The particles were then washed with hexane and again allowed to settle. Washes were repeated with isopropyl alcohol, ethanol, and finally deionised water. The particles were then filtered with a cell strainer to keep particles of diameter $>$ 40 $\mu$m. The particles were resuspended in 1 $\times$ PBS. 

The PNIPAAm particles were embedded in networks of rat tail collagen I (Corning), which was fluorescently labeled with Alexa Fluor 488 (Thermo Fisher Scientific) \cite{burkel2017}. To adhere the particles to the collagen, the particles were first treated with sulfo-SANPAH (Proteochem) as in our previous studies \cite{burkel2017}. Next, the pH of the collagen was neutralised using 100 mM HEPES buffer to a final concentration of 3 mg/mL and mixed with the PNIPAAm particles. The collagen then polymerised at 27$^\circ$C for 1 hour.   \tred{After polymerisation, the collagen formed a network of fibers that branch into multiple fibers at discrete nodes. The average distance along a fiber between nodes, which we refer to as the fiber length, was measured as in our previous work \cite{burkel2018}. Briefly, images of the fibers were segmented \cite{mashburn2012}, giving segmented areas of the void space between fibers. The average of the segmented areas was 3.9 $\mu$m$^2$. Hence, the mesh size, which is proportional the square root of the segmented areas, was 1.97 $\mu$m. Next, we accounted for the fact that some fibers in the image cross without connecting \cite{stein2008}, by selecting representative fibers and manually counting the number of unconnected crossings. The product of the mesh size and number of unconnected crossings gave the fiber length, of 19.7 $\pm$ 2 $\mu$m (mean $\pm$ standard deviation).}

\paragraph*{Microscopy and Imaging.}
Images of PNIPAAm particles embedded in collagen networks were collected using a spinning disk confocal microscope (Yokogawa CSU-X1) with a Nikon Ti-E base and a 20 $\times$ 0.75 NA air objective (Nikon) using a Zyla sCMOS camera (Andor). The particles diameters and the distances between particle centers were measured manually using ImageJ.

The temperature was controlled with an H301 incubator (Okolab) mounted on the microscope stage and controlled with an UNO controller (Okolab). To verify that the samples reached the desired temperature, we used a digital thermometer having accuracy of 0.1$^\circ$C (Fisherbrand Traceable) with its probe inside a dish with water placed next to the samples. Images were collected when samples were at 26$^\circ$C (undeformed state) and 39$^\circ$C (contracted state).

\subsection*{Model}  
\begin{figure}
    \centering
    \captionsetup[subfloat]{farskip=-0.3pt,captionskip=-0pt}
    \subfloat[]{\includegraphics[height=2.47cm,width=0.25\textwidth]{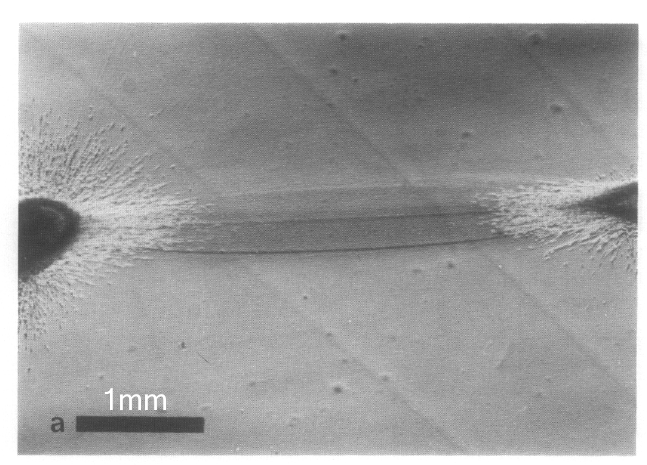}
        \label{f1a}}  
    \subfloat[]{\includegraphics[height=2.35cm,width=0.25\textwidth]{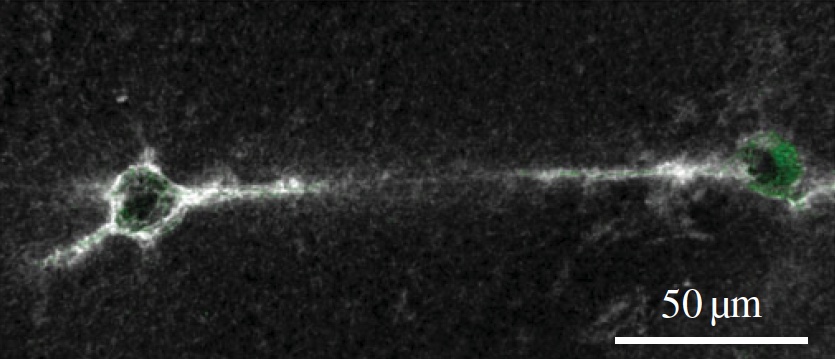}
        \label{f1b}}
    \subfloat[]{\includegraphics[height=2.43cm,width=0.22\textwidth]{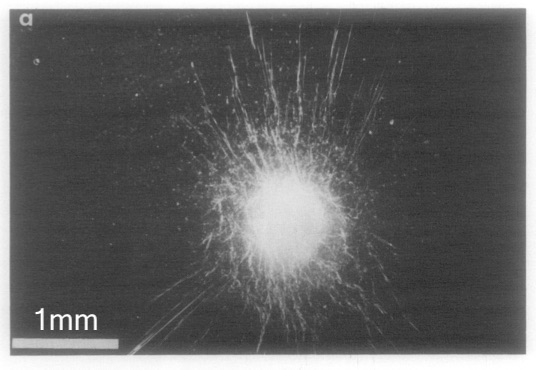}
        \label{f1c}}
    \\
       \subfloat[]{\includegraphics[width=0.3\textwidth]{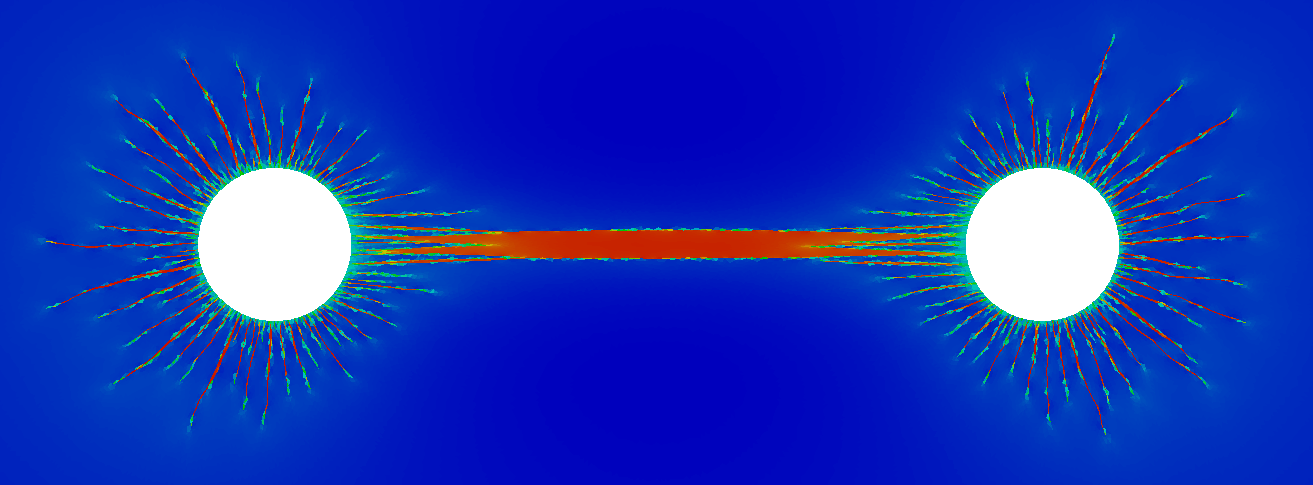}
        \includegraphics[width=0.3\textwidth]{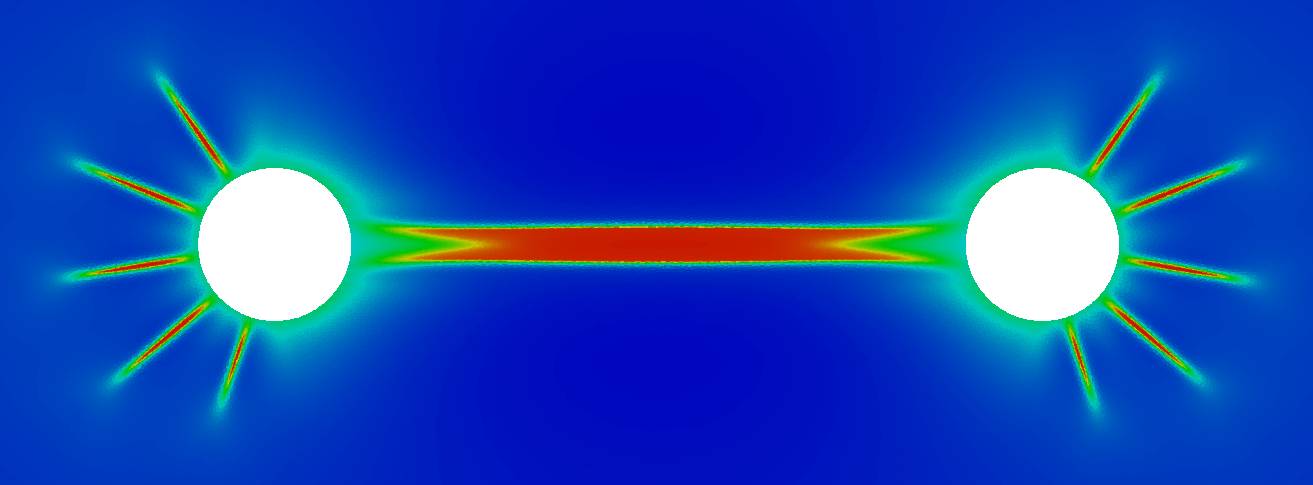}
        \includegraphics[width=0.3\textwidth]{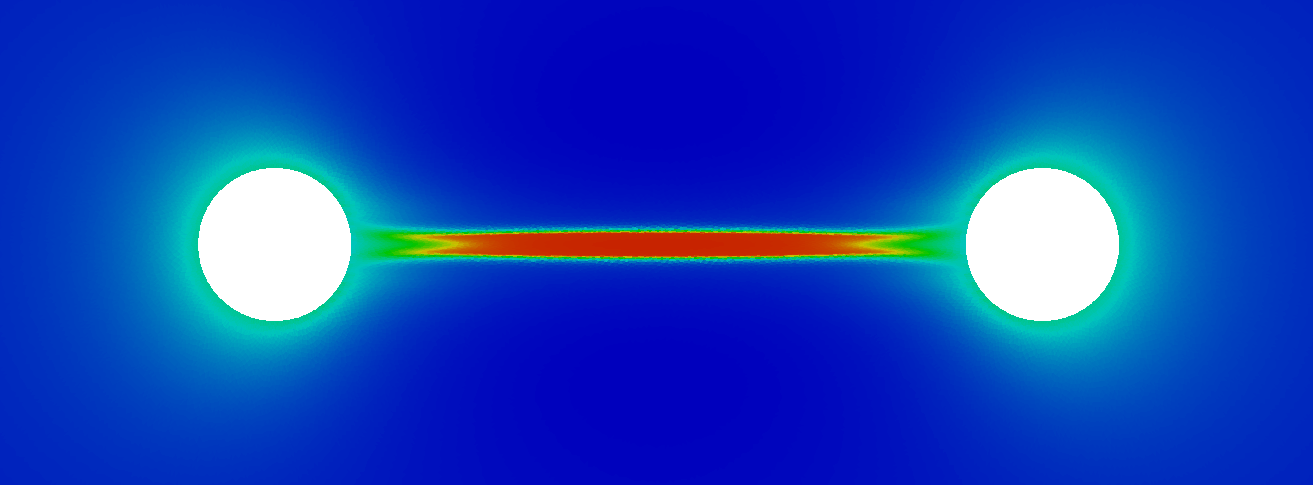}
        \includegraphics[width=0.06\textwidth]{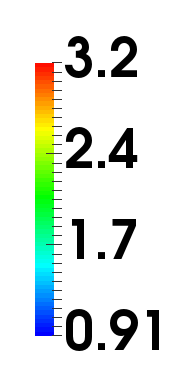}
        \label{f1d}}
    \\
    \vspace{0.3cm}
    \subfloat[]{
        \includegraphics[width=0.3\textwidth]{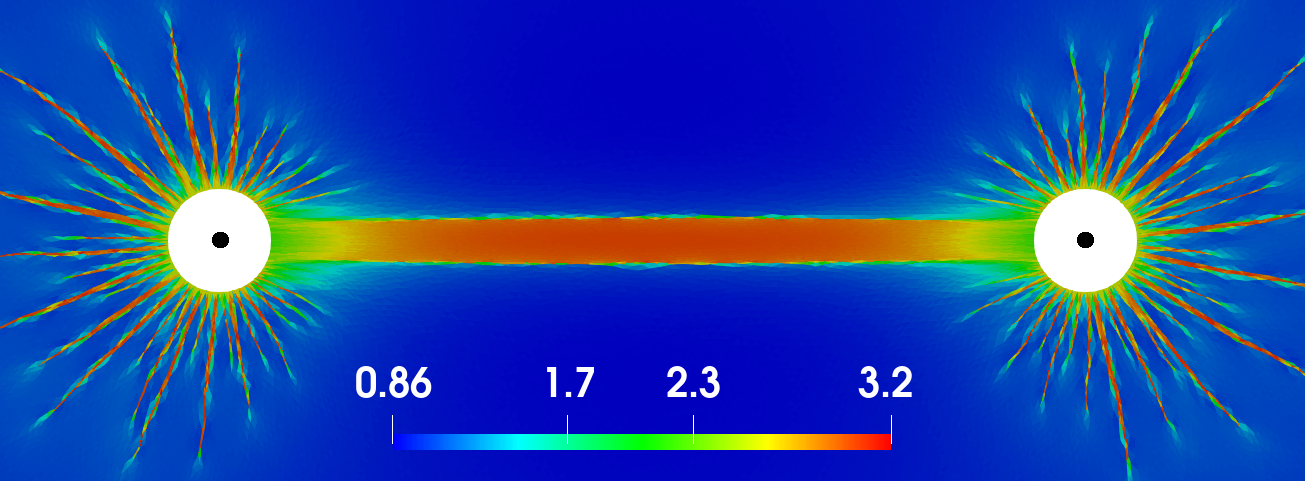}
        \includegraphics[width=0.3\textwidth]{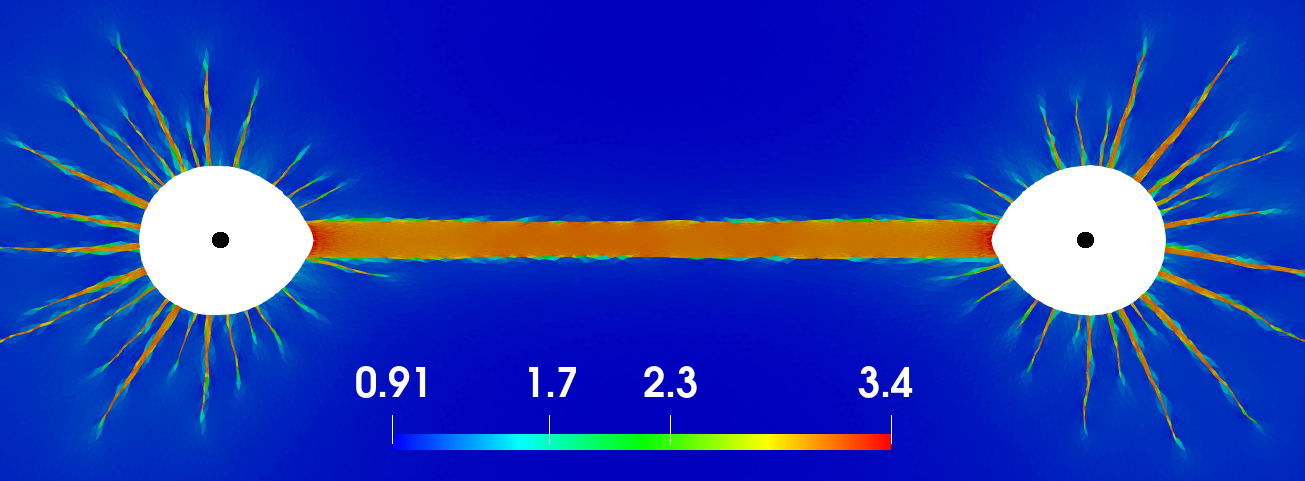}
        \includegraphics[width=0.3\textwidth]{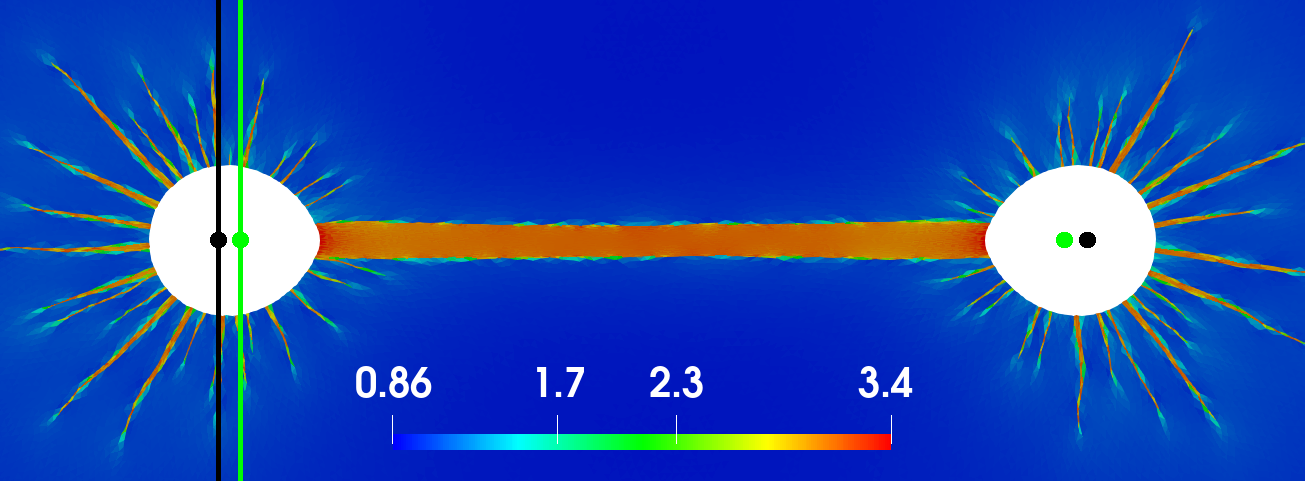}
    \hspace{1cm}
        \label{f1e}}
    \\
    \subfloat[]{ \includegraphics[width=0.3\textwidth]{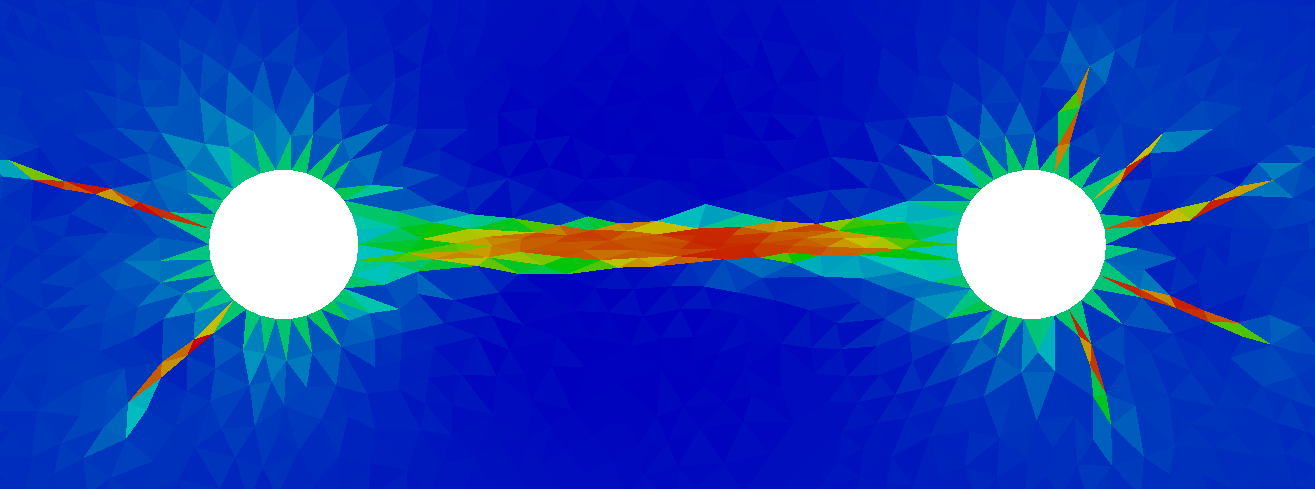}
        \includegraphics[width=0.3\textwidth]{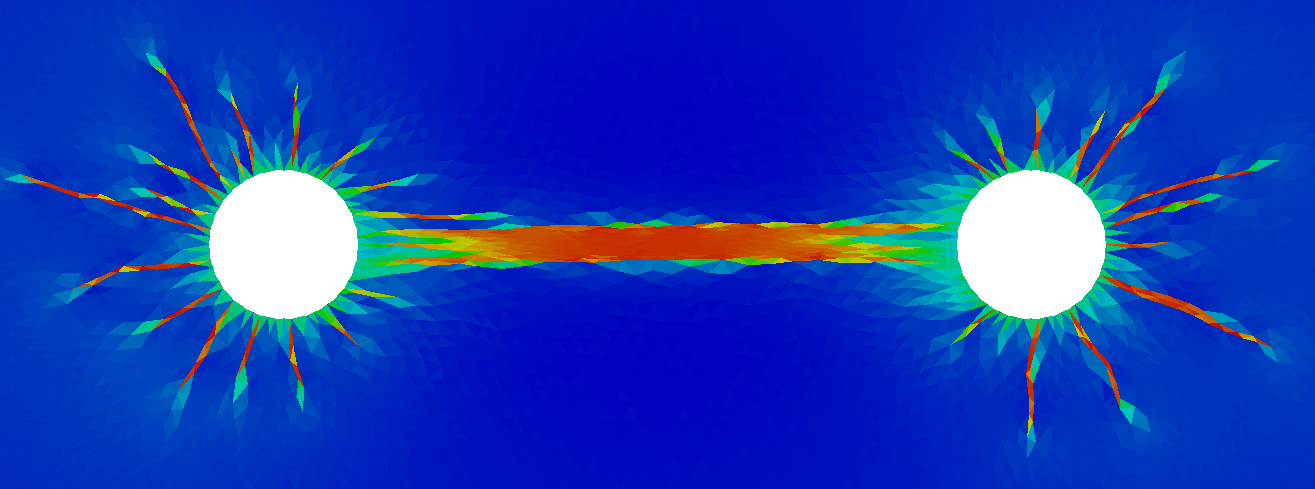}
        \includegraphics[width=0.3\textwidth]{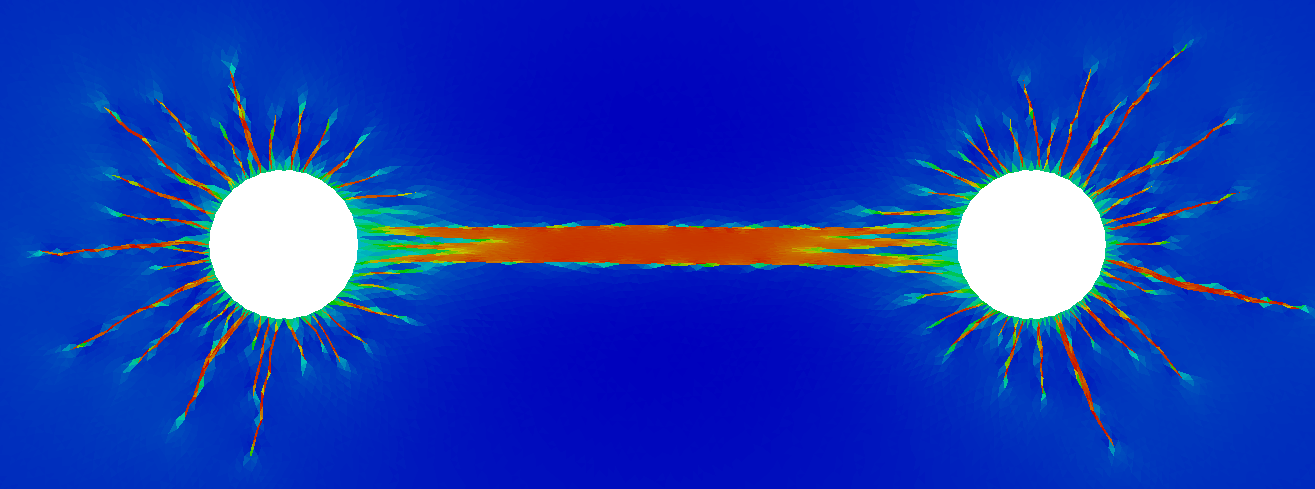}
     
        \includegraphics[width=0.06\textwidth]{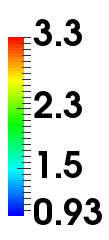}
        \label{f1f}}    
   \caption{\baselineskip10pt{\footnotesize \textbf{Tethers joining Hairy Clusters.} (a)-(c) experiments from earlier work. (de)-(f) simulations of our model. See Tables S1, S2 for parameters. (a)  Tether between mm sized explants (reproduced from \cite{stopak1982connective}) (b) Tether between $\mu$m-sized individual cells (reproduced from \cite{Notbohm2015}). (c) Radial hairs issuing from an isolated explant (reproduced from \cite{stopak1982connective}). (d) Effect of higher gradient energy, Eq. \eqref{grad}, in controlling number and fineness of hairlike radial bands. (I) $\eps=0$,  (II) $\eps=0.01R$, (III) $\eps=0.05R$, where $R$ is particle radius. (e) Shape of deformable explants and relative motion (I) $k =100$ with fixed centers. (II), $k = 1$ with fixed centers.  (III), $k = 1$  but centers are free, and move from black to green dot.
(f) Fineness of phase mixture  in  simulations depends on numerical mesh resolution (increasing from left to right).  \scd
} }
 \centering
 \label{f1}
\end{figure}

\begin{figure}
   \centering
   \subfloat[]{\includegraphics[width=0.3\textwidth]{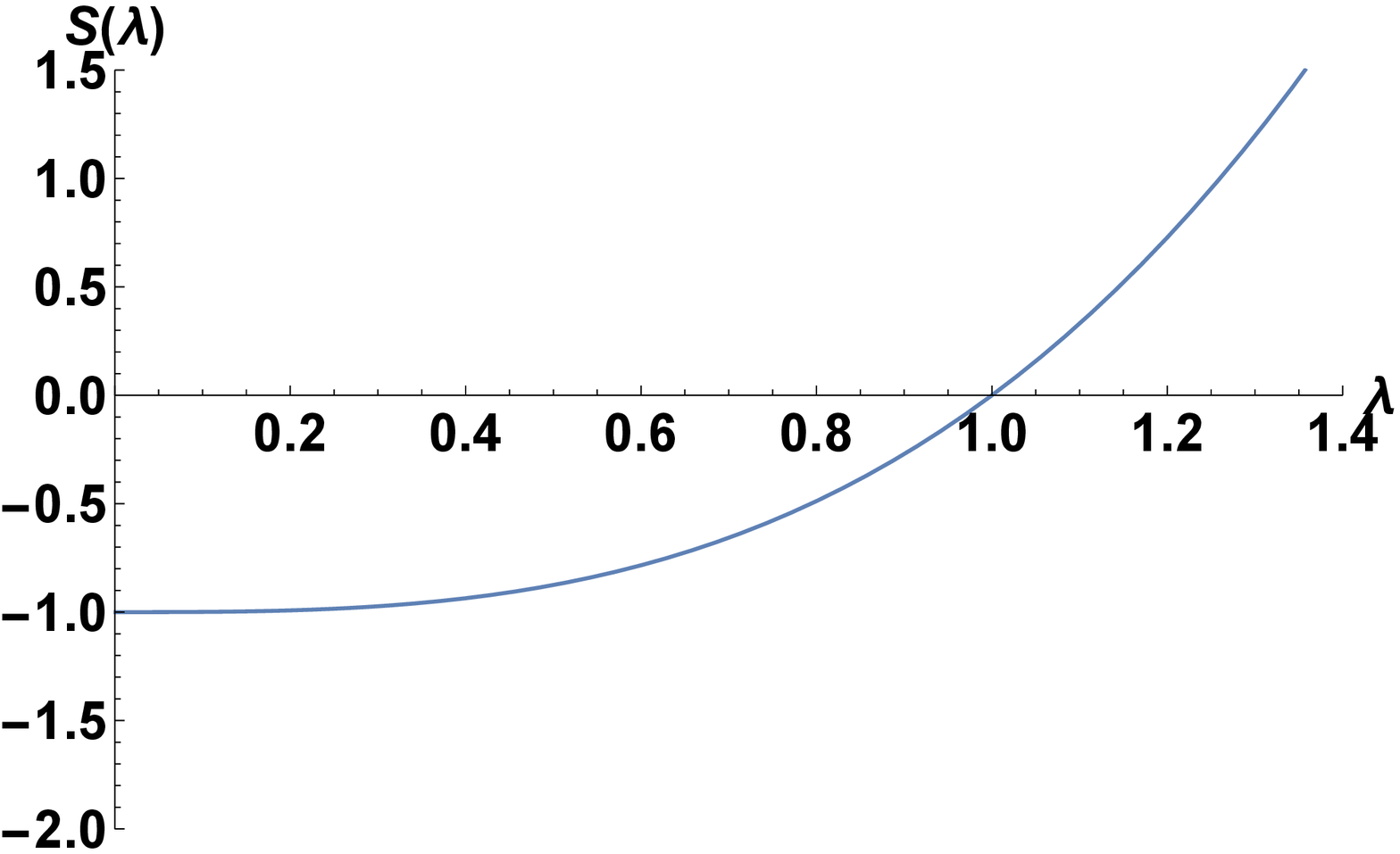}
       \label{f2a}}
   \hspace{0.2cm}
   \subfloat[]{\includegraphics[width=0.3\textwidth]{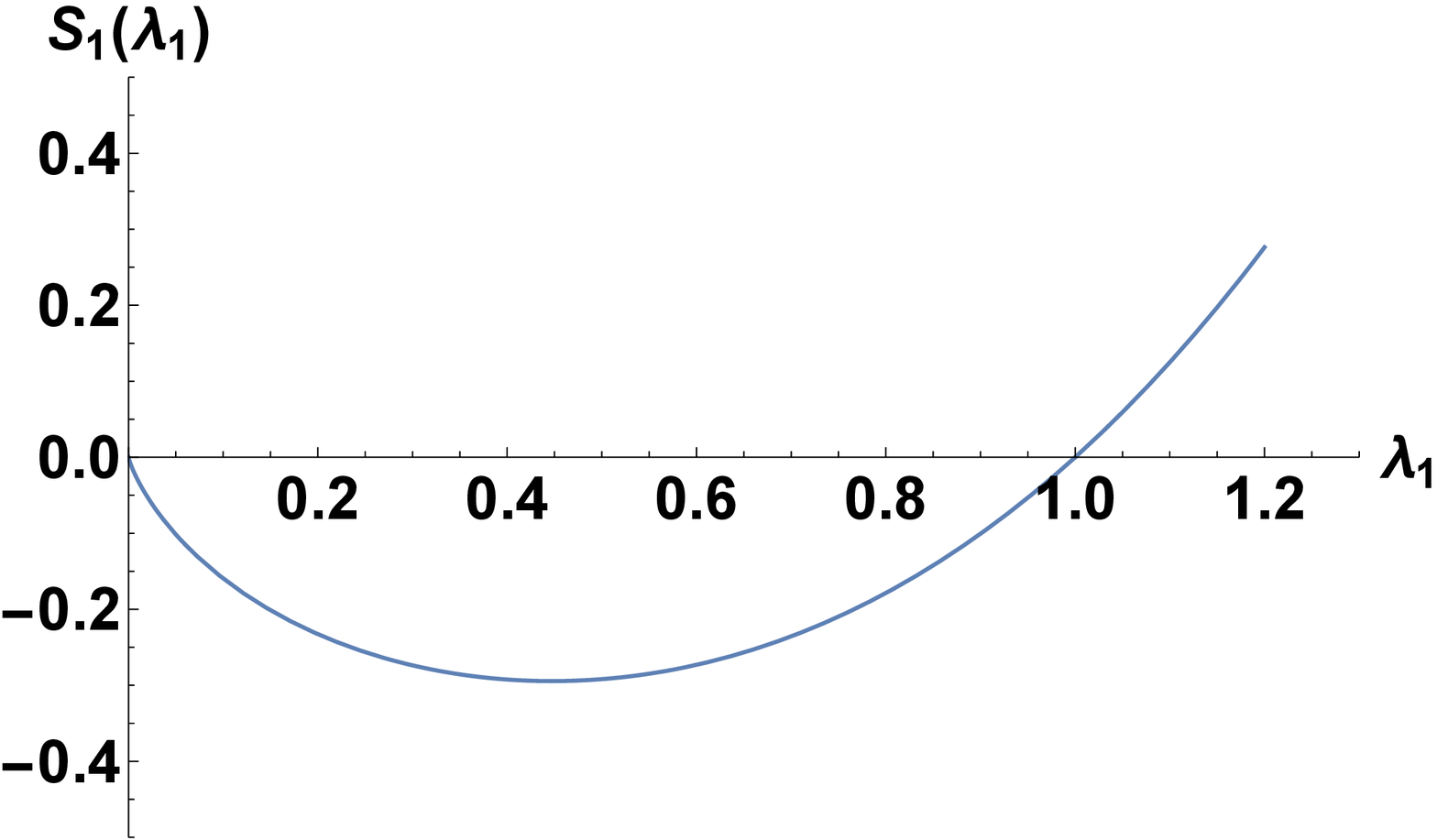}
       \label{f2b}}
       \hspace{0.2cm}
 \subfloat[]{\includegraphics[width=0.3\textwidth]{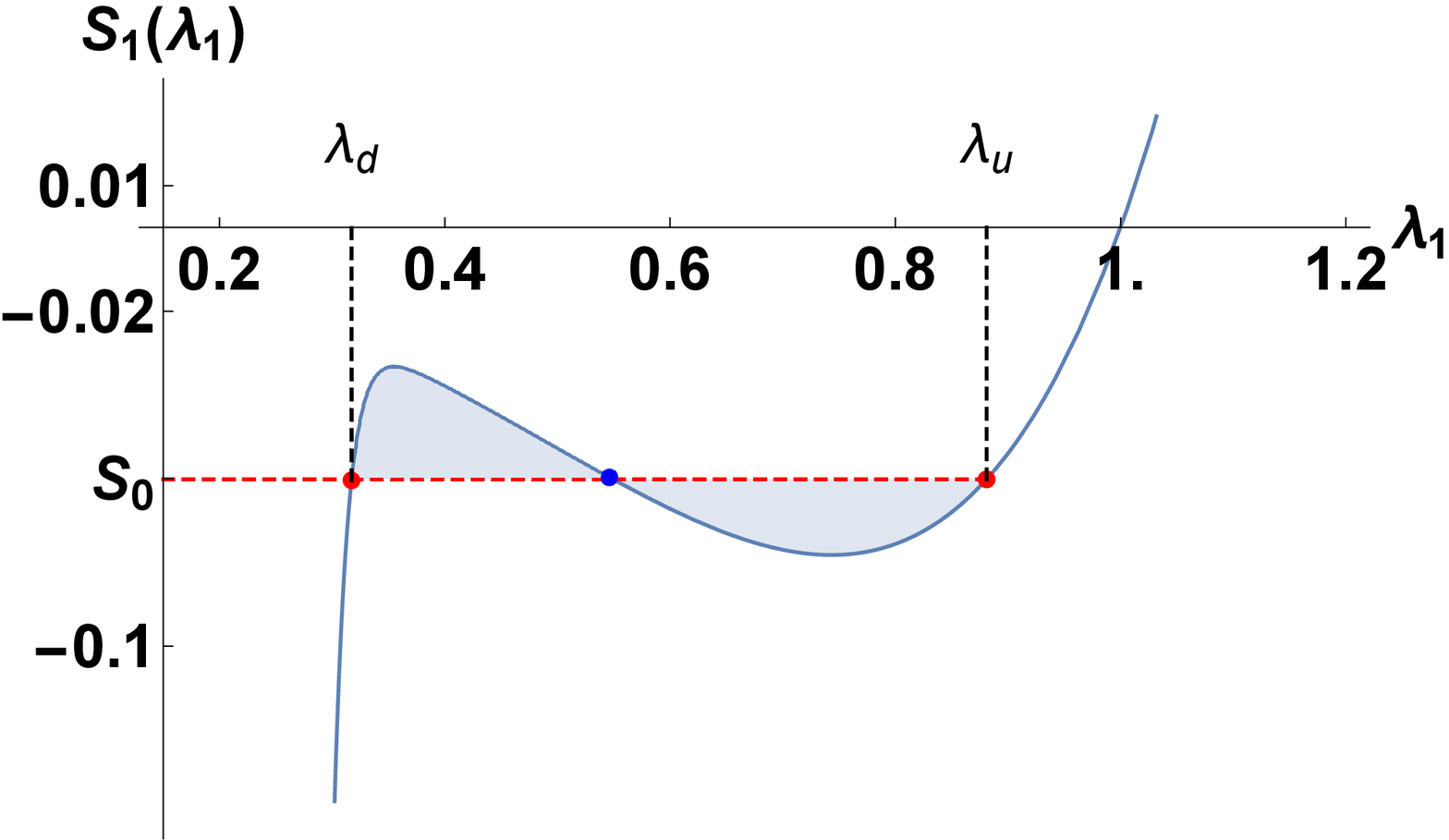}
       \label{f2c}}
       \\
   \subfloat[]{\includegraphics[width=0.3\textwidth]{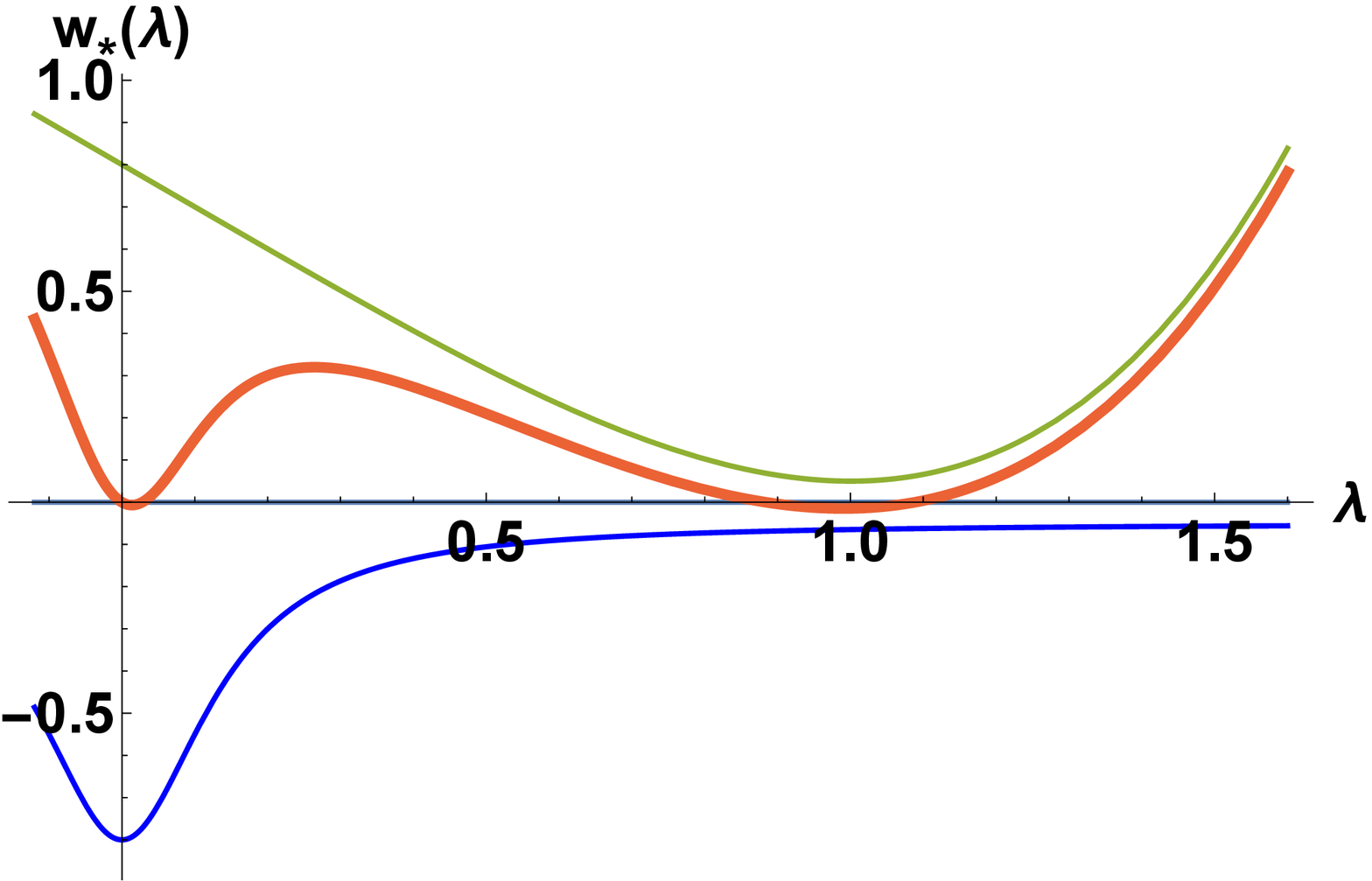}
       \label{f2d}}
    \hspace{0.2cm}
   \subfloat[]{\includegraphics[width=0.3\textwidth]{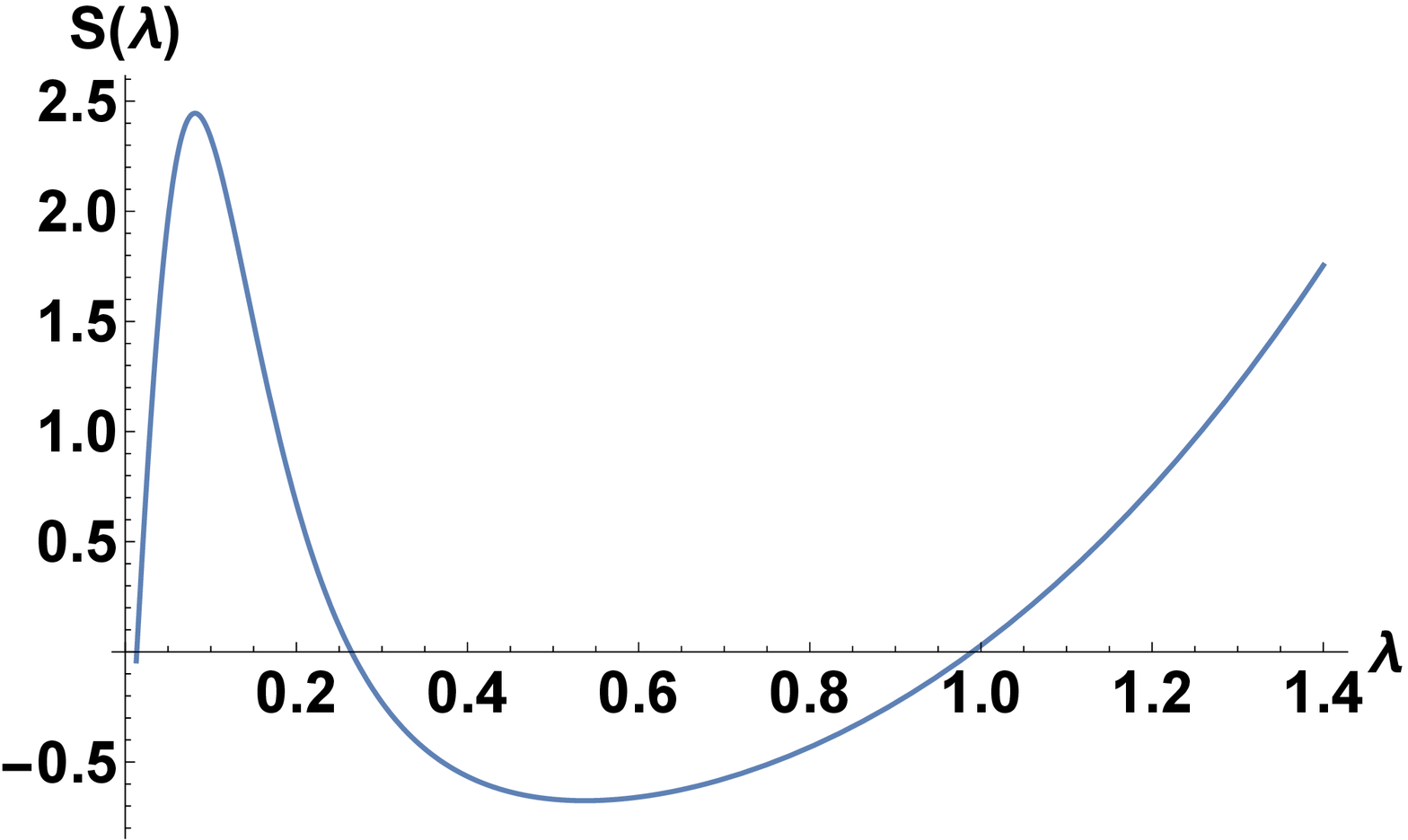}
       \label{f2e}}
       \hspace{0.2cm}
   \subfloat[]{\includegraphics[width=0.3\textwidth]{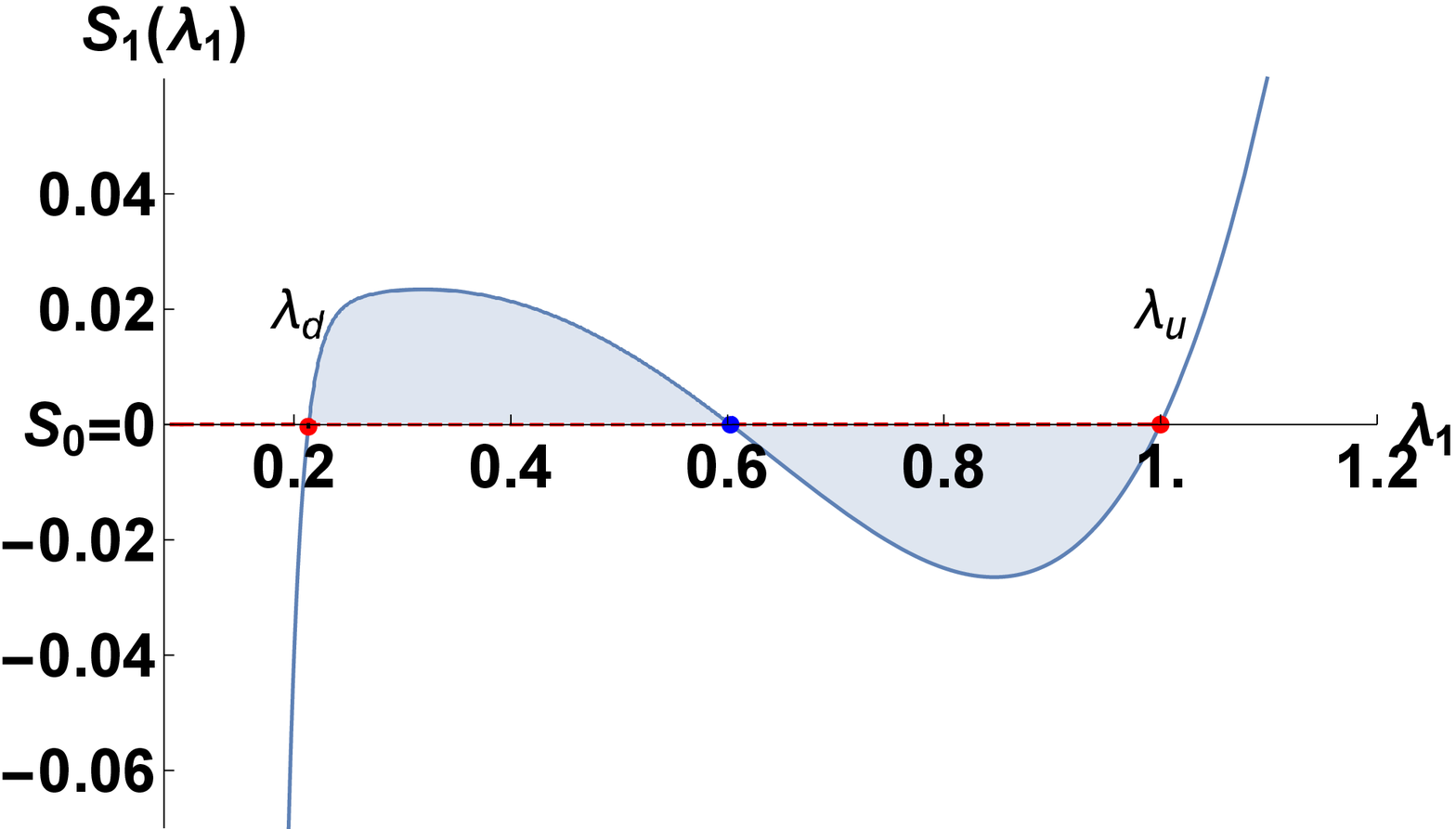}
       \label{f2f}}      \caption{ \baselineskip12pt{\footnotesize \textbf{One Well or Two? Designing an Energy Density for Fibrous ECM.} (a) Force-stretch curve of a single fiber that loses stiffness  in compression and stiffens in tension. 
       (b) Stress-stretch curve for uniaxial compression  of 2D (orientation-averaged) energy density $W$ from \eqref{int} (corresponding to (a)) has a decreasing unstable branch. Here $S_1(\la_1)=\partial W(\la_1,1)/\partial\la_1$.
       (c) As in (b) but with added energy penalty \eqref{pe} due to fiber volume, which resists extreme compression. At a compressive stress $S_0$ (red line) there are multiple stretch states, two stable (red dots) and one unstable (blue dot). (d) Adding an attraction potential (blue) to the fiber potential (green) corresponding to (a) produces a two-well bistable potential (red). (e)  Force-stretch curve corresponding to the red potential in (d). (f) Uniaxial compression curve of the corresponding orientation-averaged energy density $W^*$ (with volume penalty) shows bistability, with two stable stress-free states (red dots), and an unstable one (blue dot)  in between.}   }
   \label{f2}
\end{figure}

\paragraph*{Energy Density.}
For a single fiber, we introduce the effective stretch $\la$, which equals the distance between its endpoints divided by its undeformed, or relaxed, length.  The energy of a single fiber can be written as  $w(\la)$ as a function of effective stretch $\la$.  When the fiber is in tension, it is straight and $\la$ equals the actual stretch (strain $+1$), while $w(\la)$ equals the elastic energy due to stretching of the fiber. When it is in compression, it may be buckled, in which case the elastic bending energy of the fiber can still be expressed as a function  $w(\la)$ of the distance between its endpoints, hence of the effective  stretch $\la$.  See Supplemental Fig.~\ref{sf1a}. In order to model a 1D two well energy  $w(\la)$  for a single fiber  like the red curve in Fig.~\ref{f1d}, we start with the derivative $S(\la)=d w(\la)/d\la$, which represents force as a function of stretch. We choose a  polynomial that vanishes at $0$, $1$ and an intermediate value,
$$S(\la) = \la^5 -(a_m+1)\la^3 + a_m \la,$$
to qualitatively represent the curve in Fig.~\ref{f2e}.  Here $a_m$ is a parameter that controls the relative height of the two wells (minima) of $w(\la)$.
Integrating this with respect to $\la$  gives an energy 
\begin{equation}
w(\la)=  \la^6/6 -(a_m+1)\la^4/4 + a_m \la^2/2+1/12-a_m/4
  \label{wa}
\end{equation}
We model the ECM as a 2D nonlinear elastic continuum undergoing possibly large  deformations $ \by(\bx)$ where a particle with position vector $\bx$ in the undeformed state is mapped to deformed position $\by=\by(\bx)$.  The elastic  strain energy density of the material can be written as a function $W(\bF)$ of the deformation gradient $\bF=\nabla\by$, which is a $2\times2$ matrix.  We model the random fiber network as an isotropic material, which means that $W$ depends on $\bF$ only through the principal stretches $\la_1$, $\la_2$, whose squares are the eigenvalues of the  Cauchy-Green deformation matrix $\bF^T\bF$. Equivalently $W$ is a function of the two deformation invariants $I_1(\bF)=\hbox{tr}(\bF^T\bF)=\la_1^2+\la_2^2$ and $J(\bF)=\det\bF=\la_1\la_2$. Here $J$ is the Jacobian determinant of the deformation and the ratio of deformed density $\rho$ to undeformed density $\rho_0$   satisfies $\rho/\rho_0=1/J$. 

To connect the single fiber energy with the 2D strain energy density $W$ we follow  \cite{vainchtein2002strain,gloria2014}.  We assume the network contains a uniform distribution of fibers. A homogeneous deformation (with constant strain) is equivalent to a biaxial stretch in two orthogonal directions with stretches  $\la_1$, $\la_2$.  A fiber of undeformed length $l$ that makes an angle $\theta$ with the principal stretch axes in the undeformed state,  will have endpoints at $(0,0)$ and $(l\cos\theta,l\sin\theta)$. After deformation the latter will become  $(\la_1l\cos\theta,\la_2l\sin\theta)$. As a result, the stretch ratio of the fiber will be $\la(\theta)=\sqrt{(\la_1 \cos\theta)^2 + (\la_2 \sin\theta)^2}$, and its energy will be $w(\la(\theta))$.  Summing over all fiber orientation angles $\theta$, we find the elastic energy density of the network $$ \hat{W}(\la_1, \la_2) = \frac{1}{2\pi}\int_0^{2\pi} w(\la(\theta)) d\theta= \frac{1}{2\pi}\int_0^{2\pi} w \left(\sqrt{(\la_1 \cos\theta)^2
 + (\la_2 \sin\theta)^2} \right) d \theta,$$ which gives Eq.~(\ref{int}).  It turns out that for $w(\la)$ given by Eq.~(\ref{wa}), the integral in Eq.~(\ref{int}) can be evaluated explicitly:
\begin{equation} \tilde{W}(\bF) = \frac{1}{96} [ 5I_1^3 -12 I_1 J^2 - (1+a_m) ( 9I_1^2 - 12 J^2) + 24a_m I_1+8].
  \label{W}
\end{equation}
  Here  $I_1 = \hbox{tr}{\bF^{\rm T} \bF}=\la_1^2+\la_2^2$ and $J=\det\bF=\la_1\la_2$ are the 2D deformation invariants.
We then add a fiber volume penalty term to the energy to account for resistance of  densified fibers to complete crushing by virtue of their nonzero volume. This term increases the energy abruptly when the Jacobian (volume ratio) $J=\det\bF$ becomes less than a small positive constant $b<<1$, while it becomes negligible as $J$ increases from $b$.  Such a function is given by
\begin{equation}\Phi(J)=\exp[A(b- J)]  \label{Phi}
\end{equation}
(Supplemental Fig.~\ref{sf1b}) where $A$ is a large positive constant. 
The total energy is 
\begin{equation}W^*(\bF)= \tilde{W}(\bF)+\Phi(\det\bF).  \label{WW}
\end{equation}
\paragraph*{Model for Active Particles.} 
We consider two models, one  for a cell cluster (e.g., acinus) and another for active particles. The  model for  acini is given by Eq.~(\ref{spr}); see also Supplemental Fig.~\ref{sf1c}.
\tred{Active particles are modelled as  spheres whose radius can undergo a stress-free transformation strain due to temperature (\MM: Experiments with Active Particles), but they can also deform in response to boundary forces. This deformation is modelled minimally using linear springs on the boundary of the sphere.} Each point on the sphere circumference is connected to the matrix by a linear spring of zero natural length. 
 \begin{equation}C\{\by;\bz_1,\ldots,\bz_n\}=\sum_{i=1}^n\int_{\Gamma_i} {k\over 2}\left\vert\by(\bx)-\bz_i-(1-u_0/R)(\bx-\bar\bz_i)\right\vert^2 ds_\bx\label{spr2} \end{equation}
where $\Gamma_i$ is the boundary  of the $i^{\rm th}$ cavity ($i=1,\ldots,n$), $k$ is the spring stiffness,
 $\bz_i$ the variable (deformed) centre position of the $i^{\rm th}$ cluster, $\bar\bz_i$ is its undeformed position, $R$ the undeformed radius, and  $u_0$ the particle radius contraction, with percentage referring to $u_0/R$.
See Supplemental Fig.~\ref{sf1d}.

\subsection*{Numerical Method}  

We employ the finite element method, based on a triangulation of the domain $\Omega$.
Our approximations for the deformation are sought in the space of continuous piecewise polynomial functions of degree  two. 
Consequently, our computational methods are based  on a discrete minimisation problem  on the finite  element space. 
In order to capture areas of high densification accurately a local  mesh refinement strategy close to the areas of phase transition  is adopted. 

Challenges include  the subtle non-linear character of the problem, and the nearly singular behaviour of 
solutions in areas  where phase transitions take place. It is known that numerical algorithms can become quite
subtle exactly at these areas, and thus special care should be given to the reliable resolution of the interfaces. 

Incorporating higher gradients into the energy functional, $\eps > 0$ in Eq.~(\ref{grad}), 
introduces additional challenges, because the finite element spaces based on piecewise continuous polynomials have reduced
smoothness and are not consistent  with the standard energy setting of the model.  Furthermore, it is very 
desirable to have a computational model that works seamlessly when introducing regularisation by higher gradients.
We thus use the same discrete spaces in all models considered in our study ($\eps =0$ and $\eps >0$). To this end, 
we adapt to our problem an approach based on the discontinuous Galerkin methodology. Our approximations are still
sought on the same spaces of piecewise continuous polynomial functions over a triangulation of the domain,
however the energy functional  is modified to account for possible discontinuities of normal derivatives at the element 
faces. We  introduce a novel discrete energy functional, which includes terms accounting for the jumps of the higher gradients
at the element interfaces, as well as penalty terms which enforce weak continuity of the higher gradients and coercivity.  
To be more specific, let $\by_h$ denote a function of the discrete finite element space of piecewise polynomials
($C^0$-conforming), then the discretised functional for  has the form:
\begin{equation}
\begin{aligned}
\Psi_h[\by_h] &= \int_{\Omega} [W(\nabla \by_h) 
+ \varepsilon^2   \Bigg(
\frac{1}{2}\sum_{K \in T_h}  \int_{K} | \nabla \nabla \by_h |^2 d
\\ 
&
-\sum_{e \in E_h} \Big[ 
\int_{e}\dgal{ \nabla \nabla \by_h} \cdot \jumpop{\nabla \by_h
    \otimes n_e} + \frac{60}{h_e}\int_{e} |\jumpop{ \nabla 
    \by_{h}}|^2  \Big] \Bigg)
\end{aligned}
\label{equ:final_potential}
\end{equation}
where $E_h$ is the set of the interior facets of the triangulation,
$h_e$ is the length of the edge $e$ and
the average and  jump operators $\dgal{\cdot}$, $\jumpop{\cdot}$ are defined as follows
\begin{equation}
\begin{aligned}
\dgal{ \nabla \nabla \by_h} &= \frac{1}{2}(\nabla \nabla \by_h^+ +  \nabla \nabla \by_h^-) \\
\jumpop{\nabla \by_h   \otimes n_e} &= \nabla \by_h^+  \otimes n_e^+ +  \nabla \by_h^-  \otimes n_e^- \\
\jumpop{\nabla \by_h  } &= \nabla \by_h^+ -  \nabla \by_h^-,
\end{aligned}
\nonumber
\end{equation}    
here the superscripts $+$ and $-$ indicate functions evaluation on opposite sides of an edge $e$,
$n_e^+$, $n_e^-$ are the corresponding outward normal to the edge and $\otimes$ denotes 
the tensor product.  

The discretisation has been implemented in FEniCs \cite{AlnaesBlechta2015a}. 
For the minimisation of the discrete energy functional a parallelised nonlinear conjugate gradient method \cite{NocedalBook06}
has been developed.
The reliability of the computational experiments is guaranteed by a separate detailed mathematical study \cite{grekas}, which demonstrates the convergence of the discrete numerical solution to  the solutions predicted  by the model.

The energy functional is minimised using a parallelised version of the nonlinear conjugate gradient method \cite{NocedalBook06},  a successful  iterative method for large scale nonlinear optimisation 
problems. When ellipticity of the corresponding Euler-Lagrange equation for the unregularised problem holds, Newton's method is an efficient nonlinear minimisation technique.  
However, because  ellipticity fails in our model (and phase transition occurs) the Hessian  matrix of second derivatives of the energy is not positive definite, thus the nonlinear 
conjugate gradient is preferred, for both the unregularized and regularised  energy functionals.

\begin{figure}
    \captionsetup[subfloat]{farskip=-0.3pt,captionskip=-0pt}
    \centering
    \subfloat[]{\includegraphics[width=2.6in]{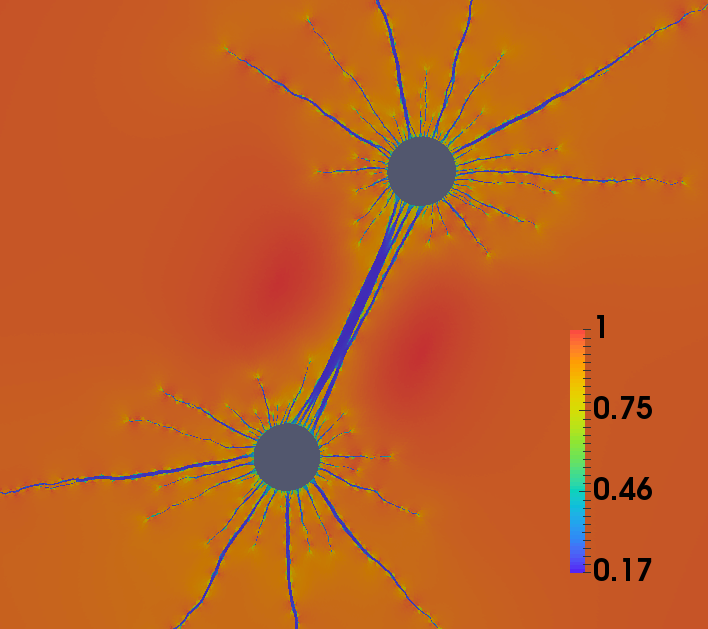}\label{ff3a}}
    \hspace{0.1cm}
    \subfloat[]{\includegraphics[width=2.6in]{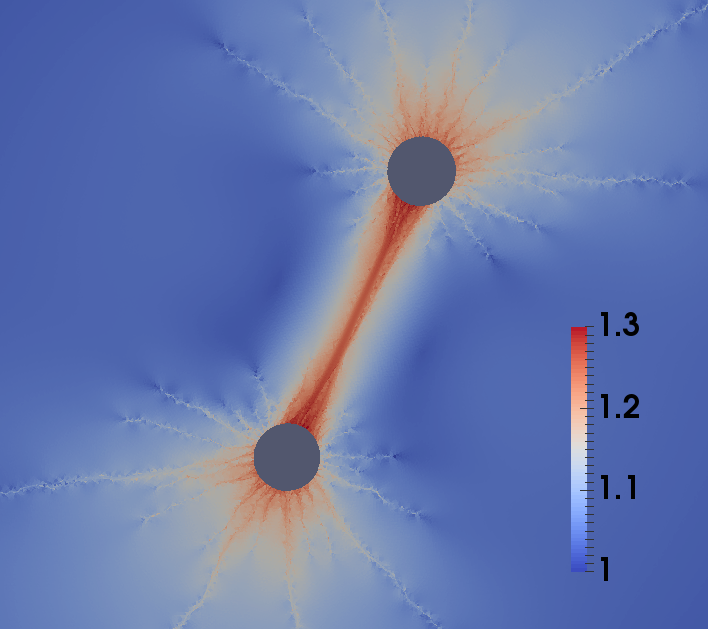}\label{ff3b}}
    \\
     \subfloat[]{\includegraphics[width=1.7in]{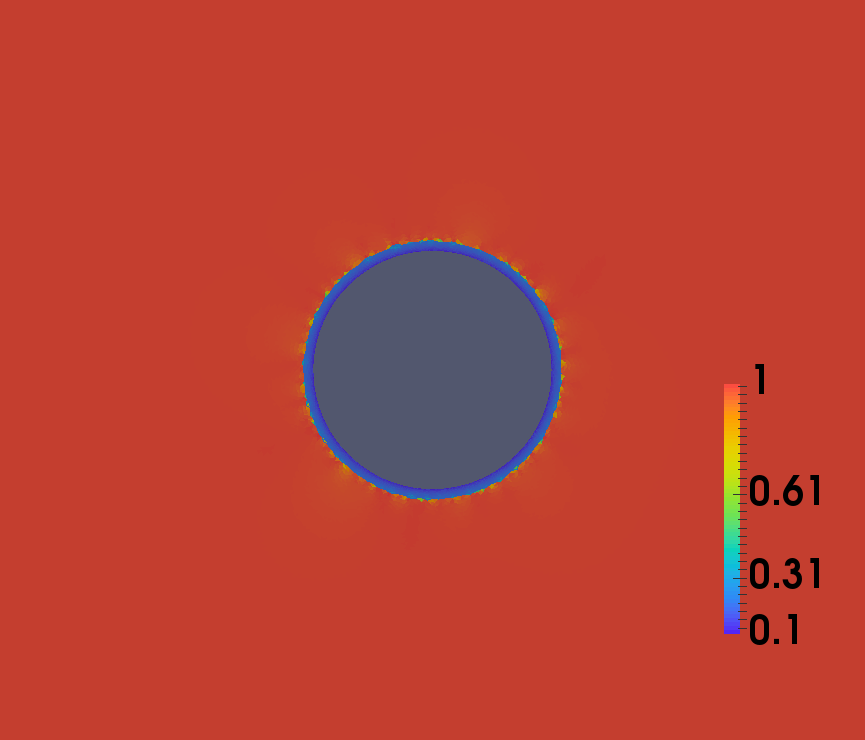}\label{ff3c}}
    \hspace{0.1cm}
    \subfloat[]{\includegraphics[width=1.7in]{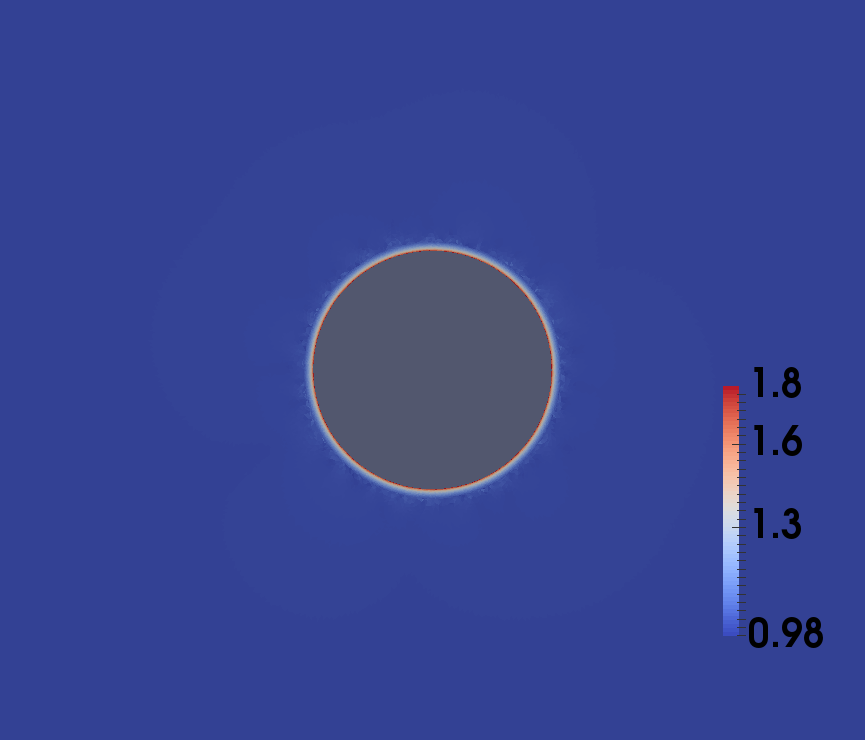}\label{ff3d}}
    \\
     \subfloat[]{\includegraphics[width=1.5in]{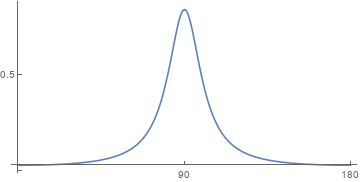}\label{sf4a}}
         \subfloat[]{\includegraphics[width=1.5in]{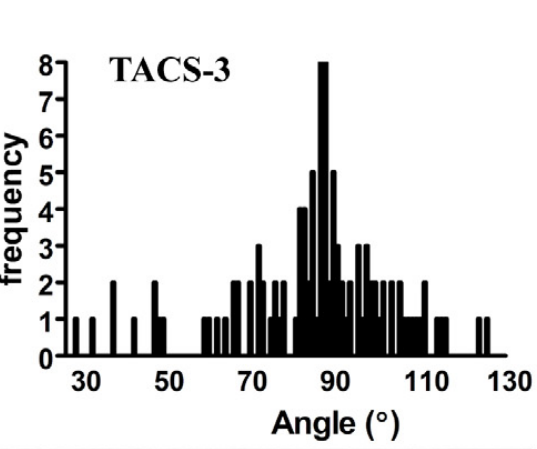}\label{sf4b}}
     \hspace{0.2cm} \vspace{-0.4cm}
      \subfloat[]{\includegraphics[width=1.5in]{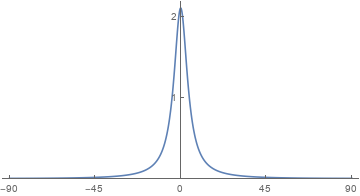}\label{sf4c}}
    \subfloat[]{\includegraphics[width=1.5in]{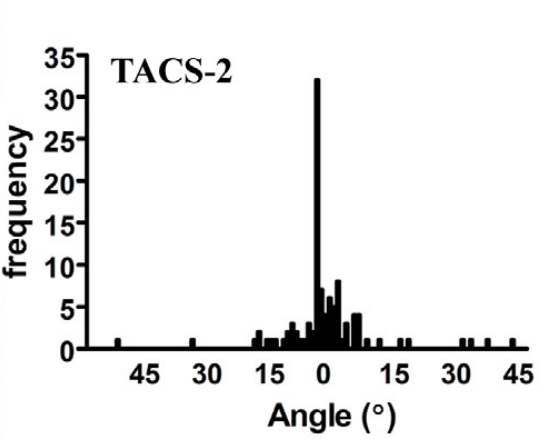}\label{sf4d}}\vspace{0.4cm}
\caption{Principal Stretches and Fiber Distributions: Principal stretches (a) $\la_1<1$ (compressive)  and (b)  $\la_2>1$ (tensile) from the simulation in  Fig.~\ref{f5e}. Note sharp change in $\la_1$ across  tether boundary indicating discontinuity of stretch normal to phase boundary, in contrast to gradual change in $\la_2$ (tangential stretch) as predicted by compatibility.  (c) and (d) same but for expanding particle, Fig.~\ref{f4e}. Scalebar: principal stretch. \textcolor{black}{(e)-(h):Deformed fiber distribution at  point where  $\la_1<1$ and  $\la_2>1$, with peak along  largest-stretch direction.  (e) Predicted distribution at a point within tether in (a),(b) $(\la_1,\la_2)=(0.2,1.06)$ with peak along the tether axis, normal to contractile-particle boundary. (f) Measured distribution from \cite{prov1} outside contractile tumor spheroid, where angles 0 and 90 are along and normal to tumor boundary, reproduced from Fig.~5 in \cite{prov1}. (g) Predicted distribution outside  simulated expanded particle in (c),(d), also Fig.~\ref{f4f}, with $(\la_1,\la_2)=(0.113,1.5)$ and peak at direction parallel to  particle boundary.  (h) Measured distribution from \cite{prov1} outside  growing tumor spheroid with peak in direction parallel to tumor spheroid boundary,  reproduced from Fig.~5 in \cite{prov1}.}}
    \label{f6}
\end{figure}

\begin{figure}
\centering
\centering
    \captionsetup[subfloat]{farskip=-0.3pt,captionskip=-0pt}
\subfloat[]{\includegraphics[width=2.0in]{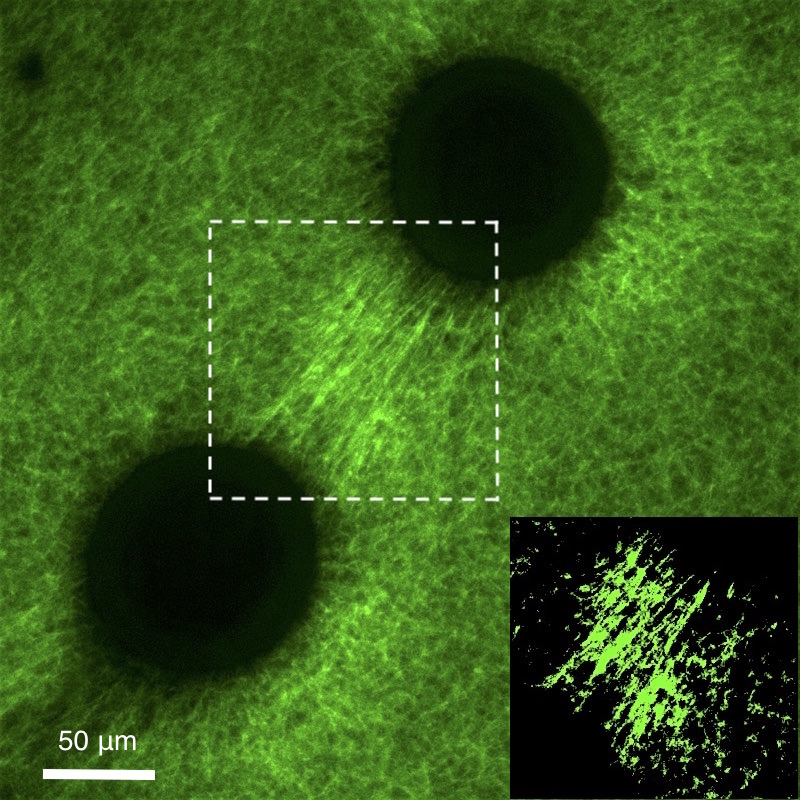}\label{f3a}}
\subfloat[]{\includegraphics[width=2.0in]{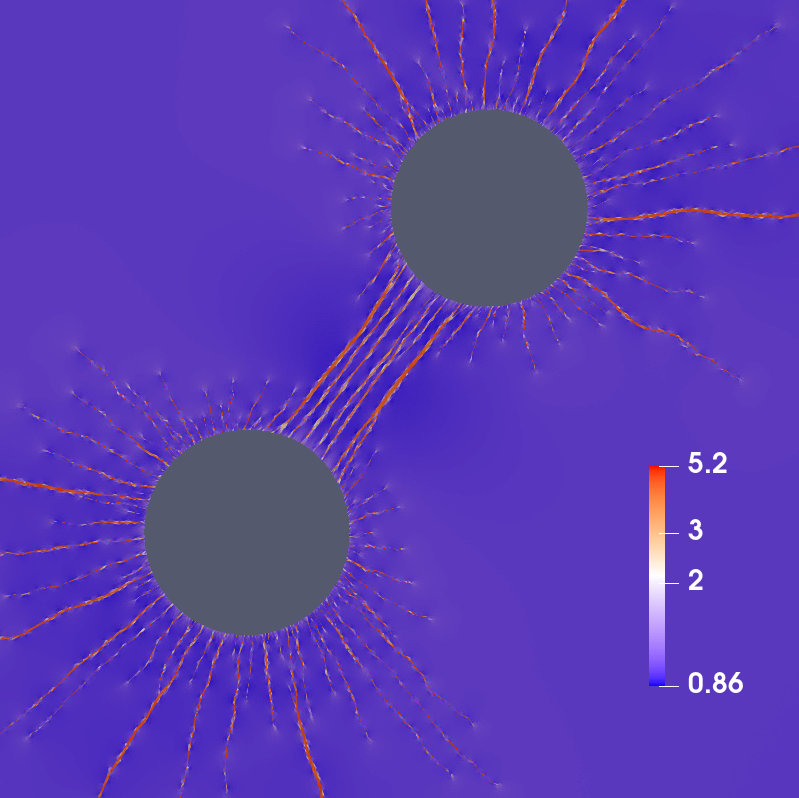}\label{f3b}}
\subfloat[]{\includegraphics[height=2.0in,width=2in]{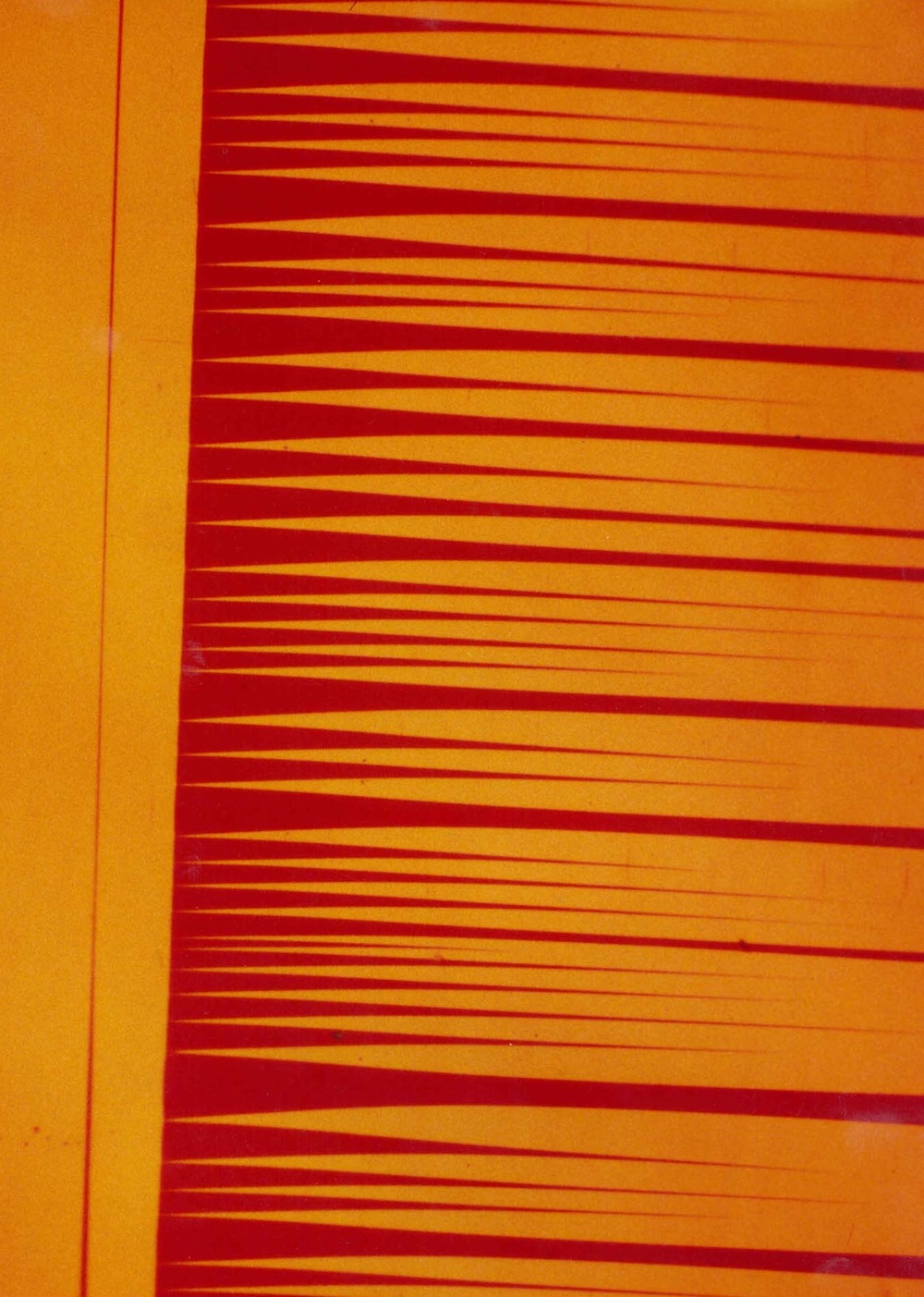}\label{f3e}}
\\
\subfloat[]{\includegraphics[height=3.06cm,width=2.5in]{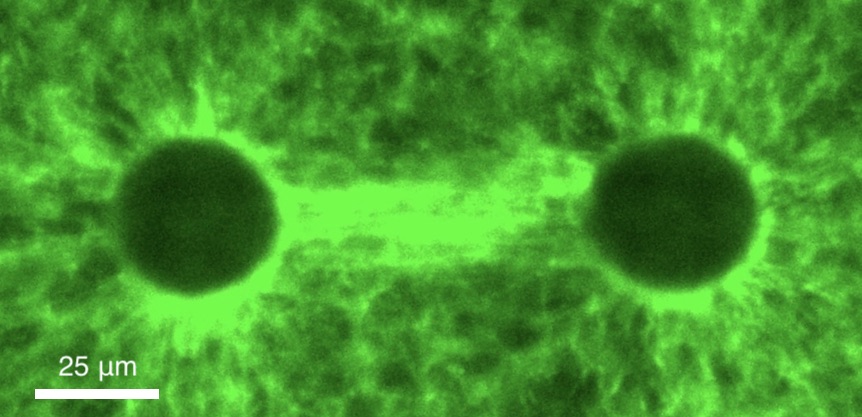}\label{f3c}}
\subfloat[]{\includegraphics[height=3.06cm,width=2.35in]{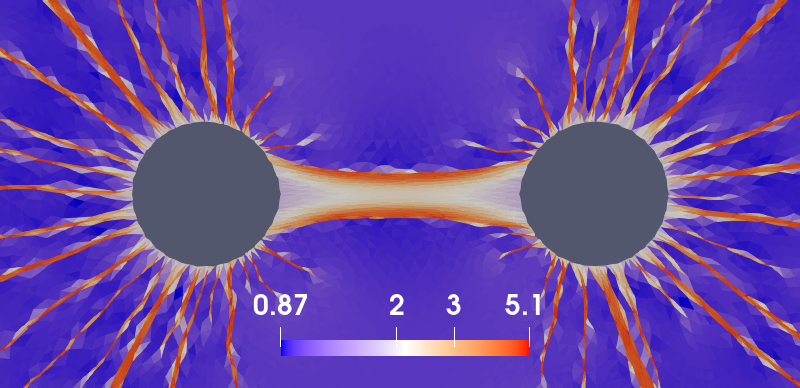}\label{f3d}}
\subfloat[]{\includegraphics[height=3.06cm]{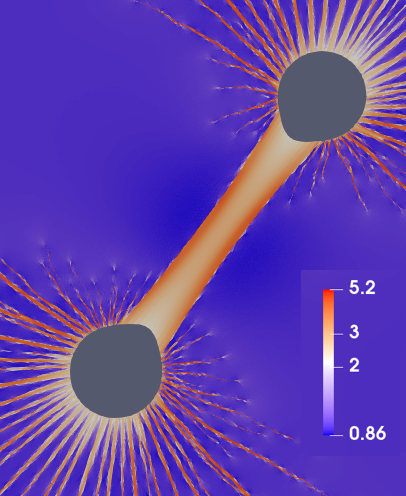}\label{f3g}}
\\
\subfloat[]{\includegraphics[width=0.5\textwidth]{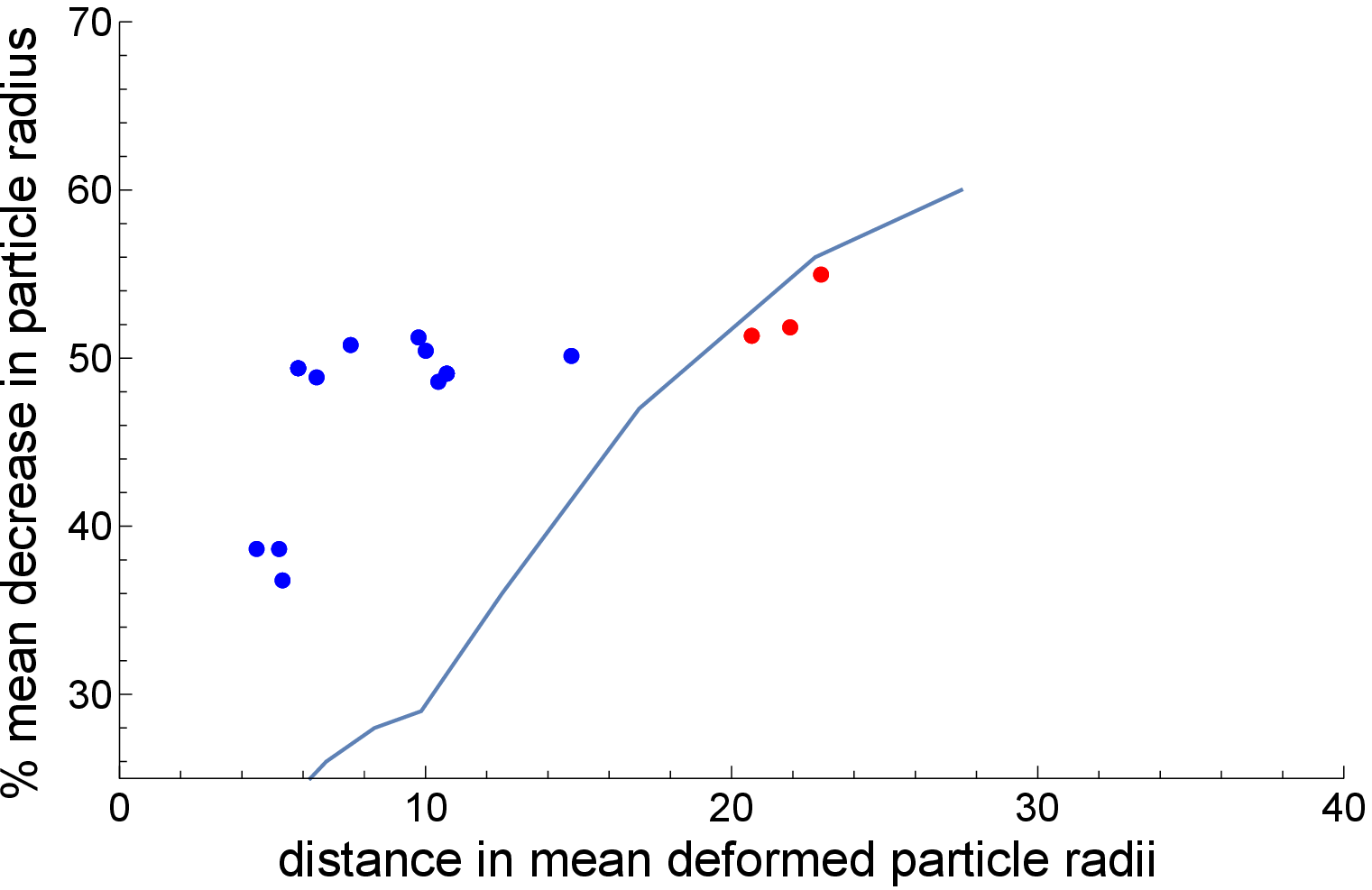}
        \label{f1h}}
  \caption{ \baselineskip12pt{\footnotesize \textbf{Splitting Hairs, Tethers and Twins.} (a) experiment and (b)  simulation of an active particle pair that contracted by $38$\%.  (a) A tether fully split into multiple thin bands.  Insert shows high-contrast version of area within dotted rectangle. (b) Simulation with initial radii, distances and contractions matched with (a) also results in split tether.
 (c) Martensitic twins  split into multiple bands; some bands have split tips near incompatible vertical boundary (experiment photo courtesy of C. Chu \& R.D. James). (d) Two smaller particles at higher 50\% contraction with full-contact tether (except torn fibers at right particle) and partially enveloping densification. (e)  Simulation of (d) shows full-contact tether with no splitting.
  (f) Hypothetical simulation of particles of (a) but extremely high contraction of 80\% (not attained by active particles) shows uniform tether and densification enveloping particles, due to compatibility of contraction with densified phase. (g) Predicting whether a tether forms between two particles.  Blue curve: separatrix constructed from simulations. Axes: \% decrease in particle radius  vs deformed distance (in deformed particle radii). Above separatrix,  tethers are predicted to form between particle pairs. No tether is predicted to form below separatrix. Our  experimental data  (each particle pair is one point) abide by the prediction: 
  \tred{blue points: tether has formed. Red points: no tether has formed. \scd}}}
 \centering
 \label{f3}
\end{figure}

\begin{figure}
    \centering
    \subfloat[]{\includegraphics[height=4.8cm,width=5cm]{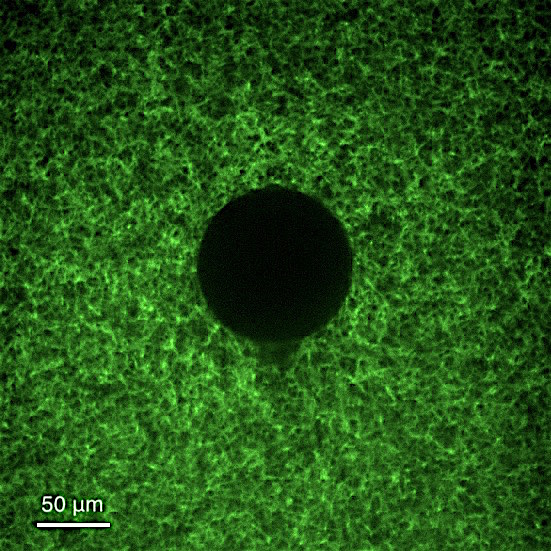}\label{f4a}}
    \hspace{0.1cm}
    \subfloat[]{\includegraphics[height=4.8cm,width=5cm]{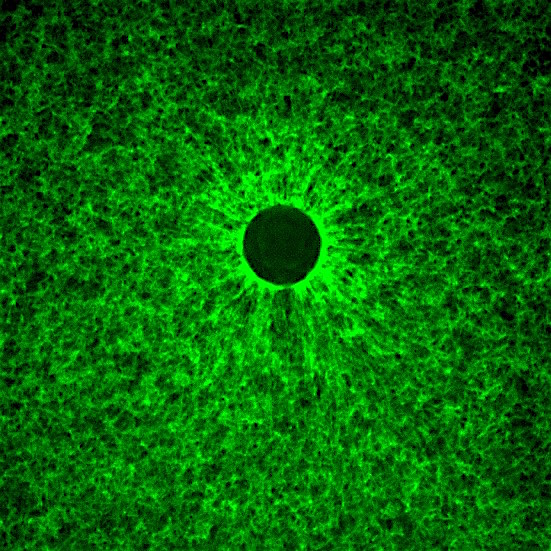}\label{f4b}}
      \hspace{0.1cm}   
    \subfloat[]{\includegraphics[height=4.8cm,width=5cm]{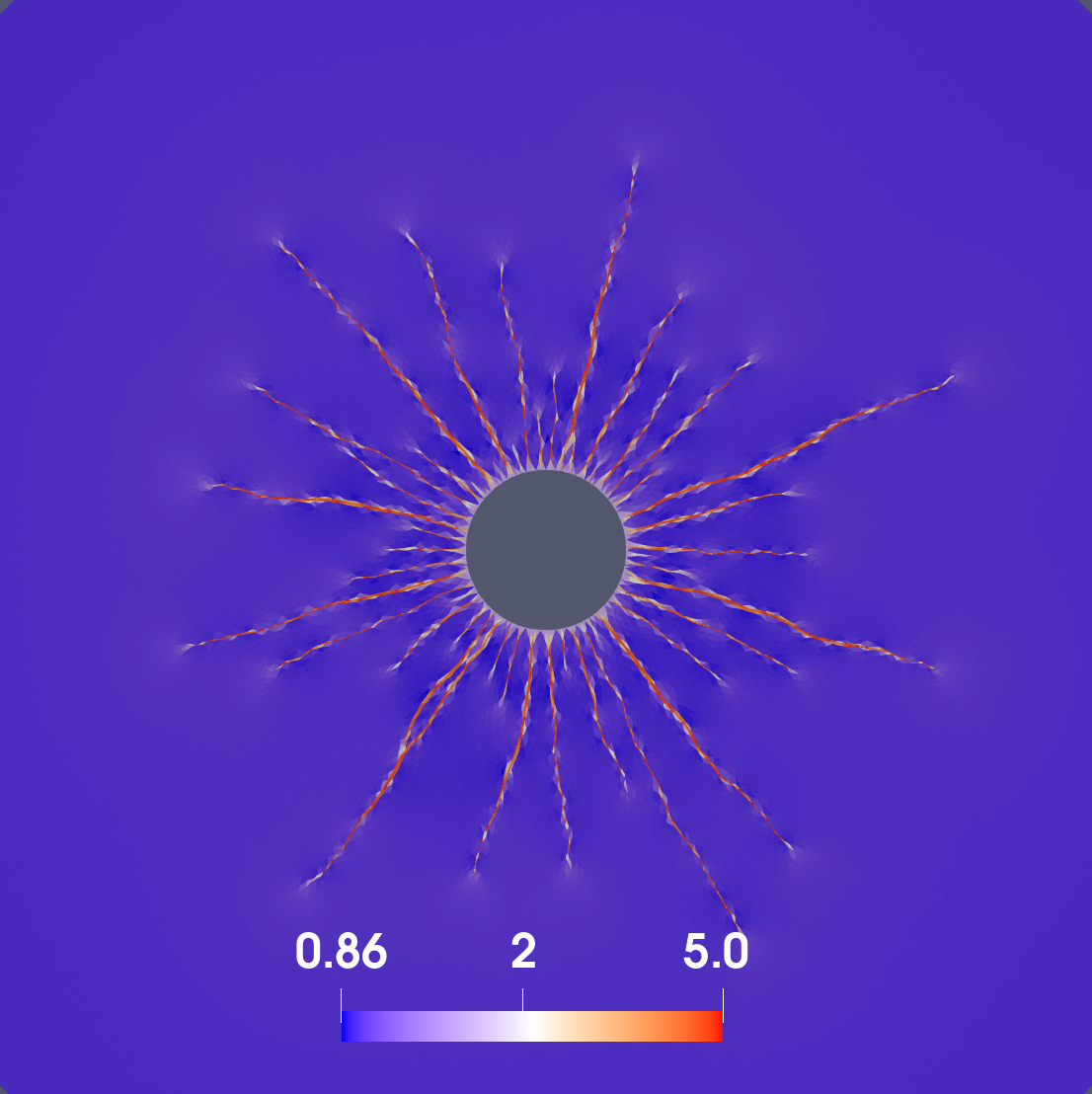}\label{f4c}}
    \\
    \subfloat[]{\includegraphics[height=4.8cm,width=5cm]{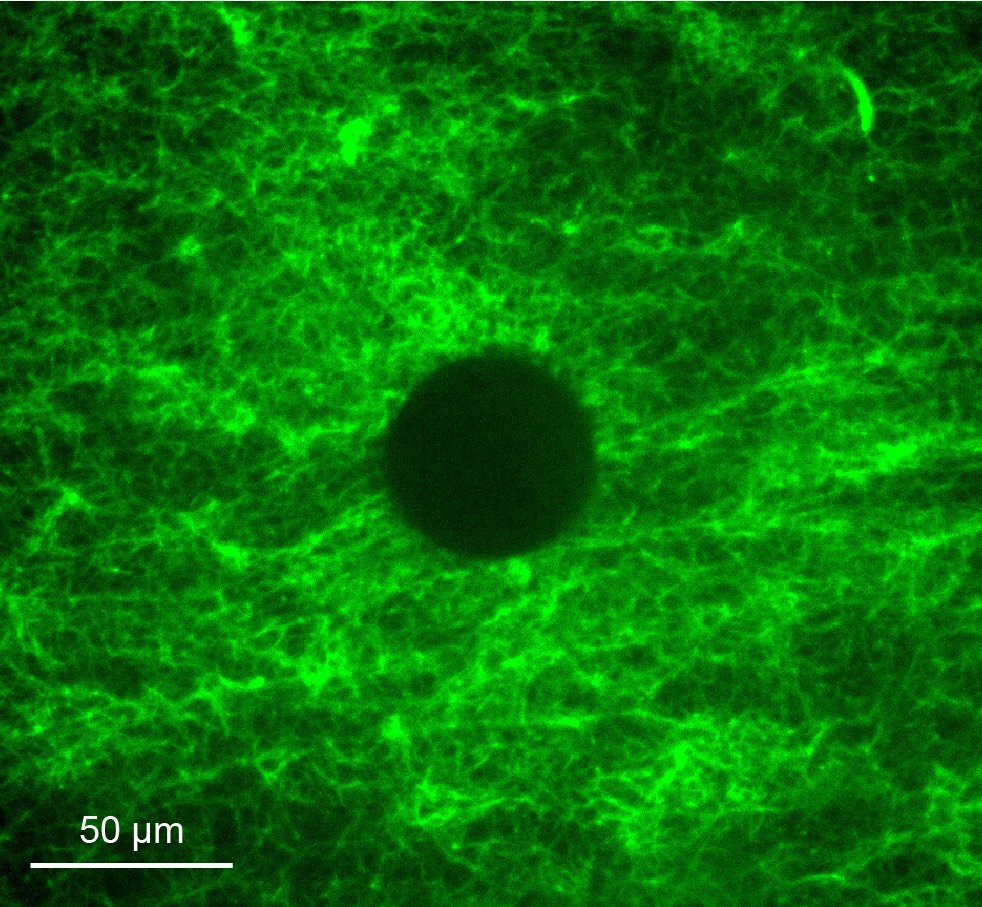}\label{f4d}}
    \hspace{0.1cm}    
    \subfloat[]{\includegraphics[height=4.8cm,width=5cm]{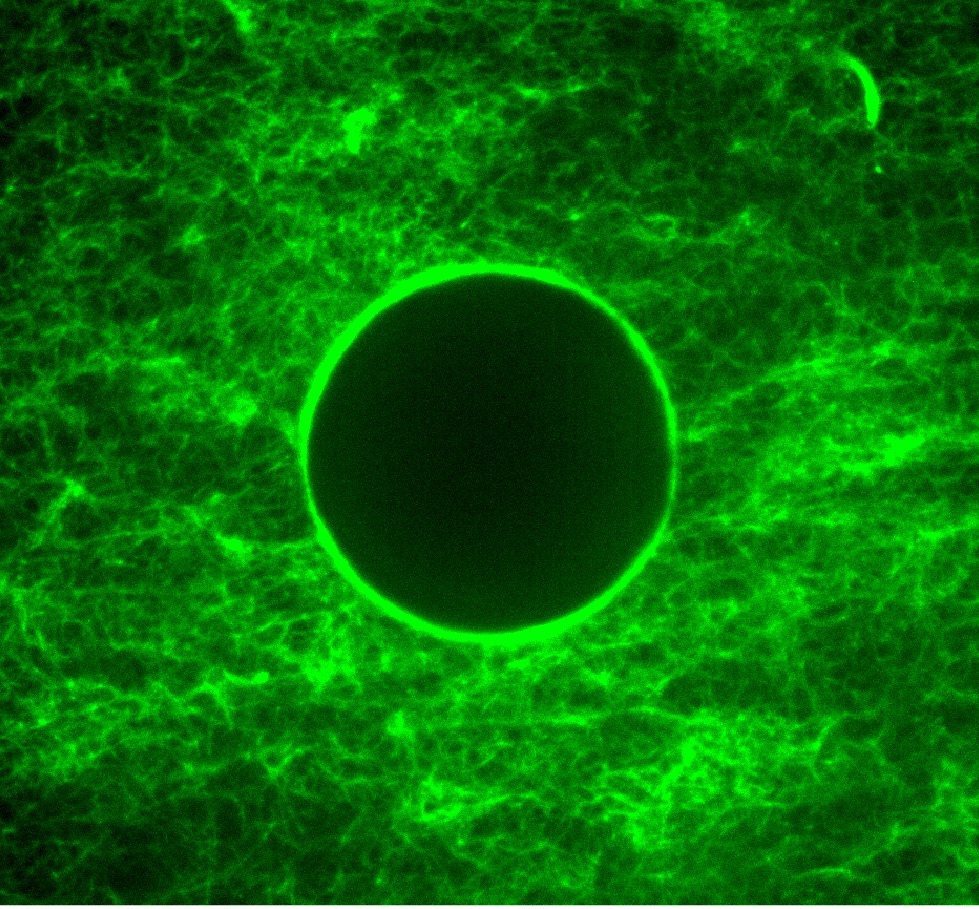}\label{f4e}}
    \hspace{0.1cm}   
    \subfloat[]{\includegraphics[height=4.8cm,width=5cm]{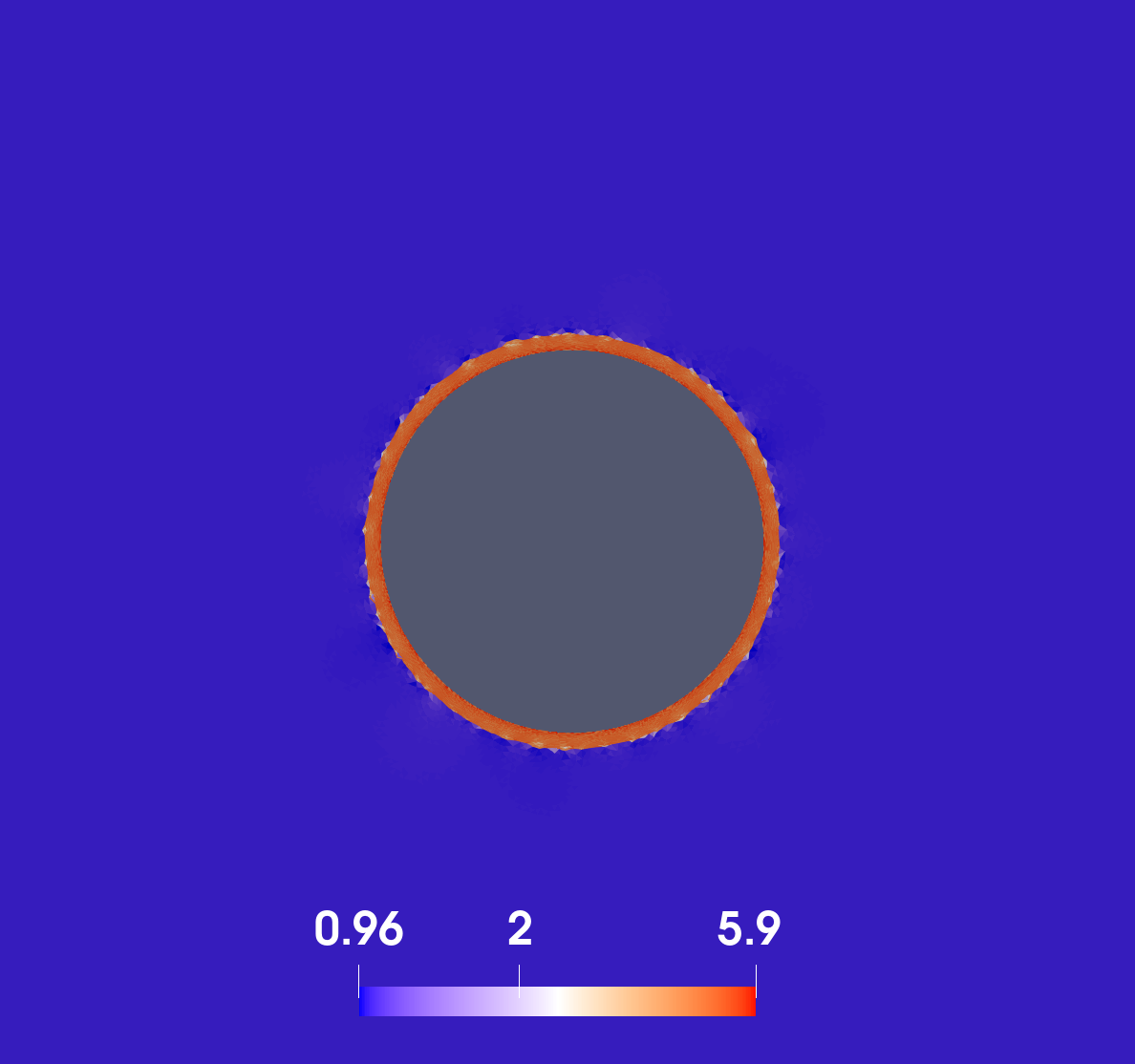}\label{f4f}}
      \caption{\baselineskip10pt{\footnotesize \textbf{Contracting vs Expanding Particles.} Experiments (green) and simulations  (purple) matching initial geometry of particles and contraction level. (a)-(c)  Contracting  particle. (a) Undeformed PNIPAAm particle. (b) PNIPAAm particle radially contracted by 50\%   (c) Simulation of contracting particle in (b). Note radial ``hairs'' and radial symmetry breaking. (d)-(f) Expanding  particle. (d) Unexpanded (undeformed) PNIPAAm particle (e) PNIPAAm particle radially expanded by  
    50\%. Note bright circumferential  densification layer. (f) Simulation  prediction for radially expanded particle in (e). Note circumferential densified phase layer surrounding particle, circular phase boundary and radial symmetry maintained. \scd}}
    \label{f4}
\end{figure}

\begin{figure}
\centering
    \captionsetup[subfloat]{farskip=5pt,captionskip=-0.5pt}
\subfloat[]{\includegraphics[width=2in]{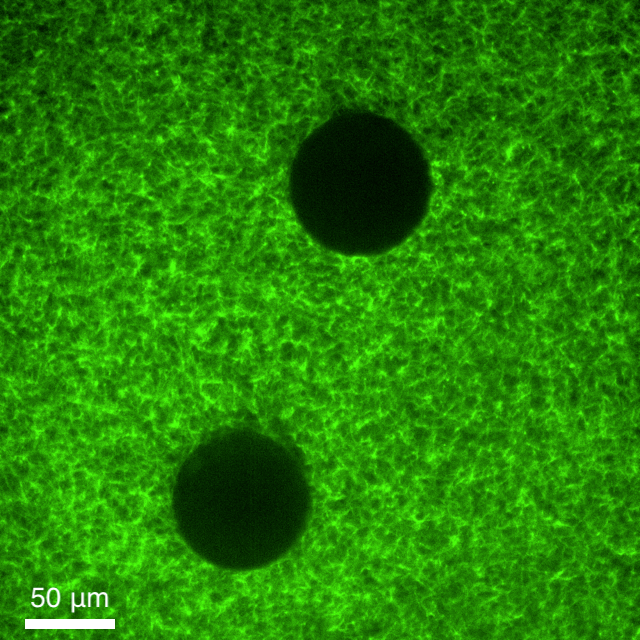}\label{f5a}}
\subfloat[]{\includegraphics[width=2in]{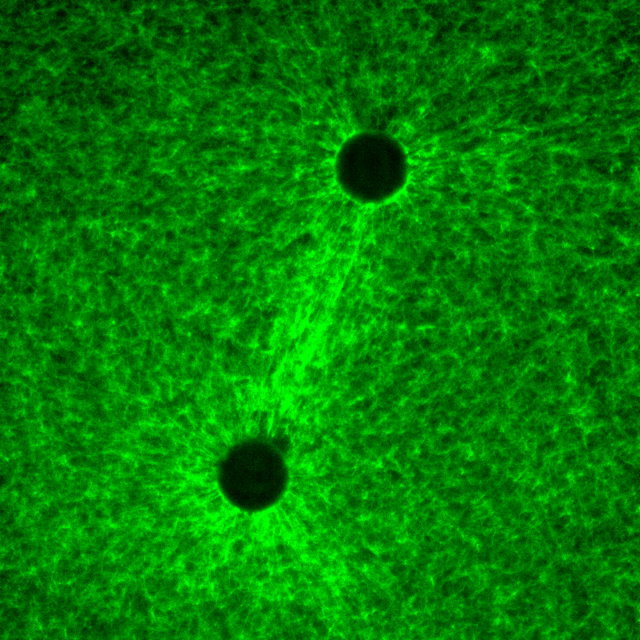}\label{f5b}}
\subfloat[]{\includegraphics[width=2in]{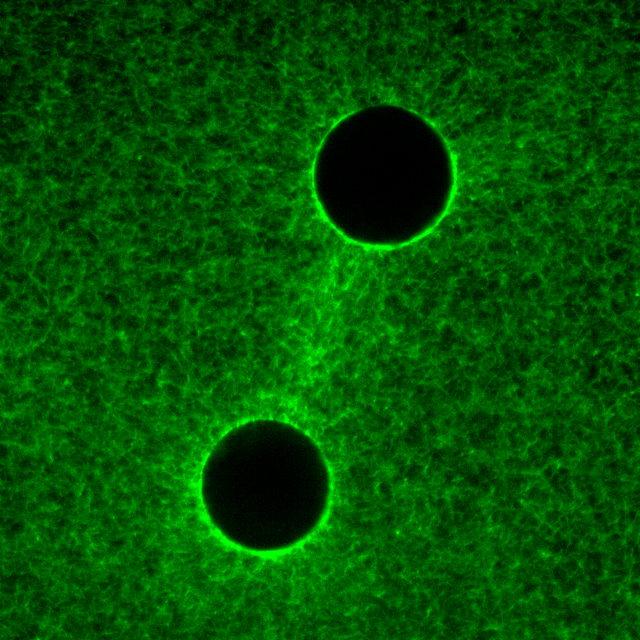}\label{f5c}}
\\
\subfloat[]{\includegraphics[width=2in,height=1.97in]{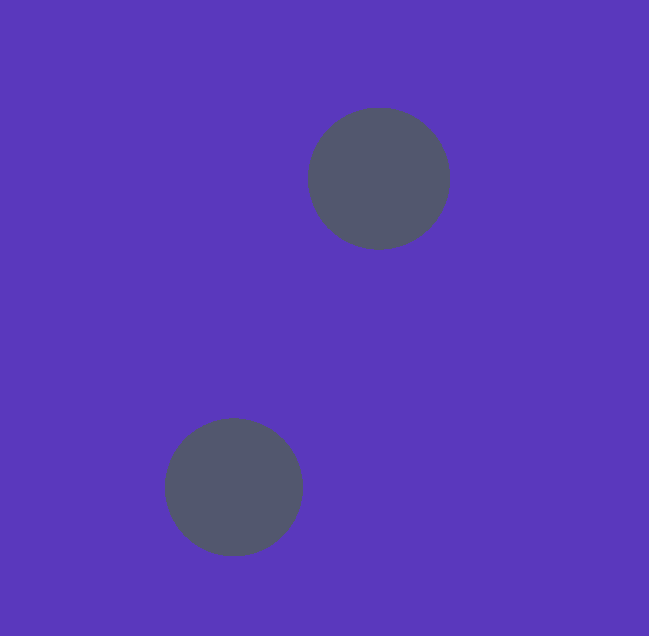}\label{f5d}}
\subfloat[]{\includegraphics[width=2in,height=1.97in]{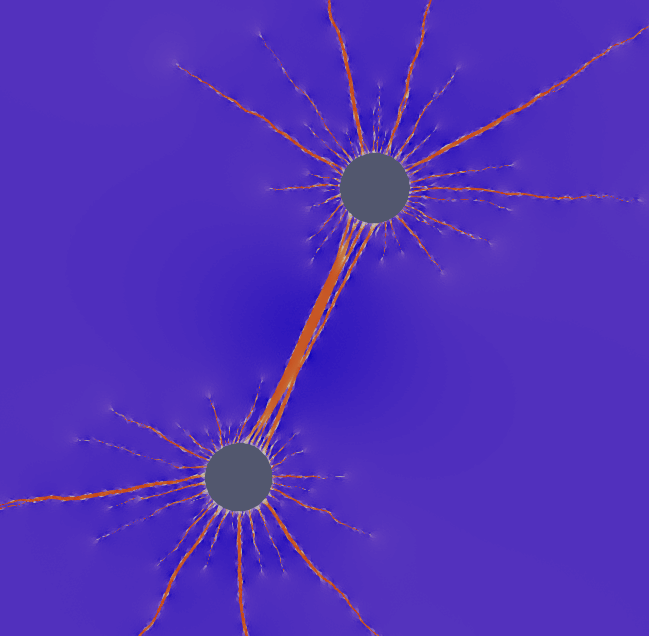}\label{f5e}}
\subfloat[]{\includegraphics[width=2in, height=1.97in]{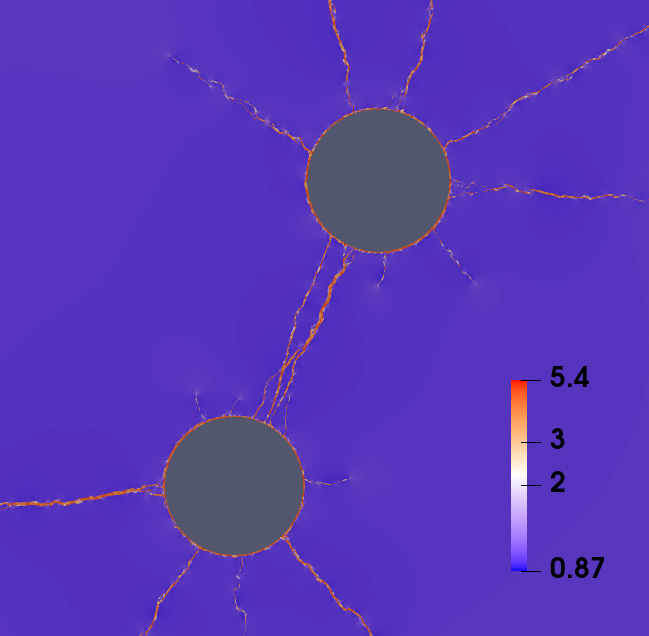}\label{f5f}}
    \vfill\break\eject
    \subfloat[]{\includegraphics[width=0.6\textwidth]{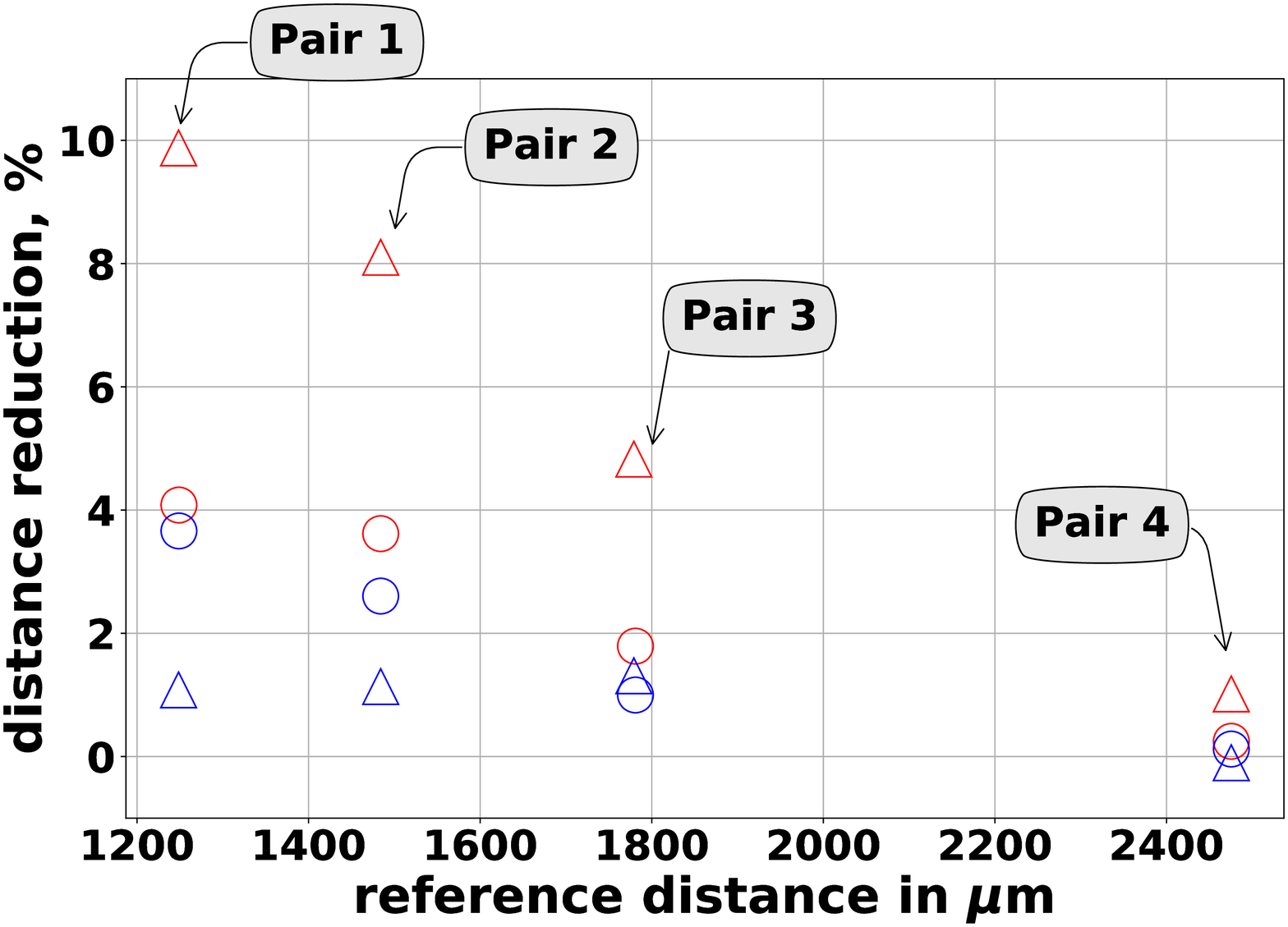}
        \label{f1g}}
      \caption{ \baselineskip12pt{\footnotesize \textbf{Reversing Contraction: Residual Densification.} (a)-(c): Experiments (green). (a) Two uncontracted active particles at $26^\circ$C (b) Contracted at $39^\circ$C (c) Re-expanded at $26^\circ$C. (d)-(f): Simulations (purple) of (a)-(c) with matched initial radii, distance between centers and radius  contraction ratios. (d) Undeformed. (e) Contracted, simulating (b). (f) Re-expanded, simulating (c). Note residual weaker tether and thinner,  fewer hairs  after reexpansion.  Also newly formed densification ring-around-the-particle in  (f) in agreement with (c). \scd (g) Relative motion of pareticle centers, for four pairs of active particles (each pair joined by a tether) during contraction (red) and reexpansion (blue).  Circles: experiments. Triangles: simulations.}}
   \label{f5}
\end{figure}

\begin{figure}
    \centering
    \subfloat[]{\includegraphics[width=4in]{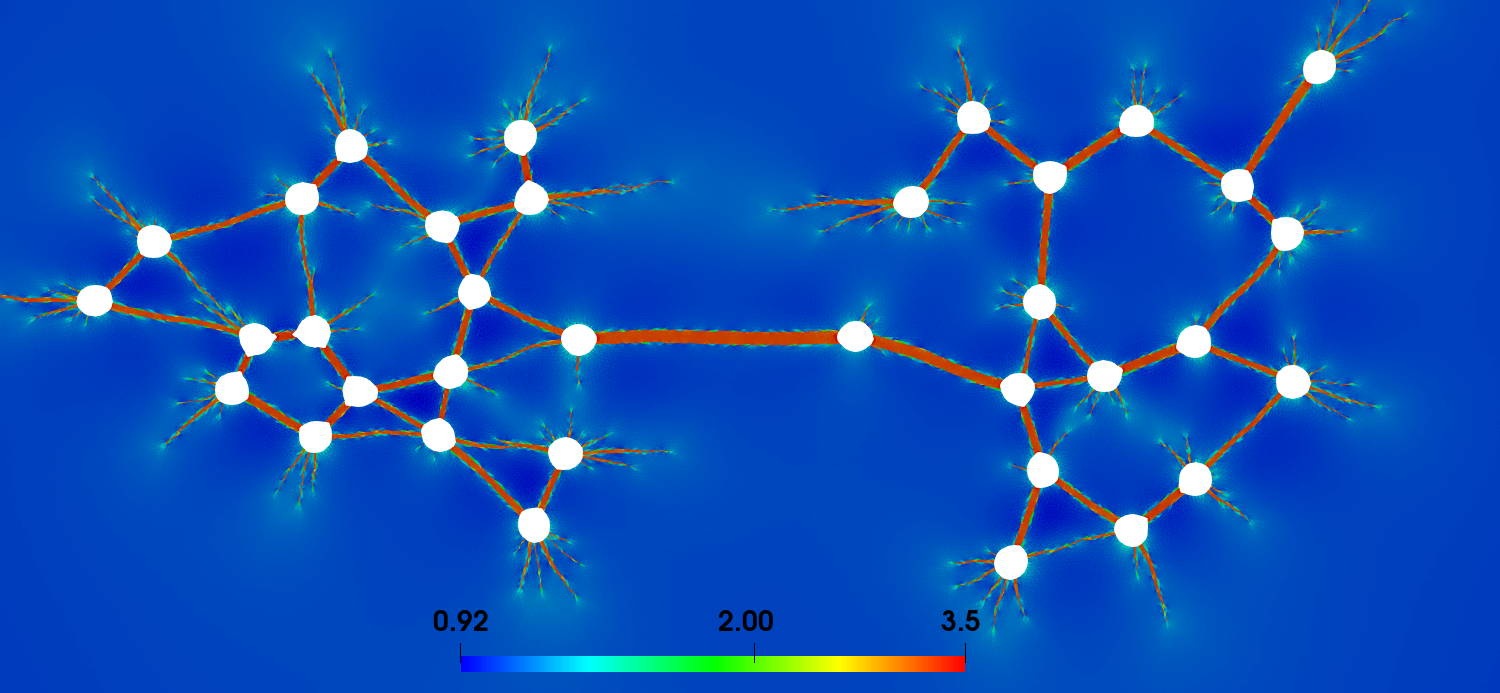}\label{f7a}}
    \\
     \subfloat[]{\includegraphics[width=4in]{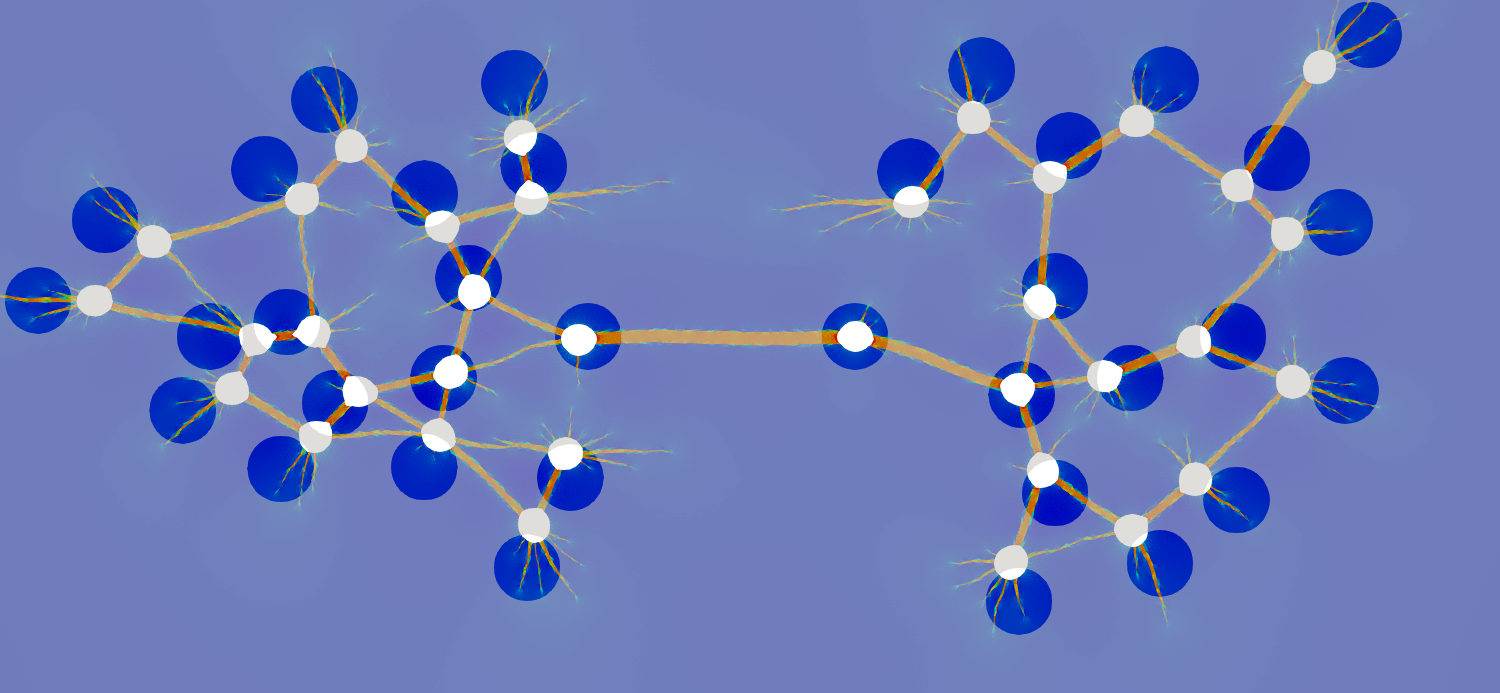}\label{f7b}}
     \\
\subfloat[]{\includegraphics[height=1.5in]{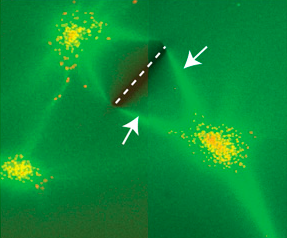}\label{f5g}}
\subfloat[]{\includegraphics[height=1.5in]{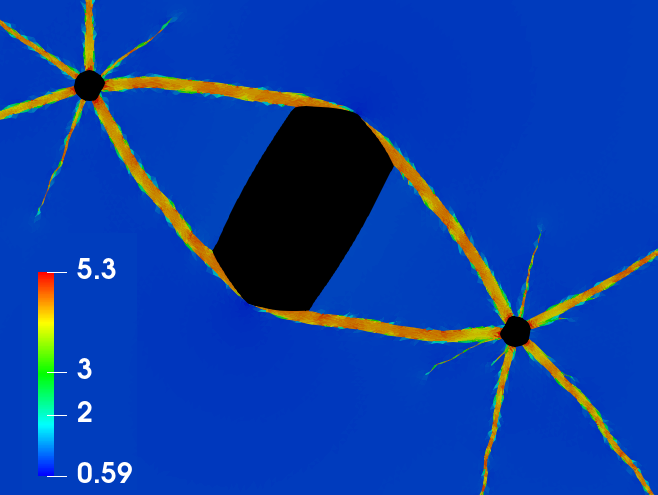}\label{f5h}}
  \caption{\textcolor{black}{Simulation with multiple acini. Their contraction causes an extensive network of tethers to form which mechanically remodels the ECM (a) deformed configuration. White: contracted acini, red: densified phase, blue: undensified phase. (b) superposed undeformed configuration (dark blue acini) prior to contraction and  deformed configuration  (white acini) showing motion and shape distortion as well as contraction. Orange: densified phase, light-blue: undensified phase in the deformed configuration. (c)-(d) A crack (dotted line) is cut across a tether between acini \cite{shi2014rapid}, the tether disappears. (c)  Two new tethers turn around the corners of the crack to bypass it (reproduced from \cite{shi2014rapid}). (d) Simulation of pair of acini with crack predicts tether bypassing the crack. \scd}}
    \label{f7}
\end{figure}

\vfill\eject

\subsection*{Author Contributions}  PR, GR and JN planned the research. GG and PR developed the model. GG and CM developed the numerical method. GG performed the simulations.  MP and JN designed the experiments. MP performed the experiments; MP and JN analyzed the data. All authors discussed and analyzed the results and contributed to writing the manuscript. {\bf Conflict of Interest.} The authors declare no conflict of interest. 

\subsection*{Funding} The work of PR, CM and GG was partially supported by the EU
Horizon 2020 Research and Innovation Programme under the Marie Sklodowska-Curie project  ModCompShock ({\tt modcompshock.eu}) agreement No 642768. The research of GG was also supported by a Vannevar Bush Faculty Fellowship.
The work of JN and MP and  was partially supported by National Science Foundation grant number CMMI-1749400. GR acknowledges the support of the National Science Foundation (DMR No. 0520565) through the Center for Science and Engineering of Materials at the California Institute of Technology. 

\subsection*{Data Accessibility}
The numerical code used in the simulations reported herein  as well as 
experimental data are available at
https://github.com/ggrekas/rsif-2020-0823.

\subsection*{Acknowledgments}
We thank Brian Burkel for assistance in microscopy. 

\bibliographystyle{naturemag}

\bibliographystyle{abbrv}

\bibliography{references}

\newpage

 \section*{SUPPLEMENTARY INFORMATION FOR: \\Cells exploit a phase transition  to mechanically remodel the fibrous extracellular matrix, $\;\;$  \textit{J. R. Soc. Interface 20200823}}
\author{Georgios Grekas, Maria Proestaki Phoebus Rosakis, Jacob Notbohm,\\ Charalambos Makridakis \& Guruswami Ravichandran}
 
 \vskip0.5cm
 \noindent \SN
 
\noindent Supplemental Figures S1-S3
 
 \noindent Tables S1, S2

\beginsupplement 
 
\section*{Supplemental Notes}
\paragraph*{Compressive Stretch Estimate.} 
To estimate compressive strains we note that in the tether  the density ratio  $\rho/\rho_0$ of deformed to undeformed density is typically in the range \cite{ban2018}  3-5 also observed in our experiments. From mass balance we have that $\rho\la_1\la_2=\rho_0$. Thus the inequality $\rho/\rho_0\ge 3$ and the fact that the tether is in longitudinal tension so that $\la_2\ge 1$  give the inequality  $\la_1<1/3$. This is roughly 70\% compressive strain $\eps_1=\la_1-1\approx-0.67$.  \tblue{This is consistent with the compressive stretch we estimated from  [\!\cite{shi2014rapid} supplemental video sm14], where the  evolving compression is recorded and the compressive strain can be obtained  from the deformation of gridlines deposited on the ECM.}

\paragraph*{Stiffness Loss.}  For a network of fibers with force-stretch relation as in Fig.~\ref{f2a}, we show that the orientation averaged stress goes to zero in the crushing limit. Consider  compression along the $y$-axis with stretch $0<\la_c<1$ combined with tension along the $x$-axis with stretch $\la_t\ge 1$. Then a fiber that makes an angle $\theta$ with the compression ($y$)-axis in the undeformed state, will make an angle $\theta_\ast$ such that $\tan\theta_\ast=(\la_t/\la_c)\tan\theta$.  Since $\la_t/\la_c>1$ we have $\theta_\ast>\theta$ and the angle of the fiber with the direction of compression increases and approaches $90^\circ$ in the crushing limit as $\la_c\to 0$.  This reorientation tends to decrease the component of the load in the compression direction a fiber can sustain. For example, consider uniaxial compressive stretch with $\la_t\ge1$ fixed  and $\la_c\to 0$. The fiber stretch is $\sqrt{(\la_c\cos\theta)^2+(\la_t \sin\theta)^2}$ The component of  fiber axial force in the $y$ direction of compression is $T=S\cos\theta_\ast$. Now $S$ is bounded in compression and as $\la_c\to 0$,  $\theta_\ast\to\pi/2$ so $\cos\theta_\ast\to 0$. As a result the component of the fiber force along the compression direction $T\to 0$.  The compressive stress is $\int T d\theta =\int S \cos\theta_\ast d\theta$.  Now $S\cos\theta_\ast$ is bounded for compression and $\cos\theta_\ast$ converges (almost everywhere) to zero (except when $\theta=0$) so bounded convergence implies that $\int T d\theta\to 0$ as $\la_c\to 0$. As a result  the material fails to sustain uniaxial compressive stress in the crushing limit, and the qualitative behaviour is necessarily as in Fig.~\ref{f2b}.

\paragraph*{Single Fiber behaviour.}
Modeling  the force-stretch relation of a single fiber depends on the post-buckling behaviour of a flexible beam, which can be quite complicated as buckling is a bifurcation/instability phenomenon. There are typically three main types of post-buckling response $S(\la)$, depending on whether  stability is maintained or lost on the bifurcating branch of buckled states \cite{simhadri2019}, and whether it is later regained if lost: (i) Stable post-buckling, where stiffness is diminished but load carrying capacity in compression is not lost  (as $\la$ decreases below $1$). An example is $S(\la) =  k(\la^3-1)$ with $k>0$ a constant (Fig.~\ref{f2a}). (ii) Unstable (collapse) post-buckling, where the stiffness becomes negative (stability is lost) with increasing compression, e.g., $S(\la) =  k(\la^5-\la)$.  (of the form shown in Fig.~\ref{f2b}). (iii) Snap-through post-buckling, where stability is first lost but eventually regained at higher compression \cite{simhadri2019}, for example $S(\la)=k\left(\la^5-3\la^3 /2+\la/2\right)$ (qualitatively as in Fig.~\ref{f2e}). 

\paragraph*{Ellipticity Loss.} The strong ellipticity condition \cite{Knowles1976} plays a central role in coherent phase transformations in solids. It is closely related to the rank-one convexity condition \cite{ball1976} and implies local stability of the material. In one dimension it is equivalent to positive slope (tangent modulus) of the (nonlinear) stress-strain curve, so it would fail for some intermediate strains where the slope is negative in Figs~\ref{f2c}, \ref{f2f}. In higher dimensions its meaning  is  more complex \cite{rosakis1990}. An elastic energy density function that is globally strongly elliptic cannot sustain strain discontinuities and phase transitions \cite{Knowles1978}. In order to determine for what values of the principal stretches strong ellipticity fails in our model energy density functions $W$ and $W^*$, we use the criteria of Knowles \& Sternberg \cite{Knowles1976}.  The result is shown in Supplemental Figs~\ref{sf2d}, \ref{sf2e}.

\paragraph*{Compatibility.} The compatibility condition for strain discontinuities states that for two different values of the deformation gradient $\bF_1$ and $\bF_2$ to occur on either side of a strain discontinuity surface, across which the displacement is continuous (such as a coherent phase boundary or twin boundary), they must be rank-one connected, namely  satisfy $\bF_2-\bF_1={\bf a}\otimes{\bf n}$ for some vector ${\bf a}$, where ${\bf n}$ is the unit normal to the surface \cite{Knowles1978}. In case $\bF_1={\bf 1}$ (undeformed state) then the principal stretches of $\bF_2$ must then satisfy the inequalities \cite{Ball1987,Ball1999} 
\beq\label{s1}\la_2\ge 1 ,\qquad \la_1\le 1\eeq
It turns out that in our model the minimal energy principal stretches  $(\la^*_1,\la^*_2)$ at the energy well corresponding to the densified phase always satisfy these inequalities.  In the specific example of the bistable energy $W^*$ used in our simulations,  $(\la^*_1,\la^*_2)=(0.2,1.06)$ (Fig.~\ref{sf2a}).  Since the energy is a symmetric function of  $(\la_1,\la_2)$, the minimum at  $(\la^*_2,\la^*_1)$ corresponds to the same state as $(\la^*_1,\la^*_2)$.  The  additional minimum  at  $(\la_1,\la_2)=(0.45,0.45)$ is not compatible with the undeformed state $(1,1)$, since it violates Eq.~(\ref{s1}). The principal stretches in our simulations are never observed to take values at or near this state, This  further illustrates the role of compatibility.

In   Fig.~\ref{f6} the phase boundary between the tether and the rest of the  ECM is roughly parallel to a principal direction of stretch. In that case compatibility requires that the stretch in the direction parallel to the phase boundary has to be continuous across it  (in this case the tensile stretch $\la_{2}$, Fig.~\ref{ff3b}, whereas the compressive one $\la_{1}$ (Fig.~\ref{ff3b}) may be discontinuous as seen in the simulation.  

\paragraph*{Principal Stretches.} An isotropic elastic energy density function  $W(\bF)$  in 2D, normally expressed in terms of the deformation gradient matrix $\bF$, can be written as a function of the principal stretches $\la_2\ge\la_1>0$ which are the eigenvalues of the right stretch tensor  $\bU=(\bF^T\bF)^{1/2}$.  Typically in a tether between similar clusters,  there is moderate tension along its axis and sever compression normal to it, thus the principal axes of stretch roughly correspond to these two directions.  Because of compatibility, the compressive stretch $\la_1$ can jump across the tether boundary, but the tensile one cannot, as it stretches the material roughly parallel to the boundary. This is encountered in experiments \cite{ban2018} and captured  in our simulations as seen in   Fig.~\ref{f6}

\paragraph*{\textcolor{black}{Fiber Alignment in the Densified Phase.}} \textcolor{black}{By considering how a homogeneous deformation changes fiber directions, it is easy to show that an undeformed uniform distribution of fibers becomes a  distribution 
\beq\label{s2}g(\varphi)=\frac{(1/2\pi)}{ ({\la_2}/{\la_1})\cos^2\varphi+({\la_1}/{\la_2})\sin^2\varphi}\eeq
where $\la_2\ge\la_1>0$ are the principal stretches and $\varphi$ is the angle between  the deformed fiber and the maximum stretch direction in the deformed state, or eigenvector of the left Cauchy-Green tensor $\bF\bF^T$ with eigenvalue to  $\la_2^2$. This distribution becomes uniform when $\la_1=\la_2$, whereas for   $\la_2>\la_1$ it has a peak at $\varphi=0$, the direction of highest tension, Fig.~\ref{f6}.  To derive this, consider a biaxial stretch with  principal stretches $\la_2\ge\la_1>0$ along orthogonal directions. Let $\varphi$ be the angle between  the deformed fiber and the deformed  direction of largest stretch. Then $\varphi=f(\theta)=\arctan(\la_2\sin\theta/\la_1\cos\theta)$ where $\theta$ is the corresponding undeformed angle.  The referential distribution is then $g_0(\theta)=d f(\theta)/ d\theta$ and the deformed distribution $g(\varphi)$ satisfies $\int g(\varphi) d\varphi =\int g_0(\theta) d\theta$, which after changing variables from $\theta$ to $\varphi=f(\theta)$ and some algebra gives Eq.~(\ref{s2}), with a constant factor ensuring that $\int_0^{2\pi} g(\varphi) d\varphi =1$.}

\textcolor{black}{In the densified phase, stretches are close to the energy minimum at  $(\la^*_1,\la^*_2)=(0.2,1.06)$. For these values the deformed fiber distribution is as in Fig.~\ref{sf4a}. This situation occurs within  tethers. Here the angle is plotted relative to the particle boundary so $\varphi=90$ is along the tether axis.}

\paragraph*{\tred{Simulation Parameters.}} \tred{The model parameters used in all simulations of our own experiments are listed in Supplemental Table S1. In (5.1) we chose the smallest even powers that will allow the S-shaped form shown in Fig.~\ref{f2e}. There are 4 model parameters. The value of $b$  in (5.3) was chosen to control the density ratio of the densified phase to roughly equal 5, which is within the range 3-6 reported by \cite{ban2018} and deduced from [\cite{shi2014rapid} supplemental video sm14]. Parameter $A$ was given a high value to ensure a steep enough barrier against total volume collapse  that at the same time does not affect the energy for larger volume ratios.  Parameter $a_m$ in (5.1) was chosen so as to result in equal heights of the energy wells (densified and undensified phases) of the bistable energy $W^*$ in (5.2), (5.4).}

\tred{The Young's Modulus of the PNIPAAm active particles with the same concentration of N-isopropylacrylamide as used here, but a lower concentration of crosslinker, was measured to be $>20$ kPa \cite{matzelle2003}. Also the modulus of acini such as the ones used by \cite{shi2014rapid} is reported in the range 1--4 kPa \cite{acinimodulus}, so that active particles are roughly an order of magnitude stiffer than cell clusters. Accordingly, in our simulations we take the bulk modulus of active particles to be ten times larger that that of acini. In terms of nondimensional spring stiffness, this translates to $k=10$ for active particles in (5.5),  whereas for acini we take $k=1$ in (2.5). In comparison, the shear modulus of the bistable energy $W^*$ for small deformations is $0.1$. This is consistent with measured values of the shear modulus of 3 mg/mL collagen in the range 80--100 Pa \cite{sharma2016}.}

The values reported in Table S1 were used in all simulations of our own experiments.
\tred{In simulations shown in Figs \ref{f3}-\ref{f5} corresponding to our own experiments, we used the experimentally measured values of initial particle radii, initial center distances and particle contractions. These are listed in Table S2.}

\tred{In Figs~\ref{f1d}-\ref{f1f}, simulations are not intended to match specific experimental data, but to qualitatively show that the basic tether/hair morphology persists for different parameter choices although its details are affected.  We use the monostable energy; see Table S1.  All other parameters are as indicated in the caption.}

\vfill\eject

\begin{figure}[h!]
    \centering
    \subfloat[]{\includegraphics[width=5in]{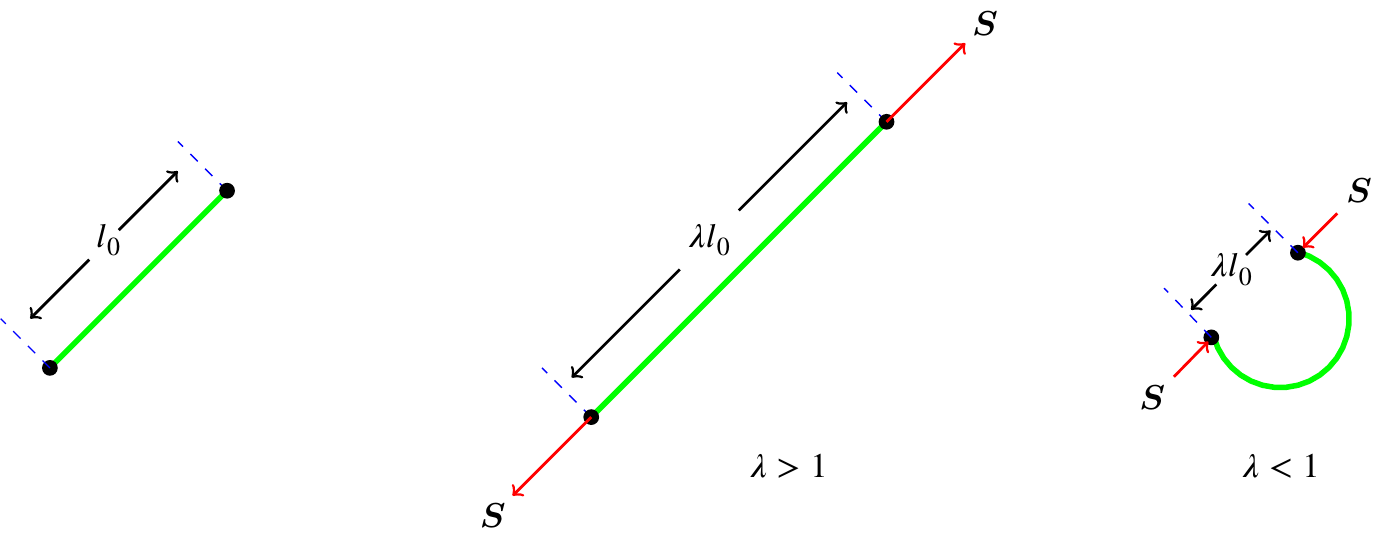}\label{sf1a}}
    \\
    \vspace{1cm}
    \subfloat[]{\includegraphics[width=2in]{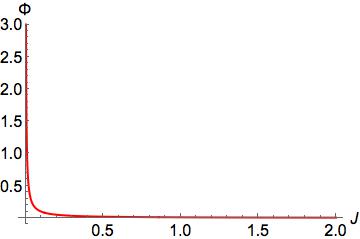}\label{sf1b}}
    \hspace{0.2cm}    
    \subfloat[]{\includegraphics[width=1.5in]{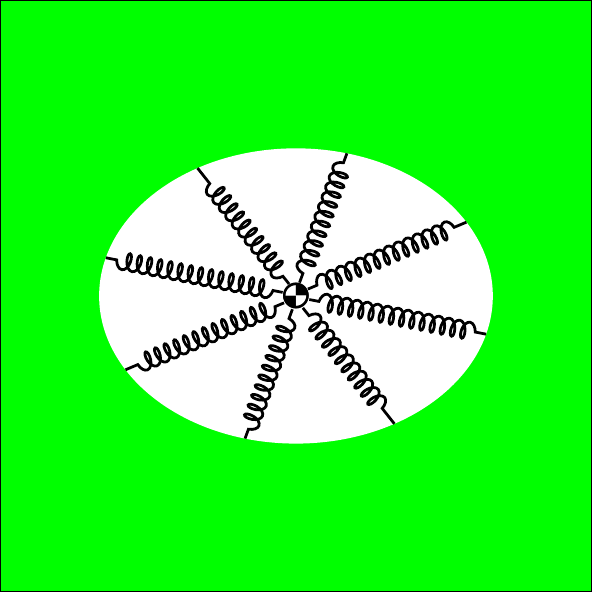}\label{sf1c}}
    \hspace{0.2cm}    
    \subfloat[]{\includegraphics[width=1.5in]{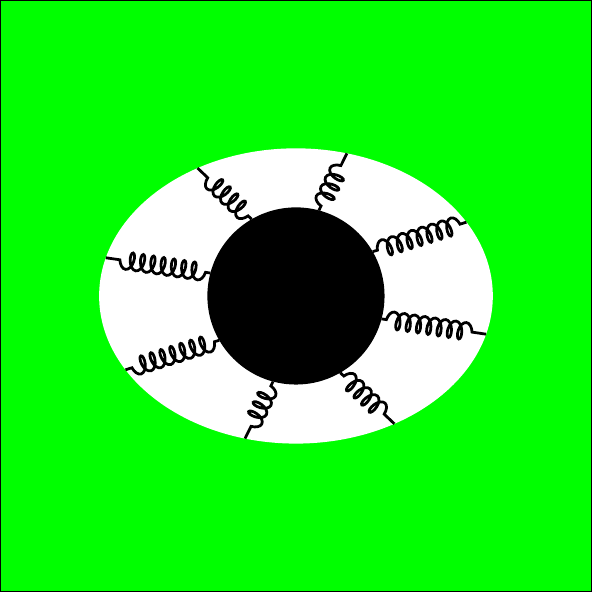}\label{sf1d}}
    \caption{Various Aspects of the Model. (a) The effective stretch $\la$ of a single fiber is the ratio of deformed to undeformed  distance $l_0$ of its endpoints. From left to right: relaxed (undeformed), under tension, buckled under compression. Red arrows represent forces. (b) Fiber volume penalty function $\Phi(J)$.(c) Soft model for clusters, explants or acini, contributing  term  Eq.~(2.5) to the total energy.  (d) Stiff model for active PNIPAAm particles contributing  term Eq.~(5.5) to the total energy.}
    \label{sf1}
\end{figure}

\begin{figure}[h!]
    \centering
    \subfloat[]{\includegraphics[width=2.5in]{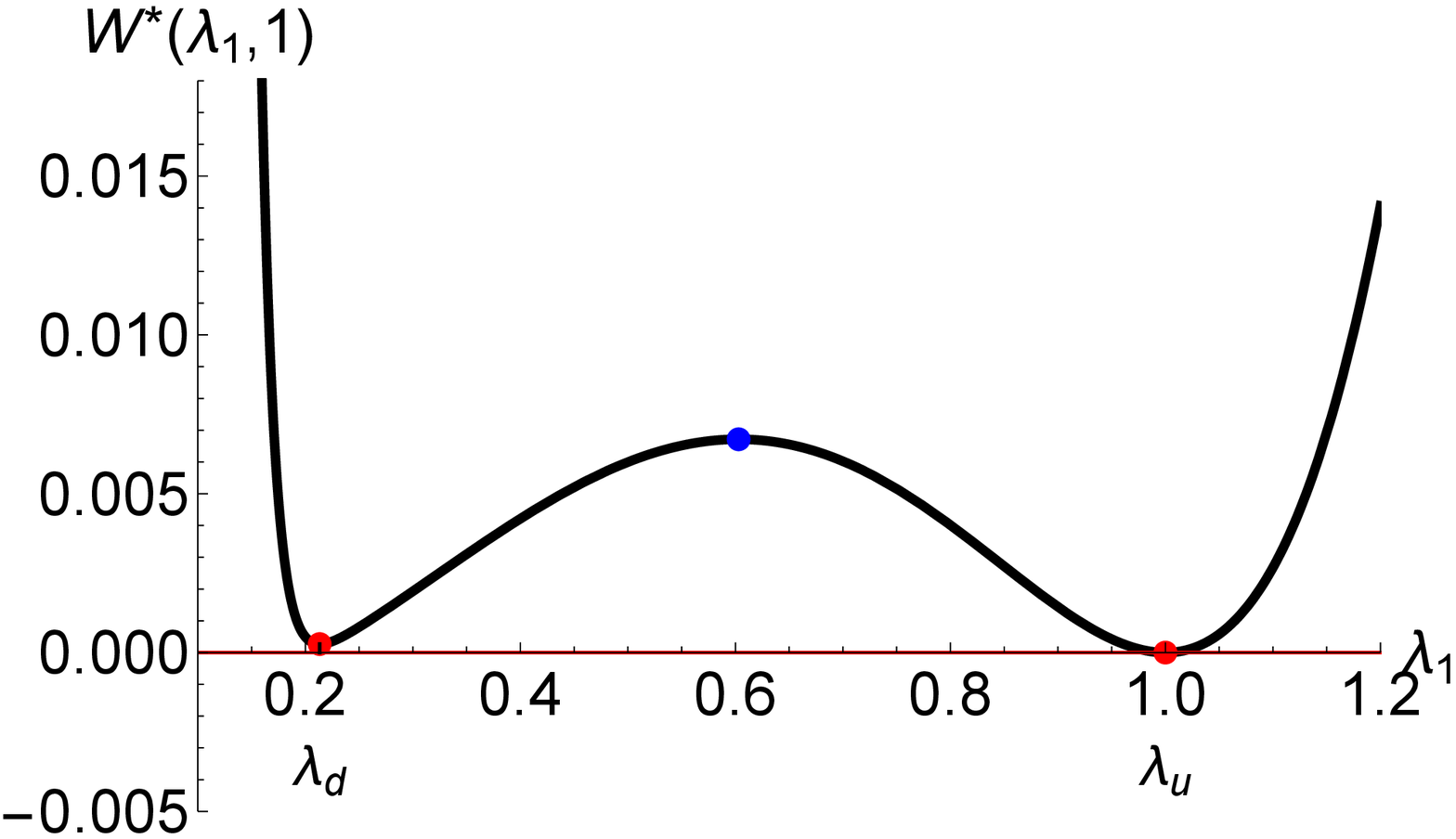}\label{sf3a}}
    \hspace{0.5cm} 
       \subfloat[]{\includegraphics[width=2.5in]{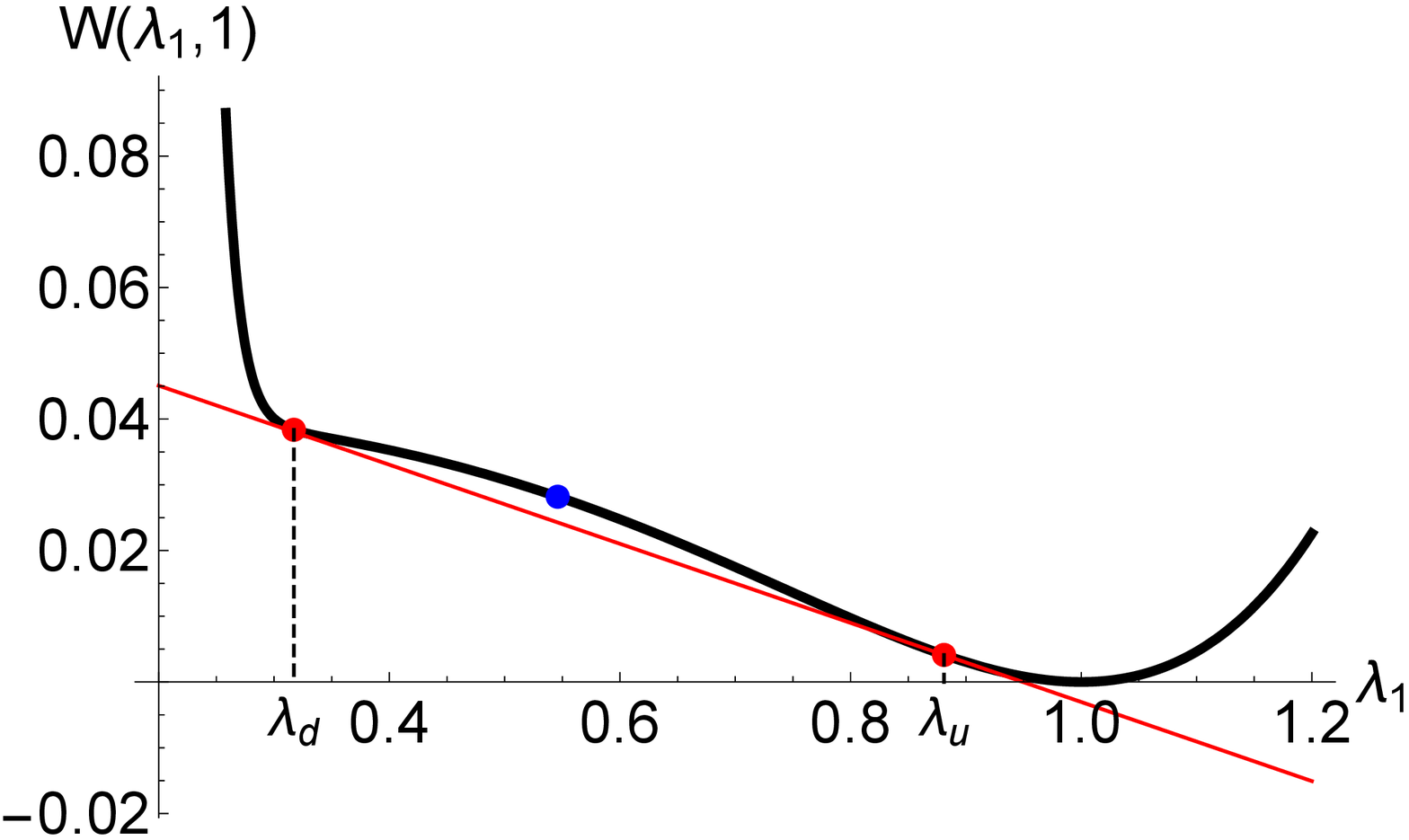}\label{sf3b}}
  \caption{\tred{Bistability and nonconvexity  in 1D uniaxial stretch (compression) of the bistable and monostable energy densities $W^*$ and $W$.
(a) The two-well structure of $W^*$ is evident from the 1D restriction to uniaxial  compression:  $W^*(\la,1)$  vs $\la$.    The two red dots correspond to densified and undensified states, $\la_d$ and $\la_u$ respectively, that are equally stable at zero stress. The blue dot  in-between is a state that also has zero stress but is unstable. (b) The monostable energy $W(\la,1)$ vs $\la$ in uniaxial stretch only has one minimum $\la=1$ observable at zero stress. At a compressive stress equal to the slope of the red line, the two red dots (tangent states) are equally stable; the blue dot is unstable. The left red state $\la_d<<1$ is densified; the one on the right  $\la_u$ is close to $1$ (undsensified).}}
  \label{sf3}
\end{figure}

\begin{figure}[h!]
    \centering
    \subfloat[]{\includegraphics[width=2in]{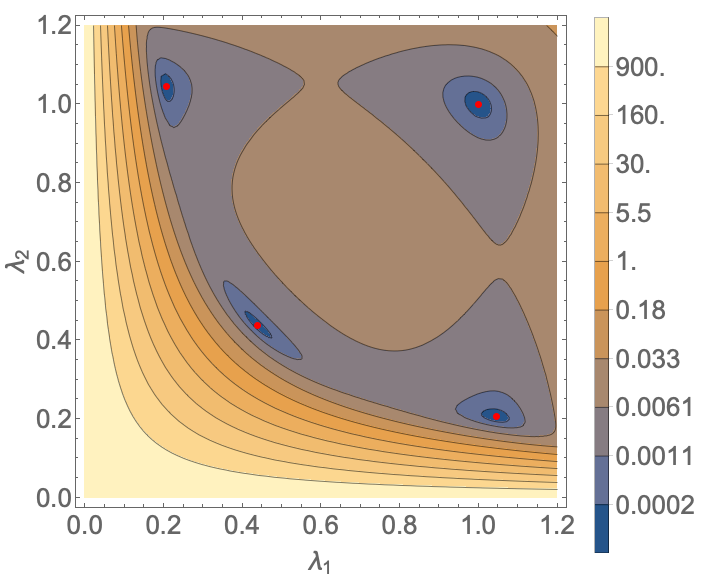}\label{sf2a}}
    \hspace{0.1cm}    
    \subfloat[]{\includegraphics[width=2in]{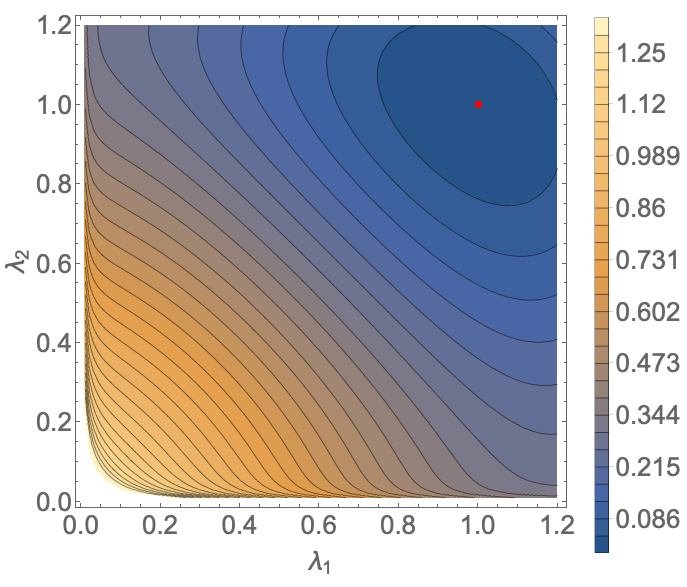}\label{sf2b}}
    \hspace{0.1cm}    
    \subfloat[]{\includegraphics[width=2in]{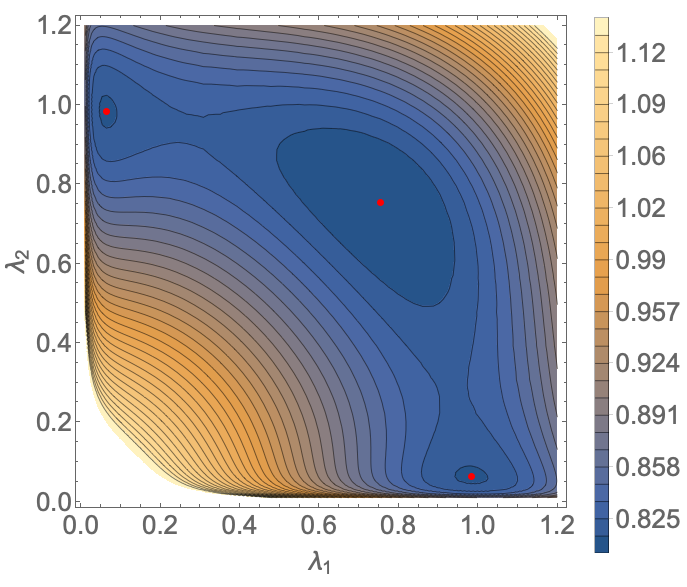}\label{sf2c}}
       \\
    \vspace{0.1cm}
    \subfloat[]{\includegraphics[width=2in]{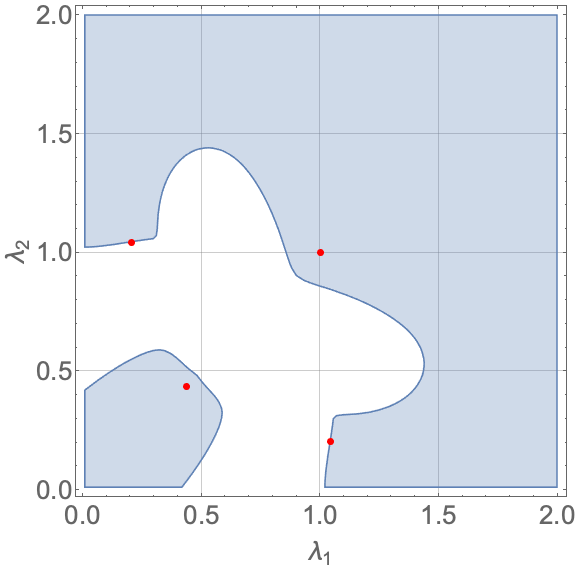}\label{sf2d}}
    \hspace{0.7cm}    
    \subfloat[]{\includegraphics[width=2in]{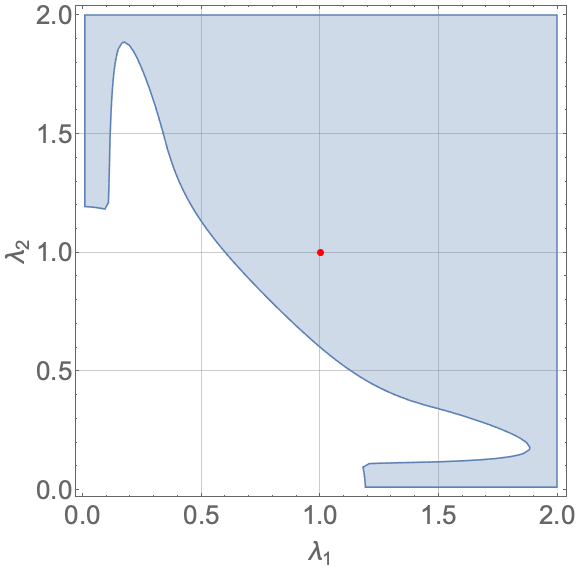}\label{sf2e}}
      \caption{Bistability and nonconvexity in 2D of the bistable and monostable 2D energy densities $W^*$ and $W$. (a)-(c) Level curves (black) and local minima (red dots) of $W^*$ and $W$  in the  $(\la_1,\la_2)$-plane. (a) The bistable energy $W^*$. (b) The monostable energy $W$. (c) The Gibbs free energy $W(\bF)-\bS\cdot\bF$  for a suitable compressive stress $\bS$. Pairs of minima of the form $(\la_1,\la_2)$ and $(\la_2,\la_1)$ correspond to the same state. (d)-(e) Domains of validity  (light blue) and failure  (white) of the strong ellipticity condition in the principal stretch $(\la_1,\la_2)$-plane: (d) of the bistable energy $W^*$ and (e) of the monostable energy $W$. }
    \label{sf2}
\end{figure}

\newpage

\clearpage
\begin{table}[h!]
  \begin{center}
   \caption{Simulations parameters for the Monostable and Bistable Elastic Energy. }
\begin{tabular}{ |p{2.5cm}||p{3cm}||p{3cm}||p{3.5cm}|   }
\hline   
 \multicolumn{4}{|c|}{ECM Model Parameters} \\
 \hline
\rowcolor{Gray} Elastic Energy  &  Equation (5.2) & Equation (5.3) &  Equation (2.4) \\
 \hline
Monostable &   $a_m = 0$ & $A =80$, $b =0.22$ & $\eps =0$ except Fig.~\ref{f1d}  see Fig. caption\\
\hline
Bistable   &   $a_m = 0.4696$ &  $A =80$, $b =0.11$ & $\eps =0$  \\
\hline 
\end{tabular}
\begin{tabular}{ |p{3.5cm}||p{4cm}||p{4.18cm}l|p{2cm}||}
\hline
 \multicolumn{4}{|c|}{Simulation Parameters} \\
 \hline
\rowcolor{Gray}Figure & Elastic energy & Cluster/Particle Stiffness  &\\
\hline 
\ref{f1d},\ref{f1f} & Monostable & $k=100$ in eq.~(2.5)& \\
\hline
 \ref{f1e}, \ref{f7a} & Monostable & $k=1$ in eq.~(2.5) & \\
\hline
 \ref{f3b}, \ref{f3d}, \ref{f3g}, \ref{f1h}, \ref{f4c}, 
\ref{f4f}, \ref{f5d}, \ref{f5h} & Bistable & $k=10$ in eq.~(5.5)&   \\
 \hline
\end{tabular}
 \end{center}
 \label{tableS1}
\end{table}

\begin{table}[h!]
  \begin{center}
   \caption{Experimental data. Distance unit: $\mu m$.}
\begin{tabular}{ |p{4.5cm}||p{3cm}||p{3cm}p{1.5cm}| }
 \hline
 \multicolumn{4}{|c|}{Experiment data shown in  Fig.~\ref{f3a}. } \\
 \hline
 Temperature& $26\, ^\circ$C & $39\, ^\circ$C & \\
 \hline
 Radius of left particle   & 75.8  & 52  &   \\
 Radius of right particle &   72.6 & 49.6 & \\
 Distance between centers &227.7 &   219.1 & \\
 \hline
 \hline
 \multicolumn{4}{|c|}{Experimental data shown in  Fig.~\ref{f4}. } \\
 \hline
 Temperature& $26\, ^\circ$C& $39\, ^\circ$C &\\
 \hline
Radius    & Fig~\ref{f4a} : 50.1 &Fig~\ref{f4b} : 25.3   &  \\
Radius    & Fig~\ref{f4e} : 38.1  &Fig~\ref{f4d} : 21.7  &   \\
 \hline
\end{tabular}
\begin{tabular}{ |p{4.5cm}||p{3cm}||p{2cm}||p{2cm}l|p{3cm}|| }
\hline
 \multicolumn{5}{|c|}{Experimental data shown in  Figs~\ref{f5a}-\ref{f5c}. } \\
 \hline
 Temperature& $\ref{f5a}: 26\, ^\circ$C& $\ref{f5b}: 39\, ^\circ$C & $ \ref{f5c}: 26\, ^\circ$C & \\
\hline
 Radius of left particle   & 38.5  &19  & 36.9 &  \\
 Radius of right particle &   39.6  & 19 &  38.8 &\\
 Distance between centers &189.3 &   185.9 & 187.4& \\
 \hline
\end{tabular}
 \end{center}
 \label{tableS2}
\end{table}

\end{document}